\documentclass[acmsmall,nonacm,screen]{acmart}

\usepackage{xspace}
\usepackage{subfigure}
\usepackage{cleveref}
\usepackage{multirow}
\usepackage{diagbox} 

\AtBeginDocument{%
  }

\acmConference[Preprint Under Review]{}{Sep,2024.}

\begin{document}

\title{Mutual Theory of Mind in Human-AI Collaboration: An Empirical Study with LLM-driven AI Agents in a Real-time Shared Workspace Task}


\author{Shao Zhang}
\authornote{Both authors contributed equally to this research.}
\email{shaozhang@sjtu.edu.cn}
\affiliation{%
  \institution{Shanghai Jiao Tong University}
  \city{Shanghai}
  \country{China}
}

\author{Xihuai Wang}
\authornotemark[1]
\email{leoxhwang@sjtu.edu.cn}
\affiliation{%
  \institution{Shanghai Jiao Tong University}
  \city{Shanghai}
  \country{China}
}

\author{Wenhao Zhang}
\email{wenhao\_zhang@sjtu.edu.cn}
\affiliation{%
  \institution{Shanghai Jiao Tong University}
  \city{Shanghai}
  \country{China}
}

\author{Yongshan Chen}
\email{chenyongshan@sjtu.edu.cn}
\affiliation{%
  \institution{Shanghai Jiao Tong University}
  \city{Shanghai}
  \country{China}
}

\author{Landi Gao}
\email{ytlkxglgld@sjtu.edu.cn}
\affiliation{%
  \institution{Shanghai Jiao Tong University}
  \city{Shanghai}
  \country{China}
}

\author{Dakuo Wang}
\email{d.wang@northeastern.edu}
\affiliation{%
  \institution{Northeastern University}
  \city{Boston}
  \country{United States}
}

\author{Weinan Zhang}
\email{wnzhang@sjtu.edu.cn}
\affiliation{%
  \institution{Shanghai Jiao Tong University}
  \city{Shanghai}
  \country{China}
}

\author{Xinbing Wang}
\email{xwang8@sjtu.edu.cn}
\affiliation{%
  \institution{Shanghai Jiao Tong University}
  \city{Shanghai}
  \country{China}
}

\author{Ying Wen}
\authornote{Corresponding author.}
\email{ying.wen@sjtu.edu.cn}
\affiliation{%
  \institution{Shanghai Jiao Tong University}
  \city{Shanghai}
  \country{China}
}

\renewcommand{\shortauthors}{Zhang and Wang, et al.}
\renewcommand{\shorttitle}{Mutual Theory of Mind in Human-AI Collaboration}
\begin{abstract}

Theory of Mind (ToM) significantly impacts human collaboration and communication as a crucial capability to understand others. When AI agents with ToM capability collaborate with humans, Mutual Theory of Mind (MToM) arises in such human-AI teams (HATs).
The MToM process, which involves interactive communication and ToM-based strategy adjustment, affects the team's performance and collaboration process. To explore the MToM process, we conducted a mixed-design experiment using a large language model-driven AI agent with ToM and communication modules in a real-time shared-workspace task\footnote{To facilitate the reproducibility of experiments and promote future research in human-AI teams, we will open-source our agent and environment implementation. In compliance with the principle of anonymity, we will include the link to our GitHub repository in the camera-ready version.}. 
We find that the agent's ToM capability does not significantly impact team performance but enhances human understanding of the agent and the feeling of being understood. Most participants in our study believe verbal communication increases human burden, and the results show that bidirectional communication leads to lower HAT performance. We discuss the results' implications for designing AI agents that collaborate with humans in real-time shared workspace tasks.

\end{abstract}

\ccsdesc[500]{Human-centered computing~Empirical studies in HCI}
\ccsdesc[500]{Human-centered computing~Collaborative interaction}
\ccsdesc[500]{Computing methodologies~Artificial intelligence}
\ccsdesc[500]{Computing methodologies~Intelligent agents}
\ccsdesc[500]{Computing methodologies~Cooperation and coordination}

\keywords{Human-AI Teams, Human-AI Collaboration, AI Agent, Mutual Theory of Mind, Large Language Models, Communication}

\begin{teaserfigure}
  \includegraphics[width=1\linewidth]{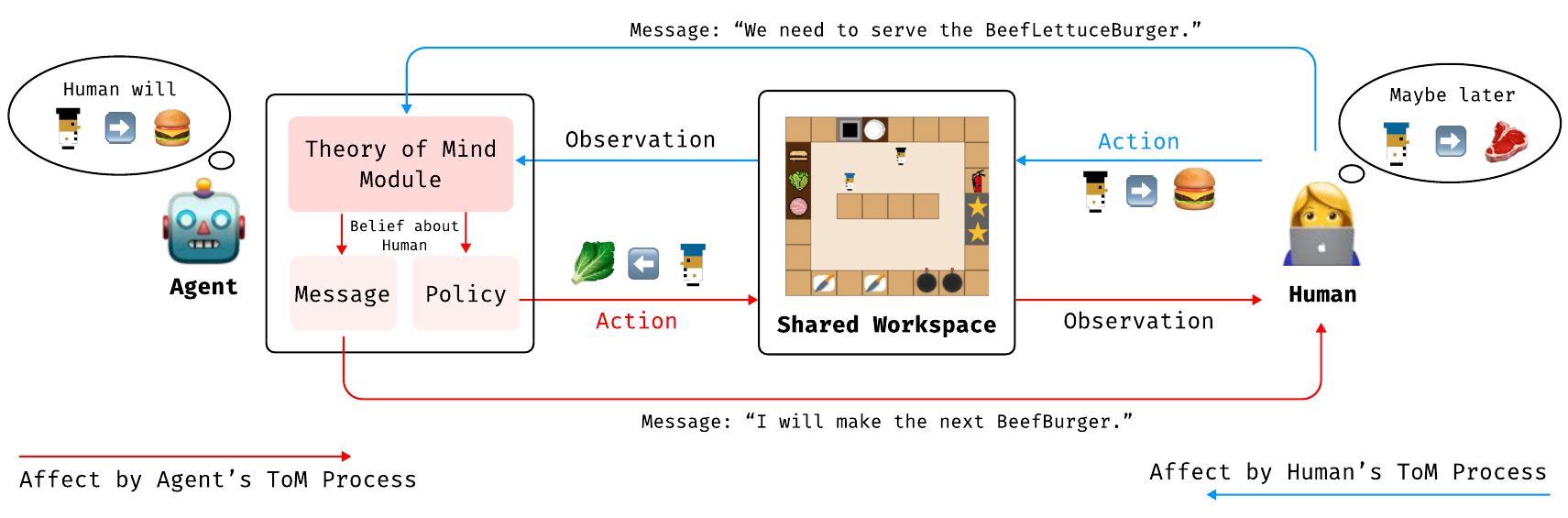}
  \caption{\textbf{The Mutual Theory of Mind (MToM) Process of Human-AI Collaboration in a Shared Workspace.} 
    We used scenarios derived from the Overcooked game to illustrate this MToM process. 
    In this example, the human controls the \textbf{black} hat chef, and the agent controls the \textcolor{blue}{\textbf{blue}} hat chef. 
    Humans and agents act in a shared workspace to complete interdependent tasks, making independent decisions while using the Theory of Mind (ToM) to infer each other's state. They observe actions as implicit communication and use messages for explicit verbal communication. We label the communication pathways shaped by ToM, as the MToM process influences explicit communication, decision-making, and behavior. Changes in agent behavior affect human inferences and decision-making, and the reverse is also true.}
  \Description{}
  \label{fig:tom}
\end{teaserfigure}

\maketitle

\section{Introduction}

\textbf{AI agent} refers to an AI-driven system or entity that can perceive the environment and make decisions based on the perceived information to achieve specific goals independently \cite{russell2016artificial}.
With the advancement of technology, an increasing number of AI agents are being deployed in various scenarios to complete tasks, including medical diagnosis \cite{vicari2003multi,iantovics2008agent,tudor2020conversational,10.1145/3613904.3642343}, scientific research \cite{m2024augmenting,aher2023using,swan2023math,guods,zhang2023geodeepshovel}, and industry applications \cite{baratta2023human,harada2023behavior,peng2023ecological}. 
As large language models continue to evolve, the range of tasks that AI agents can handle is expanding, and they are increasingly being applied in scenarios such as home robots \cite{brohan2023can,wu2023tidybot}, gaming \cite{liu2024slow,wang2024voyager}, and education \cite{zhang2024simulating,sonlu2024effects}. 
Humans and AI agents collaborate in these scenarios in the same space and can observe each other's actions, which is considered a shared workspace setting \cite{vildan2024auto,dourish1992awareness}.
These developments have also led AI agents to transition from independently performing tasks to collaborating with humans and form \textbf{human-AI teams (HATs)}, where humans and AI are recognized as unique contributors, role-specific, and working together toward a common goal \cite{o2022human}.

\textbf{Theory of Mind (ToM)} is an important ability of human beings to infer mental states, intentions, emotions, and beliefs of others for the dynamic adjustment of behaviors \cite{premack1978does,neil2018tom}, which has been applied in many AI agent frameworks \cite{DBLP:conf/acl/ZhangTWW0HTLZ024,wen2018probabilistic,langley2022theory}.
When closely collaborating with humans in the shared workspace setting, ToM plays a more critical role in the AI agent framework  \cite{10.1145/3597512.3597514,hiatt2011accommodating}.
Researchers use ToM to help AI agents understand, infer, and predict human behavior, enabling them to dynamically adjust their strategies to achieve better team performance \cite{neil2018tom,hiatt2011accommodating}. 
From a human perspective, humans also build mental models of AI agents through the ToM capability in the human-AI collaboration process \cite{gero2020mental}.
Humans typically expect AI agents to align with their anticipated capabilities or roles \cite{10.1145/3544548.3581340}, and they attribute mental states to AI agents accordingly \cite{10.1145/3527188.3561925,10.1016/j.chb.2019.04.001}.
When human beings are interacting with an agent with ToM capability, \textbf{Mutual Theory of Mind (MToM)} framework, which refers to a constant process of reasoning and attributing states to each other, is considered the analysis of the collaboration process in some studies \cite{wang2024tom,weisz2024expedient,wang2022mutual}.

In the study of human cognitive activities, ToM processes are considered closely related to communication \cite{meltzoff1999origins}. 
People use their ToM abilities to infer others' mental states and intentions, which helps them decide whether the communication is necessary and what the content and style of that communication should be \cite{fussell1998teamcomm}.
This effect is even more pronounced in team collaboration. 
People determine their plans based on their partner's movements and verbal communication, and they use communication to coordinate with their partners to achieve alignment \cite{krych2007think}.
The interactivity of communication, which refers to both parties being able to simultaneously send and receive messages \cite{doi:10.1080/00913367.2002.10673674}, might be influenced by the MToM process due to its bidirectional characteristic.
Some studies suggest that bidirectional communication is shaped by the MToM process, leading to changes in both parties' mental models throughout long-term communication \cite{wang2022mutual,wang2021mtom}.
The team's communication process, including both verbal and non-verbal communication, and individual ToM capabilities are interdependent.
In such a complex MToM process shown in \Cref{fig:tom}, it is necessary to explore verbal communication interaction and how both parties engage in their own ToM processes.

Previous research has explored the significant impact of the interactivity of communication on team collaboration, including the communication direction \cite{zahra2021direction} and style \cite{julia2022bidirectional}.
However, since communication plays a role in building human trust \cite{le2023trust,sharma2024would} and shaping social perceptions \cite{zahra2020social,zhang2020ideal}, the studies on communication on HATs did not consider that the AI agent has a ToM capability in the HAT.
Since agents and humans possess ToM abilities, psychological processes and team communication become more complex when the MToM exists \cite{wang2024tom,weisz2024expedient}.
We still lack exploration of the impact of the MToM process on HATs, particularly with regard to how communication interactivity and an agent's ToM capability influence the cooperation process within a HAT in a shared workspace setting.

In this paper, we aim to investigate the communication and collaboration process in the MToM between humans and AI agents in a shared workspace.
We first develop an AI agent equipped with ToM and communication capabilities via LLM.
Based on the widely adopted HAT collaboration benchmark, Overcooked \cite{carroll2019utility,strouse2021fcp,Yang23Cole,li2024tackling,yu23hsp,gymcooking}, we implement a practical HAT task in the shared workspace setting. 
In this task, a human and an AI agent work together in a kitchen to prepare different burgers.
In this shared workspace setting, the actions between the human and AI agent have dependencies.
Moreover, the real-time requirement of the task further asks the HAT to understand each other's intentions \cite{le2023trust}.
Our research employs a mixed experimental design of $4 \times 2$ (4 communication level $\times$ 2 agent ToM capability level). This design allows us to thoroughly explore the effects of the agent's ToM capability and communication interactivity as independent variables on HATs.
Specifically, our research questions are as follows:
\begin{itemize}
    \item RQ1: How does the MToM process including communication interactivity and individual ToM capabilities influence the overall team performance of HATs? 
    \item RQ2: How does the MToM process including communication interactivity and individual ToM capabilities affect the team collaboration process?
    \item RQ3: How do humans perceive AI teammates in the MToM process?
\end{itemize}

We conducted an online experiment (n = 68) using the shared workspace collaboration task we designed. 
The results show that HAT team performance in real-time tasks is affected by communicative interactivity, with performance under bidirectional communication being lower than under other communication interactivity conditions. 
While the agent's ToM capability did not significantly impact the team's objective performance, it did affect the collaboration process.
Moreover, participants tend to overlook verbal communication due to the operational burden of sending or receiving messages in the team collaboration process.
As for human perception, participants' understanding of the agent through ToM relied more on the agent's behavior than verbal communication. 
When the agent could coordinate its actions with humans, participants tended to believe the agent understood them.

These results indicate that due to the presence of MToM, nonverbal and implicit communication via behaviors can be just as effective as verbal communication for human-AI collaboration in real-time shared workspace tasks.
We further discussed humans' patterns of understanding agents and their perception of agents' ToM capability to provide information about the human ToM process within MToM.
We hope that adopting the MToM perspective to understand human-AI collaboration in shared workspace settings can provide new insights for future AI agent design. 
We open-source our experiment platform and LLM-driven agent\footnote{In compliance with the principle of anonymity, we will include the link to the GitHub repository in the camera-ready version.} for future study on MToM in human-AI collaboration.

\section{Related Works}

With continuous technological advancements, AI agents are gradually becoming teammates collaborating with humans in shared workspaces \cite{christoforou2020overview,faccio2023human}. 
In shared workspace settings, AI agents can independently take on some traditional human tasks and roles, contributing to the team's success \cite{carroll2019utility,devin2016implemented,vildan2024auto}.
Due to the human and the AI agent being both recognized as a unique, role-specific contributor working collectively towards a common goal, the human and AI agent form a Human-AI Team (HAT) \cite{o2022human}. 
This section presents the related works of Theory of Mind and Communication in Human-AI Teams (HATs) to provide the context of these two factors in HATs research and their relationships. 
\textbf{In our study, we consider that communication and individual ToM processes mutually influence each other, and the interactivity of communication is part of the MToM process.}
We further present the current AI technology related to the agent's capabilities in communication and ToM to provide the background of the AI agent that was used in our experiments.

\subsection{Theory of Mind in Human-AI Teams}

Theory of Mind refers to an important ability of human beings to infer mental states, intentions, emotions, and beliefs of others for dynamic adjustment of behaviors \cite{premack1978does,baron1999evolution,baron1985does}.
Using shared plans and goals as the foundation for collaborative task completion, ToM enables humans to recognize and adjust plans to achieve cooperation \cite{baron1997mindblindness,carruthers1996theories}.
These ToM processes and outcomes impact human collaborative performance and overall team performance.

The AI community uses ToM to build AI agents that can infer partners' intention \cite{wen2018probabilistic,devin2016implemented}, including partner agents and humans \cite{hiatt2011accommodating,10.1007/s10458-009-9093-x,10.1016/j.procs.2021.09.122}.
Many studies suggest that the ToM is a crucial factor for AI in achieving dynamic autonomy adjustment \cite{vildan2024auto}, which significantly impacts team performance~\cite{gero2020mental}.
ToM can help improve understanding of which tasks are suitable for which teammates \cite{10.1145/3544548.3580794,steyvers2022bayesian} and enable better task allocation among them \cite{10.1145/3544548.3580983}, leading to more effective coordination.
At the same time, researchers are also studying human perceptions and reactions when machines exhibit behaviors that suggest an understanding of human capabilities \cite{wang2024tom}.
From the human perspective, when collaborating with AI, humans with good ToM capability will naturally have a mental model of AI \cite{gero2020mental,bansal2019beyond,liang2019implicit}.
Based on the mental model, people tend to attribute mental states to AI teammates~\cite{10.1145/3544548.3580794,10.1016/j.chb.2019.04.001,10.1145/3597512.3597514} and expect AI to perform specific roles in interactions that align with their expectations and the mental models of AI teammates \cite{10.1145/3544548.3581340}.

When a human collaborates with an agent with ToM capability, the mutual Theory of Mind (MToM) framework is considered to analyze the collaboration process \cite{wang2021mtom}.
MToM refers to a constant process of reasoning and attributing mental states to each other during interaction when humans and AI have ToM capabilities \cite{weisz2024expedient,wang2022mutual}.
In past MToM studies, language communication was used as a cue for understanding \cite{wang2021mtom,wang2022mutual}. 
However, in shared workspace settings, in addition to verbal communication, there are more actions and task dependencies between individuals, leading individuals to rely on actions and behaviors to understand others' intentions.
We still need to fully understand the MToM process in HATs in shared workspace settings, including greater immediate and extensive dependence on actions and tasks between humans and AI agents.

To understand the MToM process, we use whether the agent has ToM capability to determine whether the MToM exists in HAT. 
We used LLMs to help the agent understand humans based on their historical actions and messages, allowing the agent to build a belief about the human and adjust its actions accordingly. Additionally, we considered the impact of ToM on communication. In the next section, we discuss the role of communication in HATs and how ToM influences communication.

\subsection{Communication in Human-AI Teams}

Communication plays a key role in supporting the human-human team collaboration process, including verbal and non-verbal communication~\cite{key1980relationship,fussell1998teamcomm,mcneese2015articulating,lee2012loosely,10.1145/3492832}.
Verbal communication is typically a direct and explicit exchange of information \cite{jakobson1972verbal,zhang2023comm}, while non-verbal communication is more implicit and requires understanding based on mental states and additional contextual information \cite{mehrabian2017nonverbal,argyle1972non}.

Similar to human communication \cite{lea1992paralanguage,10.1145/3613904.3642860,fleming1990multiple}, communication in HATs is also the primary factors influencing human perception on the AI teammates.
Many studies indicate that different factors of communication have impacts on humans' trust on AI agents \cite{le2023trust,sharma2024would}.
Research on the role of communication in HATs has also examined the quantity and frequency of communication~\cite{o2022human,doi:10.1177/15553434211017354}.
Communication in HATs is typically less than in the human-human team \cite{7497782}.
However, some studies have found that AI agents that disclose more information to humans and provide more explanations via non-verbal communication are not always associated with better decision-making or higher human trust on the AI agent~\cite{le2023trust}.
Communication also influences the social behavior of AI and human preferences~\cite{lu2021more}.
Some studies have explored the verbal and non-verbal communication strategies that humans prefer AI teammates to use in team environments, finding that humans tend to prefer proactive communication from teammates and expect them to respond \cite{zhang2020ideal,zhang2023comm}.

Since communication is one of the primary means of information exchange within teams, its interactivity has received significant attention \cite{zahra2021direction,oliver2019communication}. 
Interactive verbal communication is typically considered bidirectional \cite{doi:10.1080/00913367.2002.10673674}, meaning that messages can be sent and received simultaneously \cite{doi:10.1177/14614440022225751}.
\citet{zahra2021direction} reports when AI agent responds to human cues, humans perceive the AI agent as more intelligent.
In the presence of bidirectional communication, the communication style can affect the objective task load on humans. 
Specifically, communication styles that involve feedback can increase the human objective task load~\cite{julia2022bidirectional}.

Verbal and non-verbal communication is also influenced by the Theory of Mind (ToM) process in human-AI collaboration.
In HATs, humans and AI agents typically need to understand the status of their teammates~\cite{jiang2023situation}, as well as what their teammates know and plan \cite{demir2017team}, to facilitate cooperation.
Communication is regarded as an important channel to help HATs exchange information~\cite{zhang2023comm}, which can be sent in verbal and non-verbal ways.
Humans use ToM to infer the mental states and intentions of others, which directly affects whether they choose to communicate, the emotional tone of the communication, the way of communication, and the content of the communication.
The amount of communication in HATs is positively correlated with stronger shared mental models, which leads to better team performance \cite{DEMIR2020102436}.
In current HATs research, studies indicates how an agent's ToM abilities can enhance its social attractiveness in communication tasks \cite{peters2006designing,benninghoff2013theory,10.1145/3597512.3597514}.
In studying the MToM process, the mutual influence between communication and individual ToM processes makes communication an essential factor that cannot be overlooked in the MToM process \cite{wang2022mutual}.

Unlike previous studies focuses more on verbal communication with ToM, we considered both verbal and non-verbal communication processes in the MToM framework.
In our study, we explore how communicative interaction affects the MToM process. We examined HAT scenarios with bidirectional communication, unidirectional communication, and no communication. Given the  inherent presence of implicit non-verbal communication, our between group conditions are on verbal exchanges between humans and AI.
We also studied non-verbal communication by manipulating the agent’s ToM abilities as a within-subject factor, assessing how the human’s ToM perceives the agent and affects communication and decision-making. Similarly, we integrated communication into the agent’s ToM control framework. The agent’s ToM considers both human actions and language, shaping its communication and decisions.

\subsection{LLM-driven Theory of Mind AI Agents}

Considering the Theory of Mind's (ToM) importance in human-human interaction and collaboration, the AI community has developed many AI agent frameworks based on ToM to improve AI agents' capability to interact with humans and other agents \cite{wen2018probabilistic,langley2022theory}.
In multi-agent system research, researchers primarily focus on how to construct agents capable of collaborating with any unseen partners (including humans) as generalization problems \cite{wang2024zscevalevaluationtoolkitbenchmark,albrecht2018autonomous}.
Considering the problem as the Ad Hoc teamwork scenario \cite{albrecht2018autonomous}, researchers use many methods based on ToM to help agents adapt to different partners, including intention sharing \cite{guan2023efficient}, classifying types of human teammates \cite{zhang2023fast}, using latent variables for inference \cite{ma2022elign}, and employing human behavior data to model humans \cite{carroll2019utility}.
Most human modeling approaches assume that human behavior patterns are static and unchanging, leading to limitations in the AI agent's ToM process, as it cannot make real-time inferences about humans~\cite{lindner2022humans}.

In the past, agents built using reinforcement learning (RL) could only achieve team coordination through non-verbal communication \cite{ijcai2021p0466,le2023trust} or network parameter sharing between agents \cite{wang2023order}, making it difficult to communicate with humans. 
Some works have attempted to enable language communication with the RL agents with language model \cite{9889428,brohan2023can,meta2022human}, but still facing the grounding problem \cite{kenthapadi2024grounding}.
This issue also made it challenging to study communication processes in many HATs research using real AI technology \cite{zhang2023comm,DEMIR2020102436}. 
With the emergence of large language models (LLMs), AI agents' communication capabilities have improved significantly.
The advent of techniques such as Chain of Thought (CoT) \cite{wei2022chain} and ReAct \cite{yao2022react} has further advanced the capabilities of LLMs in reasoning and decision-making \cite{wang2024voyager}. 
Autonomous agents driven by LLMs are beginning to emerge, capable of performing real-world tasks, such as gaming \cite{wang2024voyager,bara-etal-2021-mindcraft,gong-etal-2024-mindagent}, online shopping \cite{ma2023laser,feng2024extremely,yao2022webshop}, and housekeeping \cite{han2024llm,li2024llm}.
Many conversational agents have also incorporated ToM into their frameworks to help understand human intentions \cite{bara-etal-2021-mindcraft}, thus improving the user experience \cite{wester2024theory,wang2021mtom}.
Since simulations involving multiple LLMs cooperating have confirmed the LLMs' ability to understand human behavior and adjust their actions accordingly \cite{park2023generative}, LLMs have also been used to adapt human behavior via ToM \cite{strachan2024testing,bara-etal-2021-mindcraft,wester2024theory,sumers2023cognitive}.

Many studies propose frameworks for LLM-driven agents to facilitate the collaboration between LLM-driven agents and humans.
MindAgent \cite{gong-etal-2024-mindagent} constructs an LLM-based agent to achieve multi-agent coordination in Minecraft and cooperation with humans in a kitchen environment.
\citet{liu2024slow} used two LLMs to control the agent's slow-mind and fast-mind thinking, enabling real-time communication and cooperation with humans in a simulated Overcooked game.
ToM capabilities enabled by LLMs have also been used for competitive operations in board games, helping AI agents achieve victory in the poker game \cite{DBLP:conf/acl/ZhangTWW0HTLZ024}.

These technological foundations provide a viable pathway for real AI agents to be involved with communication and ToM capabilities in HAT research.
In our study, we build real LLM-driven AI agents based on GPT-4o mini \cite{openai2024gpt4omini} with communication and ToM capabilities to experiment with humans.
Our study can serve as a complement to previous research that used Wizard-of-Oz \cite{zhang2023comm,cooke2020framework} or rule-based agents \cite{vildan2024auto}, by exploring the performance and impact of real AI agents.

\section{Cooperative Task Setup and Formalization in Shared Workspace}
\label{sec: task}

To better validate the impact of MToM and communication on the cooperation process in HATs, we designed a shared cooperative workspace and tasks based on Overcooked. 
Overcooked is an important simulation environment used for human-AI collaboration \cite{carroll2019utility,strouse2021fcp,Yang23Cole,li2024tackling,yu23hsp,gymcooking} derived from the Overcooked video game\footnote{\url{https://www.team17.com/games/overcooked/}}. 

In this section, we introduce our environment layout design, the definition of cooperative tasks, the communication system design, and the specific metrics we use to measure the cooperation process.
The environment layout design and task design are closely related to MToM and the interactivity of communication. 
We explain the specific connections between the design and MToM, as well as communication.
We also formalize the task process to help understand the design of the agent and our experimental procedure.

\subsection{Layout Design}
\begin{figure}[htp]
    \centering
    \subfigure[The Layout of the Cooperative Workspace as a Kitchen.]{
    \includegraphics[height=0.35\linewidth]{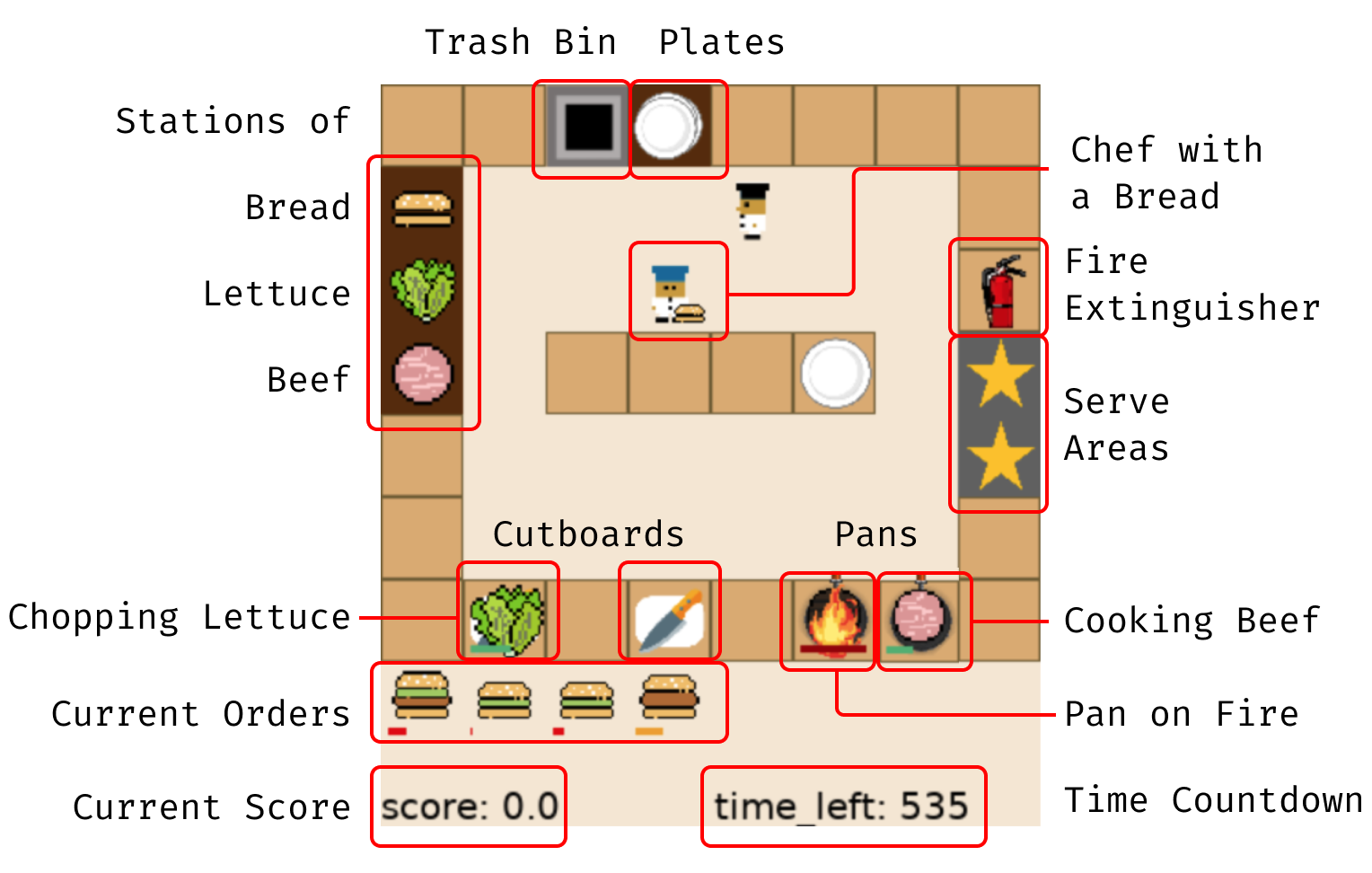}
    \label{fig:game}
    }
    \hfill
    \subfigure[Communication System in Our Task.]{
    \includegraphics[height=0.35\linewidth]{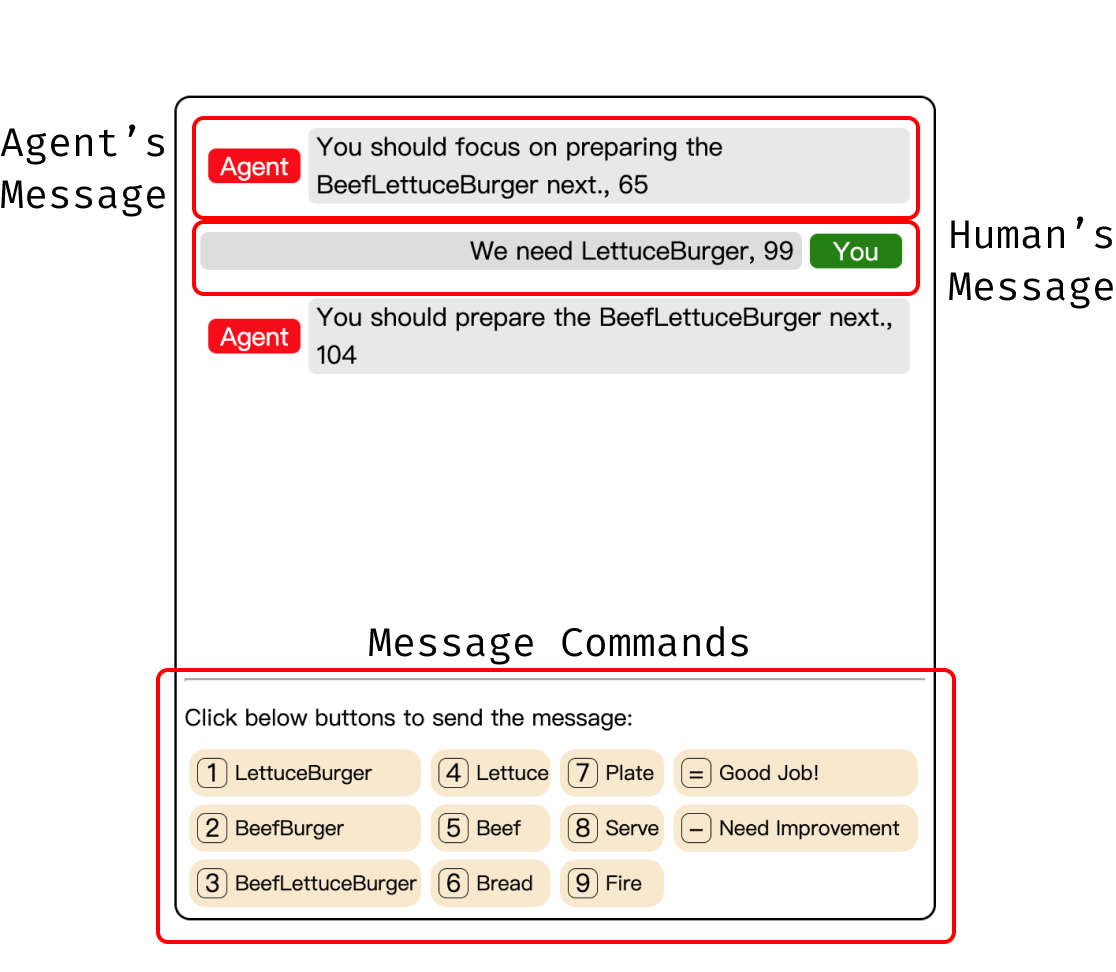}
    \label{fig:message}
    }
    \caption{\textbf{Game Layout and Communication System of the Task.}}
    \label{fig:layoutandcomm}
\end{figure}

In our study, we set the shared workspace as a kitchen.
Since we focus on the MToM process, especially the communication and individual ToM process in the HAT cooperation process, we arrange the space so that both parties must coordinate their movements to avoid collisions. 
The layout requires communication and action coordination within the team.
We redesign the basic layout \textit{Counter Circuit} from the original overcooked-ai environment \cite{carroll2019utility}, and implement it in the gym-cooking enviroment \cite{gymcooking}.
The layout features a ring-shaped kitchen with a central counter, elongated counter, and a circular path between the counter and the operational area, which is simulating a real kitchen.
In this configuration, pots, cutting board, ingredients (bread, beef, and lettuce), and serving spots are positioned in four distinct directions within the operational area, as shown in Figure \ref{fig:layoutandcomm}. 
Although the layout does not forcibly require cooperation, players may find themselves obstructed by narrow aisles, prompting the need for coordination to maximize performance.
At the same time, the circular space creates potential movement patterns (moving clockwise or counterclockwise), which can be used by humans or agents with ToM to infer their teammates’ behavior and help facilitate coordination.

\subsection{Task Design}

\begin{figure}[h]
    \centering
    \subfigure[LettuceBurger]{
    \includegraphics[width=0.4\linewidth]{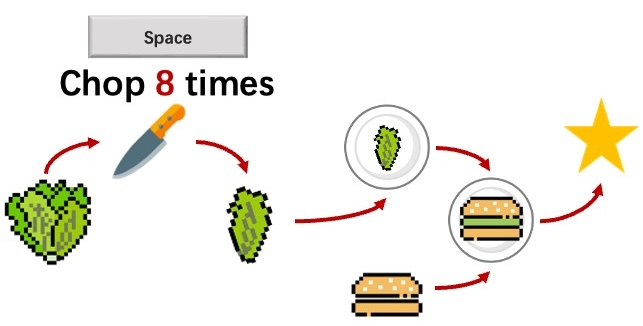}
    \label{fig:vegan}
    }
    \subfigure[BeefBurger]{
    \includegraphics[width=0.4\linewidth]{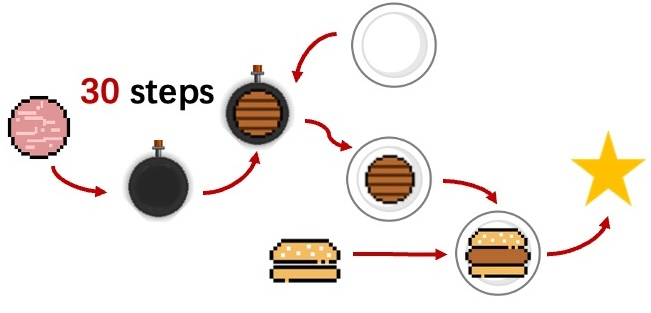}
    \label{fig:meat}
    }
    \subfigure[BeefLettuceBurger]{
    \includegraphics[width=0.4\linewidth]{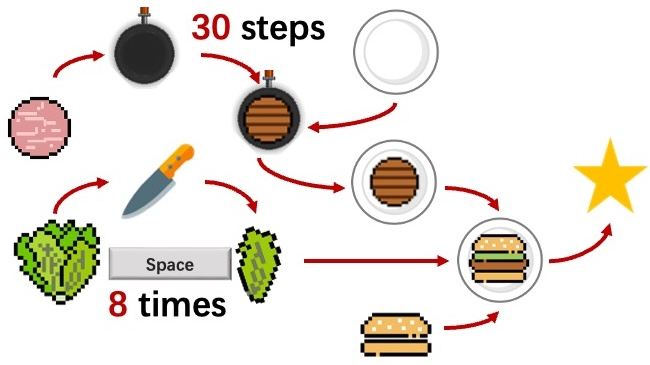}
    \label{fig:allinone}
    }
    \subfigure[Overcooked Beef]{
    \includegraphics[width=0.4\linewidth]{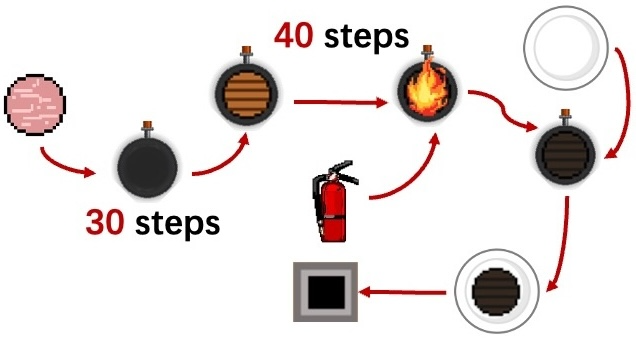}
    \label{fig:fire}
    }
    \caption{\textbf{Game Mechanism.} (a), (b), and (c) are the rules for preparing and serving burgers. (d) demonstrates the mechanism of overcooked beef and the rules for handling the fire caused by overcooked beef.}
    \label{fig:diff_level}
\end{figure}

Since we aim to explore the MToM process, we incorporate a mechanism in the burger-cooking task that requires communication for coordination. 
Ingredients for the burgers overlap, and without explicit indication, it is difficult for individuals to infer which burger their teammate is working on through a single action.
Therefore, such a task design requires the players to communicate and infer the teammate's intention on the ongoing task to prevent redundant work and effectively complete orders.
As shown in Figure \ref{fig:diff_level}, we design three burgers: LettuceBurger, BeefBurger, and BeefLettuceBurger.
Humans and agents jointly control the chefs and handle the continuously incoming burger orders in our environment. 
The task provides three ingredients for cooking burgers: bread, beef, and lettuce.
The lettuce needs to be chopped, and the beef needs to be cooked. All ingredients are assembled into a burger on a plate, with no specified assembly order. 
A LettuceBurger requires bread and chopped lettuce, a BeefBurger requires bread and well-cooked beef, and a BeefLettuceBurger requires bread, chopped lettuce, and well-cooked beef.
Each order is only available for a limited time, indicated by a countdown on the interface. 
To coordinate their teamwork effectively, both the human and the agent need to pay attention to the remaining time of each order.

In addition to the cooperation, we introduce additional failure events that will damage performance in the cooperation process.
If well-cooked beef is not promptly removed using a plate during cooking, the pan will catch fire. 
The pan becomes unusable until a team member uses a fire extinguisher to put out the fire.
Overcooked beef must be removed using a plate; otherwise, the pan will remain occupied by the overcooked beef, rendering it unusable for further cooking.
This process does not directly result in a score penalty (i.e., there is no explicit punishment), but it will impact the team's cooperation process.

\subsection{Communication System Design}
\label{subsec: human_comm}

We design a communication system for HATs within the task to control the interactivity of communication. 
In the game, we implement a dialogue module where humans could click buttons to send messages in a dialog box. 
The buttons in the communication system (as shown in Figure \ref{fig:layoutandcomm}) are designed to represent the items needed for the task, including burgers, all ingredients, and plates. 
The needs for all items are expressed as ``We need (specific item)'' to the agent.
Additionally, we include options for handling special situations (such as extinguishing fires). 
Beyond task-related information, there are also two buttons to express emotions. 
The human participants can see messages sent by the agent in the same dialog box.

\subsection{Formulation}
\label{subsec: formulation}

We formulate this scenario as a two-player decentralized Markov decision process (DEC-MDP) \cite{bernstein2002complexity}. The DEC-MDP containing one agent and one human can be formalized as $<\mathcal{S}, \{\mathcal{A}^{i}\}, \{\mathcal{A}^{h}\},\rho,\mathcal{T},r>$, where $\mathcal{S}$ is the state space, $\rho: \mathcal{S}\mapsto [0,1]$ is the distribution of the initial state $s_0$.
$\mathcal{A}^{i}$ and $\mathcal{A}^{h}$ are the action spaces of the agent and the human, and $\mathcal{A}=\mathcal{A}^{i} \times \mathcal{A}^{h}$ is the joint action space. 
$\mathcal{T}: \mathcal{S}\times \mathcal{A} \times \mathcal{S} \mapsto [0,1]$ denotes the transition probability and $r: \mathcal{S}\times \mathcal{A} \mapsto  \mathbb{R}$ is the reward function.
At time step $t$, the agent and the human take action $a^{i}_{t}$ and $a^{h}_{t}$ simultaneously.

\textbf{State}. Both the agent and the human have full access to the game states and each other's actions.
Players can directly see the status of all items in the game interface, such as the location where items are placed and their current state (e.g., beef cooking in a pan).
Players can also view the remaining game time and current score through the information displayed. 
The remaining time for each order, the progress of chopping lettuce, the process of cooking beef, and the process of extinguishing a fire are shown through progress bars.
All actions taken by teammates, the teammates' location, and the items they are holding are fully visible to each other.
Players can also view messages in real-time, including both the messages they have sent and the messages they have received, along with the corresponding timestamps.

\textbf{Action}. In this environment, the actions that the human and the agent can take to control the chefs include moving up, down, left, and right, as well as ``interact''. 
All activities such as picking up items, serving dishes, and extinguishing fires are considered as ``interact'' actions.
The specific interaction rules are illustrated in Figure \ref{fig:diff_level}. 
We denote the actions to control the chefs as $\mathcal{A}^{\text{control}}$.
The agent and the human share the same $\mathcal{A}^{\text{control}}$, while the communication actions are designed individually for them, denoted as $\mathcal{A}^{\text{comm},i}$ and $\mathcal{A}^{\text{comm},h}$ respectively. The communication action space for the agent $\mathcal{A}^{\text{comm},i}$ consists of any possible sentences less than 10 words from an LLM, while the communication action space for the human $\mathcal{A}^{\text{comm},h}$ consists of 11 message templates.
Details about the communication system are presented in \Cref{subsec: human_comm} and \Cref{subsec: agent_comm}.

\textbf{Reward}. The scores for completing the three different types of orders vary and serving the wrong burger or missing an order will result in a penalty. The specific rewards are detailed in \Cref{tab:reward}.

\begin{table}[]
    \centering
        \caption{\textbf{Rewards in Game.}}
    \begin{tabular}{l|c}
    \toprule
        \textbf{Event} & \textbf{Rewards} \\
        \midrule
        Serve a LettuceBurger & +15\\
        Serve a BeefBurger & +20\\
        Serve a BeefLettuceBurger & +25\\
        Serve a Wrong Burger (or Something not a Burger) & -10\\
        Miss an order & -10\\
\bottomrule
    \end{tabular}

    \label{tab:reward}
\end{table}

\subsection{Objective Metrics}
To capture the dynamics of the cooperation process and observe the impact of MToM and communication on collaboration, we defined a set of task-related metrics to measure team performance and the team collaboration process.

\textbf{Task Score.} 
We define the team's objective performance as the team's score. The game is set to be completed in 500 time-steps, and the specific reward calculation method is shown in Table \ref{tab:reward}.

\textbf{Contribution Rate.}
We first define key task events $KE$ to capture which team member completes specific tasks. 
Based on the burger-making process, each of the three types of burgers involves certain essential and non-repeatable events. 
For example, making a BeefBurger consists of 5 key events: Cook Beef, Use Beef, Use Bread, Use Plate, and Serve, and each key event will only be completed once.
The completion of key events is marked by specific ``interact'' actions, which we define as Key Actions.
We map the key actions with the key events in \Cref{tab:event_action}.
In this mapping, each key event completed by the player is counted once as the player's contribution to the overall performance.
Based on these non-repeatable key events, we can attribute the contribution to the two players by counting the key events they completed in preparing each properly delivered burger.
We define the agent's contribution ratio $CR^{i}$ as:
$CR^{i} = \frac{KE^{i}}{KE^{i}+KE^{h}} \times 100\%$, where $KE^{i}$ and $KE^{h}$ represent the key events completed by the agent and the human respectively.

\begin{table*}
    \renewcommand{\arraystretch}{1.5} 
    \caption{\textbf{The mapping from key event to key actions.}}
    \centering
    \resizebox{\linewidth}{!}{
    \begin{tabular}{l|l}
    \toprule
    \textbf{Key Events}  & \textbf{Key Actions} \\
    \midrule
    Cook Beef & \textcircled{1} Get Beef from station Put onto Pan \\
    Use Beef & \textcircled{1} Plate well-done Beef from Pan \\
    Prepare Lettuce & \textcircled{1} Get lettuce from station \textcircled{2} Put onto Cutboard \textcircled{3} Chop Lettuce \\
    \multirow{2}{*}{Use Lettuce}      & 
                                        \textcircled{1} Plate Lettuce Done from Cutboard \textcircled{2} Plate Lettuce Done from Counter \textcircled{3} Put onto Plate with BeefBurger                     
                                        \\
                                        &
                                        \textcircled{4} Put onto Plate with Bread \textcircled{5} Put Lettuce onto Plate \textcircled{6} Put Lettuce onto Plate with Beef
                                        \\
    \multirow{2}{*}{Use Bread} &         
                                        \textcircled{1} Get Bread from Station \textcircled{2} Plate Bread from Counter \textcircled{3} Put onto Plate with BeefLettuce
                                        \\
                                        & 
                                        \textcircled{4} Put onto Plate with Lettuce \textcircled{5} Put Bread onto Plate \textcircled{6} Put Bread onto Plate with Beef
                                        \\

         Use Plate & \textcircled{1} Get Plate from Station \\
                  Serve & \textcircled{1} Deliver Burger \\
    \bottomrule
    \end{tabular}

    }
    \label{tab:event_action}
\end{table*}

\textbf{Message Count.}
To study the impact of MToM on the communication process, particularly human communication preferences, we record the types and quantities of messages humans choose to send when they can communicate.
We count the number of sent messages as \textit{Message Count}.

\textbf{Failure Count.}
We use cooperation failures in the task as an auxiliary analysis of team performance. We categorize three events as cooperation failures: missing an order, serving a burger or other item not required by the current orders, and the pan catching fire. We separately record the occurrence of each event and summarize them as \textit{Failure Count}.

\section{LLM-driven Agent with Theory of Mind and Communication Module}\label{agent}

\begin{figure}
    \centering
    \includegraphics[width=0.8\linewidth]{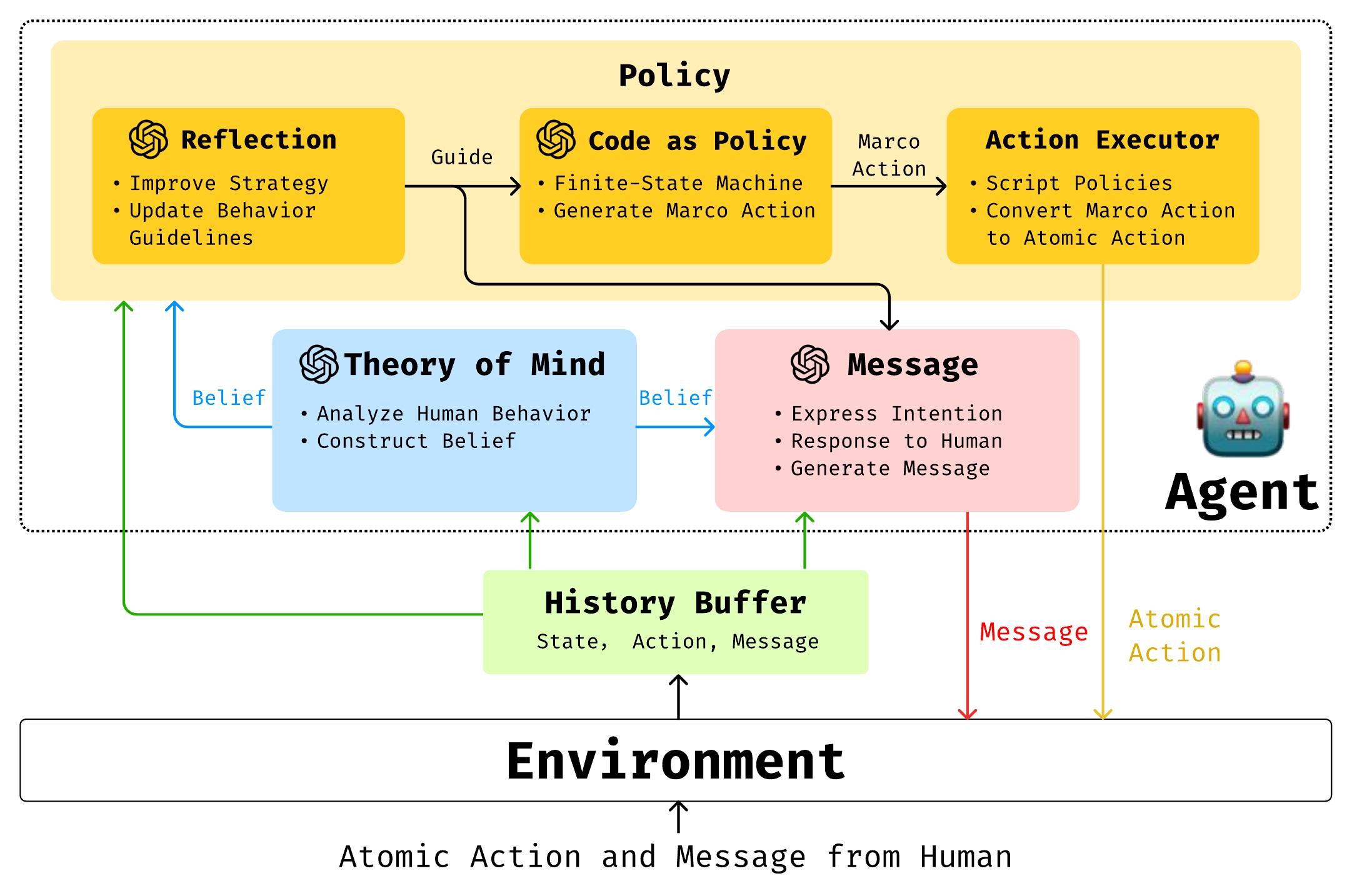}
    \caption{\textbf{Agent Framework.} The framework shows how the LLM-driven agent with Theory of Mind and communication capability takes action and sends messages to the human player. We use a history buffer to save the game history, including game state, actions, and player messages. The Theory of Mind module uses history as input to analyze human behavior. The Policy and Message modules also have the history input to understand the whole picture of the game. We summarize the process of generating action and message: (1) The Theory of Mind module analyzes human behavior and messages, then generates beliefs about humans and provides a guide for adjusting the strategy for better team coordination and communication. (2) The Policy module uses the belief from the Theory of Mind module and the history to improve the agent's strategy by continually updating behavior guidelines. It outputs an action to control the agent. (3) The Message module uses the history, the inferred belief from the Theory of Mind module, and the guidelines from the Policy module to generate the message that aligns with the agents' actions and intentions.  }
    \label{fig:framework}
\end{figure}

This section introduces our agent implementation based on GPT-4o mini \cite{openai2024gpt4omini}.
As shown in Fig. \ref{fig:framework}, we design three main modules for the agent, including \textbf{Theory of Mind}, \textbf{Policy}, and \textbf{Communication}.
The Theory of Mind module infers the intentions of human partners based on human behavior, summarizing these behaviors to guide further the agent's strategies for cooperating better with humans. 
The Policy module controls how the agent interacts in our environment and continually updates the agent's behavior policy to improve performance.
The Communication module helps the agent receive human messages to adjust the agent's behavior and send messages to the human.

\subsection{Theory of Mind Module}

With a ToM capability, individuals can form hypotheses about the mental states of others as a belief through their actions and communication history, thus understanding and predicting others' behaviors \cite{premack1978does}.
Belief in Theory of Mind refers to an individual's cognition of things, which further influences their behavior \cite{baron1985does,neil2018tom,wen2018probabilistic}. 
For example, in our Overcooked environment, if the human player believes that ``my partner will cook the beef,'' the human player might focus on other tasks, assuming that their partner will take care of the beef.
Based on the Theory of Mind mechanism, we design the Theory of Mind module, which enables the agent to establish the belief about the human, including the tendency, convention, and plan, using the human partner's behavioral history and communication messages.

Based on the formulations in \Cref{subsec: formulation}, we denote the history from time-step $0$ to time-step $t$ of the game that the agent perceives as:
\begin{align*}
\mathcal{H}_{0:t} = 
\{ 
&(s_{0}, a_{0}^{\text{control},i}, a_{0}^{\text{control},h}, a_{0}^{\text{comm},i}, a_{0}^{\text{comm},h}, r_{0}),\\
& \quad \quad \quad \quad \quad \quad \quad \quad \quad \vdots \\
&(s_{t}, a_{t}^{\text{control},i}, a_{t}^{\text{control},h}, a_{t}^{\text{comm},i}, a_{t}^{\text{comm},h}, r_{t})
\}.
\end{align*}

\begin{figure}
    \centering
    \includegraphics[width=0.8\linewidth]{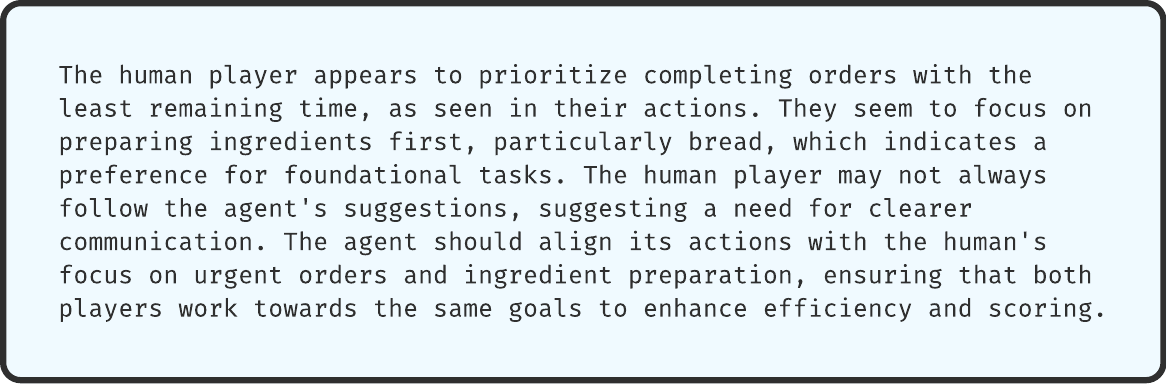}
    \caption{\textbf{An example of Belief Output from the Theory of Mind Module.} }
    \label{fig:belief}
\end{figure}

As shown in \Cref{fig:framework}, the Theory of Mind module uses the history $\mathcal{H}_{0:t}$ as an input to summarize the history, infers the conventions and tendencies of the human and explains how the agent's policy can be adjusted to coordinate better with the human player. 
The Theory of Mind module outputs the belief in natural language, as shown in \Cref{fig:belief}. 
The entire process can be formalized as follows:

\begin{equation}
    \label{eq:humaninfer}
    b^{n} = \text{LLM}\left(\mathcal{H}_{0:t_{n}}, b^{n-1}\right)~,
\end{equation}
where $t_{n}$ is the time-step when the $n$-th belief inferences is performing. The Theory of Mind module executes every $75$ time-steps.

\begin{figure}
    \centering
    \includegraphics[width=0.8\linewidth]{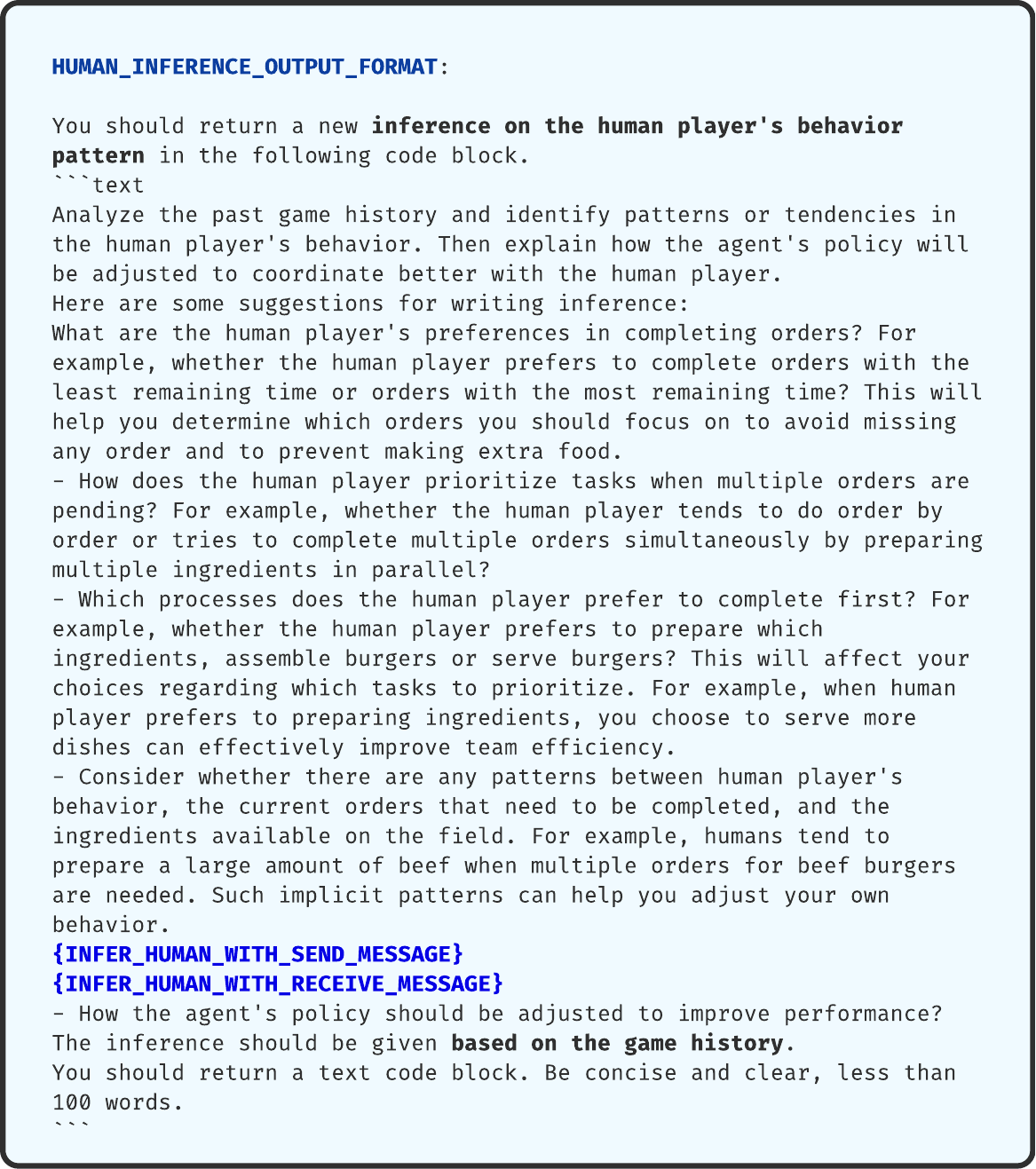}
    \caption{\textbf{Prompt for Theory of Mind Module.}}
    \label{fig:humaninfer}
\end{figure}

\begin{figure}
    \centering
    \includegraphics[width=0.8\linewidth]{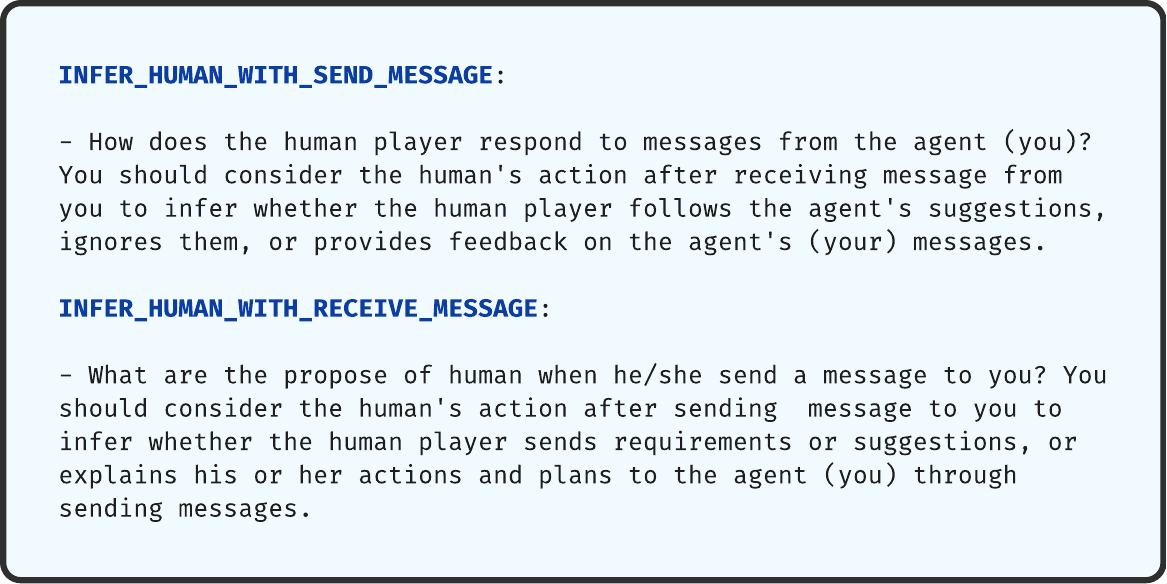}
    \caption{\textbf{Prompts as Variables in the Theory of Mind Module.} Top is INFER\_HUMAN\_WITH\_SEND\_MESSAGE and button is INFER\_HUMAN\_WITH\_RECEIVE\_MESSAGE.}
    \label{fig:humaninfer_message}
\end{figure}

The core prompts used in the Theory of Mind module are shown in \Cref{fig:humaninfer,fig:humaninfer_message}.
To implement the within-group condition in the experiment, i.e., whether the agent has ToM capabilities, we implement the Theory of Mind module as optional, and we can control whether the ToM results are included in the decision-making process, as shown in \Cref{fig:framework}.
When the condition is that the agent has no ToM capabilities, the ToM inference process (\Cref{eq:humaninfer}) will not be executed, and no ToM results are used in the Policy module and the Communication module.

Considering the communication conditions in our experiments, we use two variables, INFER\_HUMAN\_WITH\_SEND\\\_MESSAGE and INFER\_HUMAN\_WITH\_RECEIVE\_MESSAGE to control whether the agent can send message to human and whether the AI agent can use the received message to infer human in Theory of Mind module, as shown in \Cref{fig:humaninfer_message}. 
For example, if the conditions are set so that both human and agent cannot send messages and the agent has ToM capability, the INFER\_HUMAN\_WITH\_SEND\_MESSAGE and INFER\_HUMAN\_WITH\_RECEIVE\_MESSAGE will be set to empty strings to make the agent do not consider the human message and the human response to the agent's messages in the ToM process.

\subsection{Policy Module}
Our task is time-sensitive, requiring the agent and human player to make decisions and adjust their policies in time to avoid missing orders and overcooking the dishes.
Due to the real-time requirement, the agent can not directly prompt the LLM to generate its next action, which results in unacceptable latency.
Besides, LLMs are better at high-level reasoning and planning instead of generating detailed control signals \cite{wei2022chain}. 
Therefore, the agent leverages the LLM to generate high-level plans, which we call ``macro actions.'' 
Meanwhile, the agent prompts the LLM to reflect and update some behavior guidelines iteratively. 
The agent then transforms the generated plans into actable actions, including moving and interacting with objects, which we call ``atomic actions,'' through an action executor.

As shown in \Cref{fig:framework}, the Policy module consists of three parts: Code-as-Policy Generator, Policy Reflection, and Action Executor.

\subsubsection{Code-as-Policy Generator}
Given the latency in API calls with GPT-4o mini and the real-time decision-making requirements of our task, we implement an initial policy with predefined rules. This policy is further empowered by a \textit{code-as-policy} \cite{DBLP:conf/icra/LiangHXXHIFZ23} generator, which plays a crucial role in enabling the agent to effectively handle real-time emergencies.
Our initial policy is structured upon a Finite-State Machine (FSM) framework. 
This framework is instrumental in allowing the agent to operate within a set of defined states, transitions, and actions, providing a clear structure for its operations.
Each state represents the agent's particular context or situation, while environment dynamics, fixed 25-time-step intervals, and human message input trigger transitions.
The FSM structure allows the agent to complete tasks by switching between predefined states, allowing it to function without relying on external API responses.

\begin{figure}
    \centering
    \includegraphics[width=0.8\linewidth]{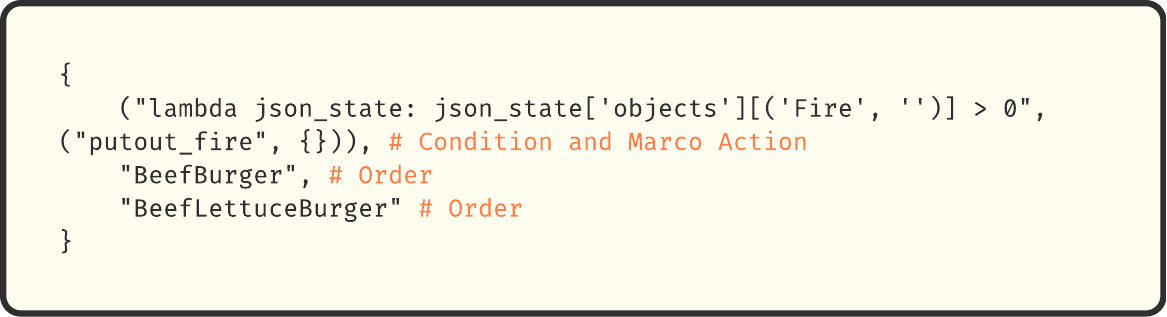}
    \caption{\textbf{Generated Code-as-Policy Example.}}
    \label{fig:code_example}
\end{figure}

However, it is impractical to exhaustively define all possible states and transitions using a finite-state machine-based rule policy, especially in complex and dynamically changing environments. To address this limitation, we incorporate a LLM to enhance the policy. When the FSM falls short in covering specific complex or unexpected scenarios, such as human messages, the LLM can generate flexible and temporary strategies, allowing the system to handle unforeseen situations. The LLM generates JSON codes to optimize the initial policy, given the recent game history and the inferred belief about the human. As demonstrated in \Cref{fig:code_example}, the LLM generates two types of code snippets: 1) Condition and Macro Action: The macro action will be executed if the condition is satisfied, where the condition is used to avoid outdated decisions caused by latency; 2) Order: The agent will prioritize this order.
The macro action generation process can be formalized as:

\begin{align*}
    \tau_{t} &= \text{LLM}\left(\mathcal{H}_{t-\lambda:t}, b^{n}\right) \\
    ma_{t} &= \pi^{m}\left( s_{t}, \tau_{t} \right)~,
\end{align*}
where $b^{n}$ represents the latest belief about human that is updated $n$ times, $\pi^{m}$ represents the latest agent policy that is updated $m$ times, $\tau_{t}$ is the generated code-as-policy and $\lambda$ is the interval the code-as-policy generator executes, which is $25$ in our experiment.

\subsubsection{Policy Reflection}

\begin{figure}
    \centering
    \includegraphics[width=0.8\linewidth]{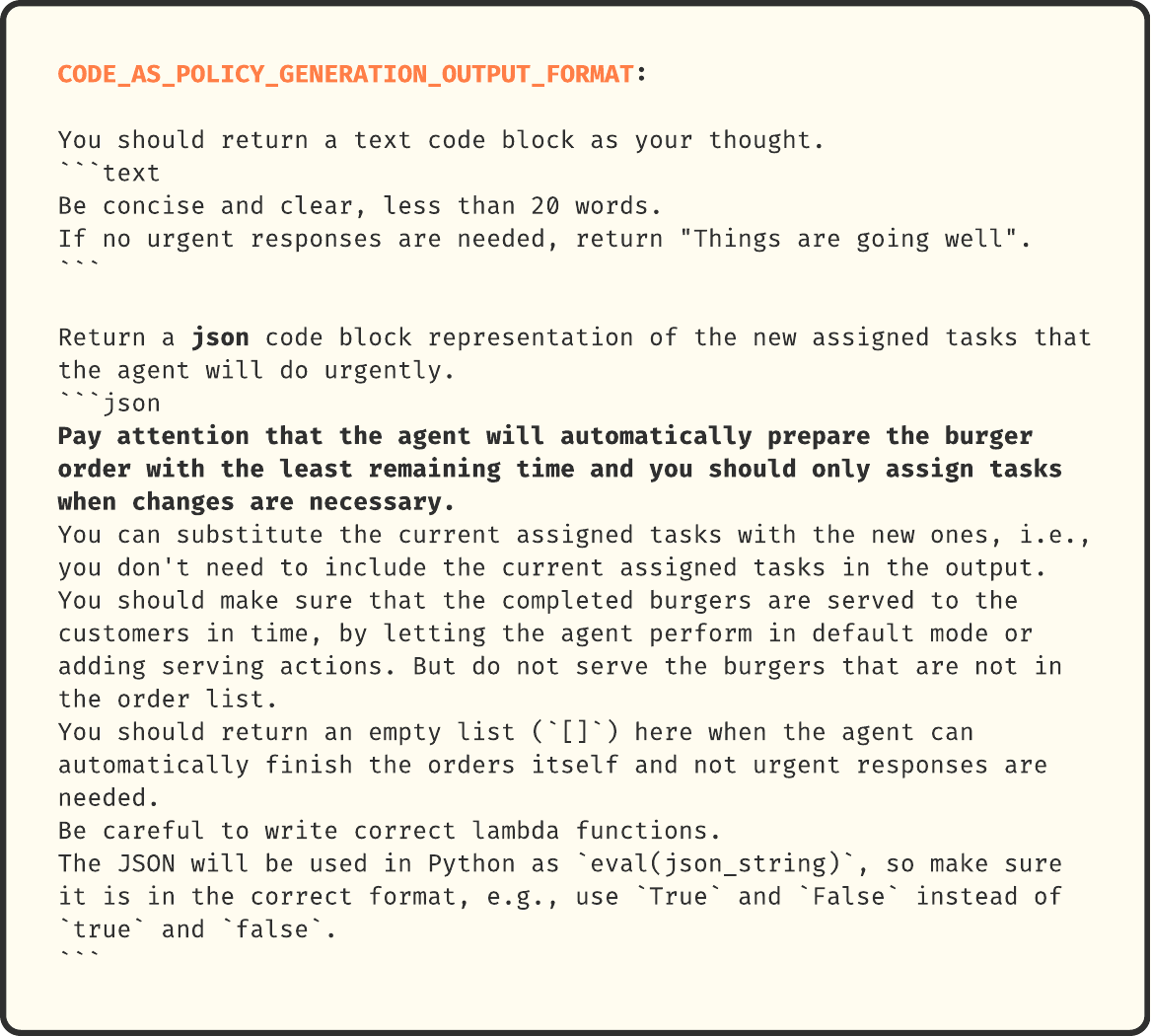}
    \caption{\textbf{Prompt for Code-as-Policy Generator.}}
    \label{fig:code_as_policy_prompt}
\end{figure}

Equipped with the FSM-based initial policy and Code-as-Policy generator, the agent already has the basic capability to interact with humans in our environment. 
However, the agent needs to improve its policy in such a long-horizon interaction process for higher performance.
Using the low-overhead in-context learning approach, the agent maintains and iteratively updates a ``Behavior Guideline,'' which summarizes the improvements to current policy.
After a ``Behavior Guideline'' is generated, the agent incorporates the ``Behavior Guideline'' into the current policy.
The entire process can be formalized as:

\begin{align*}
    \mathcal{B}^{m} &= \text{LLM}\left(\mathcal{H}_{0:t_{m}}, b^{n}, \mathcal{B}^{m-1}\right), \\
    \pi^{m} &= \pi^{m-1} \cup \mathcal{B}^{m}~,
\end{align*}
where $m$ means the reflection process executes $m$ times, $b^{n}$ is the latest inferred belief about human, $\mathcal{B}^{m}$ is the ``Behavior Guideline'' that is updated $m$ times.

\begin{figure}
    \centering
    \includegraphics[width=0.8\linewidth]{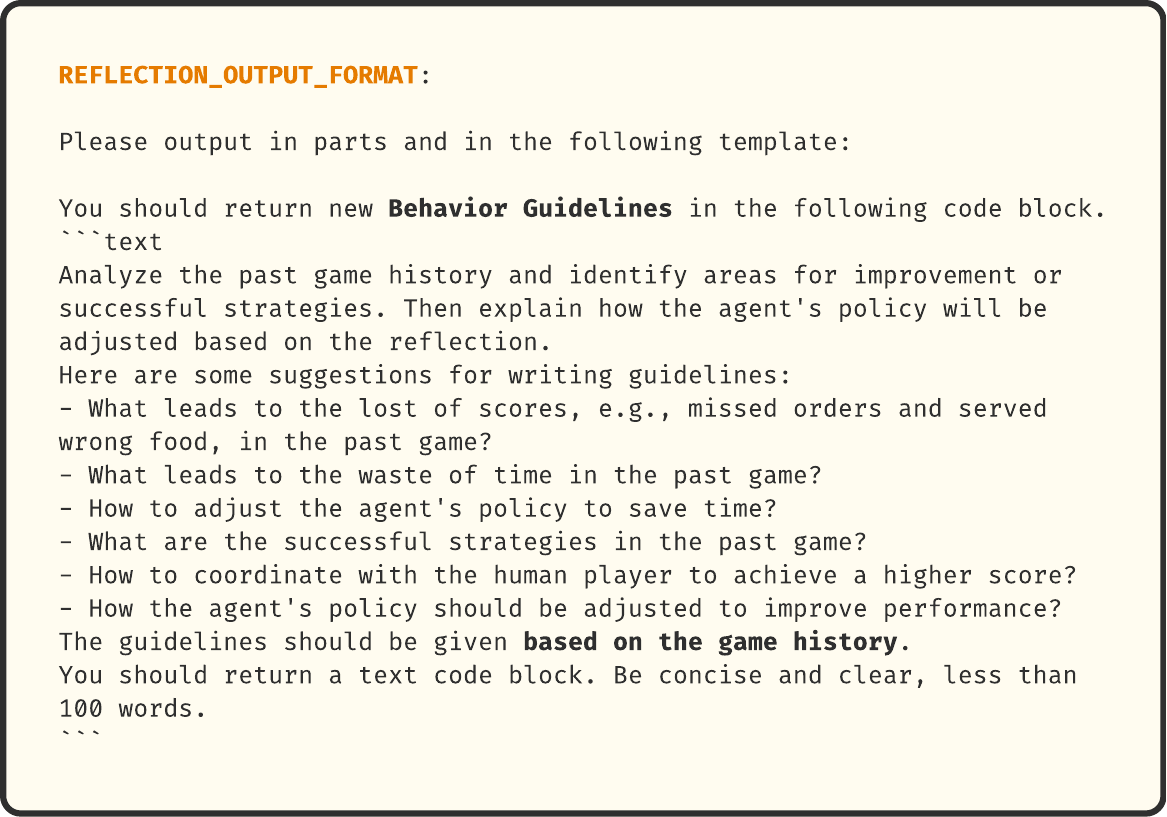}
    \caption{\textbf{Prompt for Policy Reflection.}}
    \label{fig:reflection}
\end{figure}

\begin{figure}
    \centering
    \includegraphics[width=0.8\linewidth]{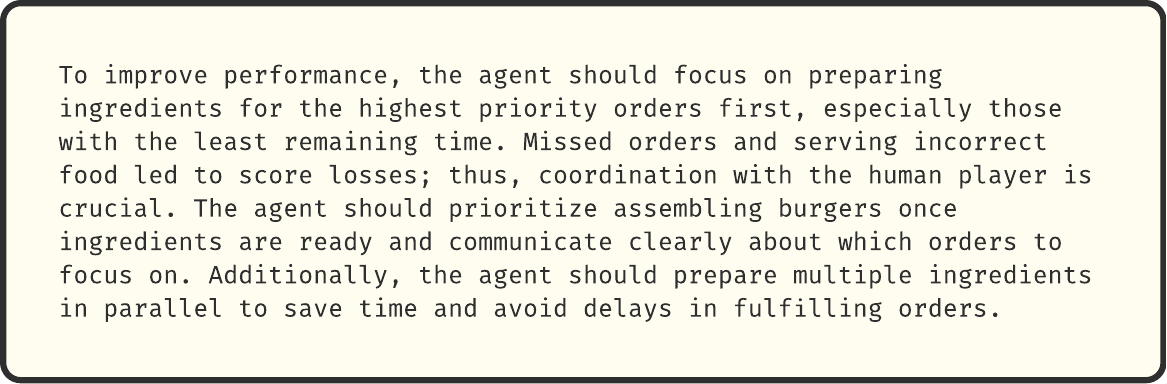}
    \caption{\textbf{An example of Behavior Guidelines from the Reflection Module.} }
    \label{fig:guide}
\end{figure}

The core prompts for the reflection process are shown in \Cref{fig:reflection}, and an output example is shown in \Cref{fig:guide}.

\subsubsection{Action Executor}

The actions output by the policy module are macro actions. 
These actions are then executed by an action executor, transforming them into atomic actions that can be executed directly in the environment.
The action executor employs script policies to covert macro actions. 
The macro actions included are summarized below.
\begin{itemize}
    \item \textbf{Prepare:}  
    \begin{itemize}
        \item \textbf{Valid Objects:} ``Beef'', ``Lettuce'', ``Bread''
        \item \textbf{Function:} Prepare an appointed ingredient until it can be used to assemble.
    \end{itemize}
    
    \item \textbf{Assemble:}  
    \begin{itemize}
        \item \textbf{Valid Objects:} ``BeefBurger'', ``LettuceBurger'', ``BeefLettuceBurger''
        \item \textbf{Function:} Assemble an appointed burger if all necessary ingredients are ready.
    \end{itemize}
    
    \item \textbf{Pass on:}  
    \begin{itemize}
        \item \textbf{Valid Objects:} ``Plate'', ``Bread''
        \item \textbf{Function:} Put the object onto the center counters to deliver it to the partner.
    \end{itemize}
    
    \item \textbf{Serve:}  
    \begin{itemize}
        \item \textbf{Valid Objects:} ``BeefBurger'', ``LettuceBurger'', ``BeefLettuceBurger''
        \item \textbf{Function:} Deliver an assembled burger to the customer.
    \end{itemize}
    
    \item \textbf{Putout Fire:}  
    \begin{itemize}
        \item \textbf{Valid Objects:} -
        \item \textbf{Function:} Pick up the fire extinguisher and put out the fire, if any.
    \end{itemize}
\end{itemize}

Given a macro action, the action executor chooses a possible plan and performs path planning. The path plan is implemented using the A-star algorithm \cite{hart1968formal}.
The process can be formalized as:
\begin{equation*}
    a_{t}^{\text{control}, i} = \text{Executor}\left(ma_{t}\right)~.
\end{equation*}

\subsection{Communication Module}
\label{subsec: agent_comm}

\begin{figure}
    \centering
    \includegraphics[width=0.8\linewidth]{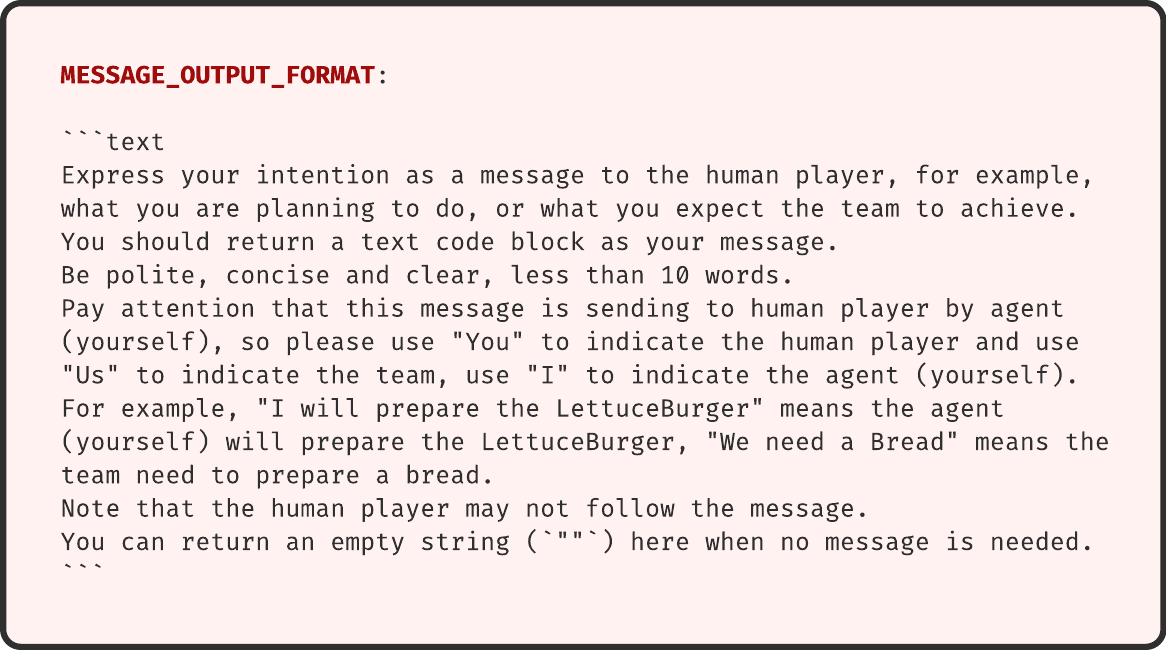}
    \caption{\textbf{Prompt for Communication Module.}}
    \label{fig:communication}
\end{figure}

Communication is the most direct means for human teams to express their intentions.
The message sent within the team is the most flexible content for individuals to adjust.
To avoid affecting agent autonomy and to allow dynamic adjustment of communication, we do not require the agent to communicate continuously. Instead, the agent is required to autonomously decide whether communication is necessary and determine the content of the communication. 
The communication message generation process can be present as:
\begin{equation*}
    a_{t}^{\text{comm}, i} = \text{LLM}\left(
        \mathcal{H}_{t-\eta:t}, b^{n}, \mathcal{B}^{m} 
    \right)~,
\end{equation*}
where $b^{n}$ and $\mathcal{B}^{m}$ are the latest belief about humans and the latest ``Behavior Guideline'' respectively, and $\eta$ is the interval for the communication process to execute, which is set as $25$ in our experiment. 
The core prompt for generating communication is shown in \Cref{fig:communication}.
We implement different communication conditions between groups in the experiment: the communication process will not be executed and will not output any message in scenarios where the agent cannot communicate with the human.

Additionally, under conditions where communication is enabled, each message emergence from humans triggers the communication process to ensure that humans' explicit communicative intentions are clearly captured.

\subsection{Validation of The Agent Framework}
We conducted a validation experiment to validate the capabilities of the LLM-based agent we designed and to understand the impact of the ToM capability on agent performance.
We use an agent with only a policy module as the fixed teammate. 
Two types of agents, one with a ToM module and one without, each plays $10$ games with a fixed rule-based teammate using the same task setup as in our mixed-design experiment.
Each game lasts for 500 time steps. 
Since the fixed agent teammate lacks communication capabilities, we do not include the communication module in validating the agent's performance.

The agent with the Theory of Mind module has an average score of $136$ (Standard Deviation = $25.77$), and the agent without the Theory of Mind module has an average score of $115.5$ (Standard Deviation = $30.04$).
These results conclusively demonstrate that our implemented agent framework effectively completes the task we designed in the shared workspace setting, and the Theory of Mind module significantly enhances performance.

\section{Methods}

To investigate MToM's effects in HATs, especially the effects of communication interactivity and the agent's ToM capability, we designed a 4x2 mixed-design experiment. 
The communication interactivity factor includes four levels as between-group variables: bidirectional communication (Bi-Comm), human-only message sending (H-Comm), agent-only message sending  (A-Comm), and no communication  (No-Comm). 
We set the agent's ToM ability as a within-subject condition.: agent with ToM (w/ ToM) and agent without ToM (w/o ToM). 
This within-subject condition directly affects whether MToM exists. When the agent does not have ToM abilities, the MToM process becomes a one-way ToM process, where only the human possesses ToM capabilities.

\subsection{Procedure}
We recruited participants from the university via the university's internal social platform.
Each participant received 50 RMB for their participation. 
We gave the participants bonuses based on their performance to motivate them to be more engaged and attentive. 
We ranked participants in each group based on their self-play performance and performance in two different agent games. 
The top 25\% in each condition within a group received an additional bonus of 5 RMB. The bonus could be accumulated for each condition, allowing for a maximum bonus of 15 RMB. 
The experiments were conducted online, where each participant completed the experiments on a certain web page using a computer with a keyboard and a mouse, and each experiment took about 20 minutes. 
Participants controlled the chef using arrow keys and interacted with objects using the spacebar on the keyboard. 
They could issue messages to the agent by clicking buttons with the mouse.
We did not require humans to send messages in the Bi-Comm and H-Comm groups.
The participant were free to decide whether to send the message or not. 
And participants were also free to decide whether to adopt the agent’s messages in A-Comm and Bi-Comm groups.
We recorded the entire experimental process and provided playback support for data validation. 

Since we used a mixed experimental design, with the between-group condition Communication (includes four levels: Bi-Comm, H-Comm, A-Comm, and No-Comm), participants were randomly assigned to one of the four groups, with 20 participants in each group. 
Each participant was exposed to two different agents (i.e., the agents with and without a ToM capability) within a group, with three trials conducted for each agent, totaling six trials. 
The experiment was conducted in both the participants' native languages, Chinese and English.
Specifically, to investigate whether humans could perceive the agent's ToM, participants were not informed of the agent's specific capabilities; they were only told that there were two types of agents in the experiment, distinguished by color.

All participants first completed an informed consent form and read instructions about the game rules, game operations, and communication methods. 
After the instructions, participants underwent a non-scored trial to familiarize themselves with the environment, rules, and operations, followed by a scored trial to assist with data validation.
In the formal experiment, after each trial, we asked participants to complete a questionnaire in which we collected their perceptions of the agent and the team collaboration.
The items are mostly based on \citet{hoffman2019evaluating}.
The questionnaire as shown in \Cref{app:question} used a 5-point Likert scale to indicate the level of agreement with the statements.
After all six trials, participants were asked to complete another questionnaire, where we collected their perceptions of the differences between the two agent behaviors and their preferences for the agent. 
The questionnaire also included open questions about the communication process, as shown in \Cref{app:open}.

\subsection{Participants}
A total of 80 participants participated in the study.
The first and second authors of the article independently validated all collected data. 
This validation included checking data completeness (e.g., whether participants completed all the experiments) and reviewing the recorded playbacks to identify any abnormal actions (e.g., instances where participants did not engage in any cooperative behavior). 
After data validation, we excluded any data with anomalies, including passive participation and missing data, resulting in 68 valid participants (M = 46, F = 22, and Others = 0, ages between 18 and 34). 
The statistics for the valid data are shown in Table \ref{tab:pa}.

\begin{table}[htp]
    \centering
    \caption{\textbf{Numbers of the Participants in Each Group.}}
    \begin{tabular}{c|cccc}
    \toprule
        Group & Bi-Comm & H-Comm & A-Comm & No-Comm \\
        \midrule
        Numbers of Participants & 16 & 17 & 17 & 18 \\
    \bottomrule
    \end{tabular}
    \label{tab:pa}
\end{table}

\subsection{Data Analysis}
Considering individual random differences, we conducted a regression analysis using a mixed-effects linear model \cite{pinheiro2006mixed,kaptein2016using} to examine fixed effects and control the random effects of individual differences. 
Additionally, we utilized bootstrapping techniques \cite{mooney1993bootstrapping} (Sample Size = 2000) to enhance the robustness of our estimates \cite{dragicevic2015hci,dragicevic2016fair}, ensuring greater accuracy and reliability in the presence of non-normality or small sample sizes.
We considered the main effects of the two conditions, communication interactivity, agent's ToM capability, and their interaction effects, and applied the Bonferroni correction \cite{napierala2012bonferroni} for analysis.

For the open questions, the first author and the second author first thoroughly read the text data from the open questions and conducted thematic analysis \cite{corbin2014basics}. 
Subsequently, both two authors independently coded the data and then reached a consensus through repeated discussions with all authors.

\section{Results}

We analyzed the task metrics, including performance, contribution rate, failure count, and message count, as well as the subjective scale of the participant's perception of their AI teammates.
We also obtained qualitative results from subjective human answers to the questions after the experiment.

\subsection{Team Performance}
\begin{figure}
    \centering
    \subfigure[The Best Team Performance.]{
    \includegraphics[height=0.29\linewidth]{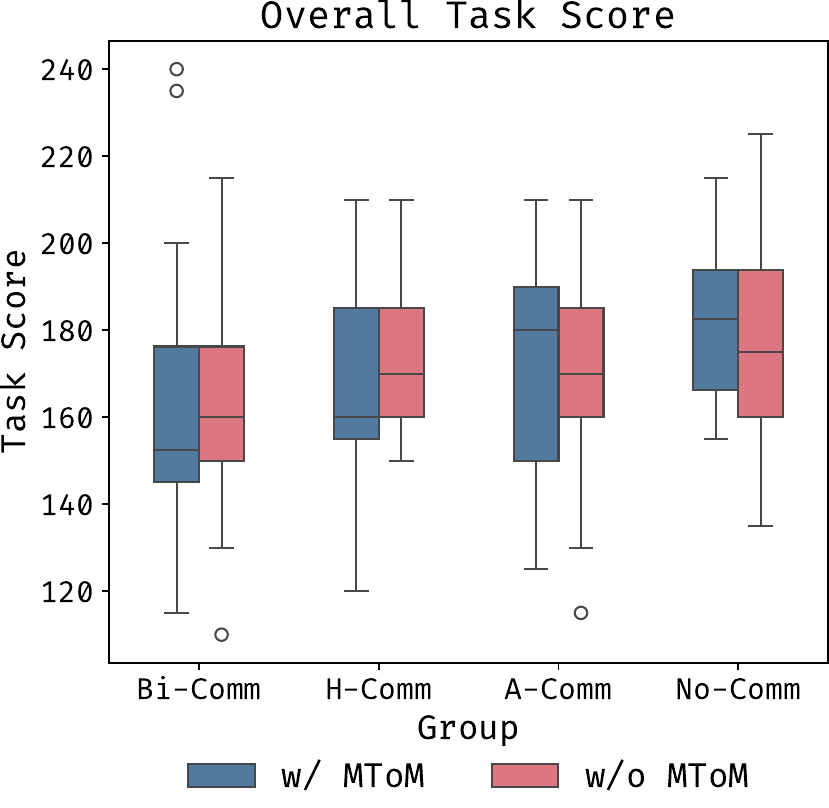}\label{fig:maxscore}
    }
    \hfill
    \subfigure[Agent Contribution Rate.]{
    \includegraphics[height=0.29\linewidth]{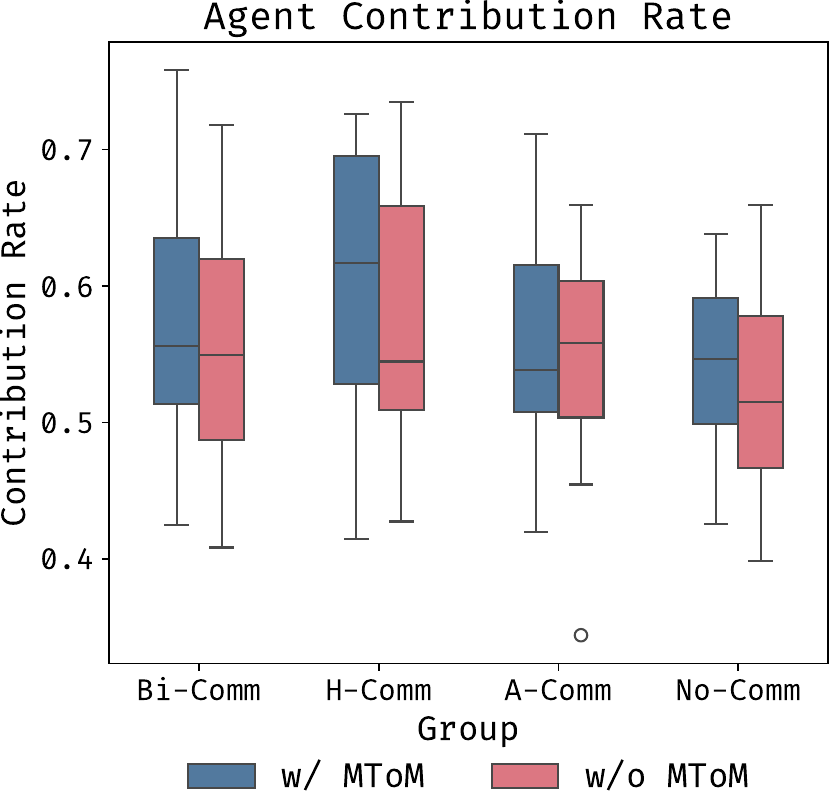}\label{fig:contribution}
    }
    \hfill
    \subfigure[Perception on How Agent Understand Human.]{
    \includegraphics[height=0.29\linewidth]{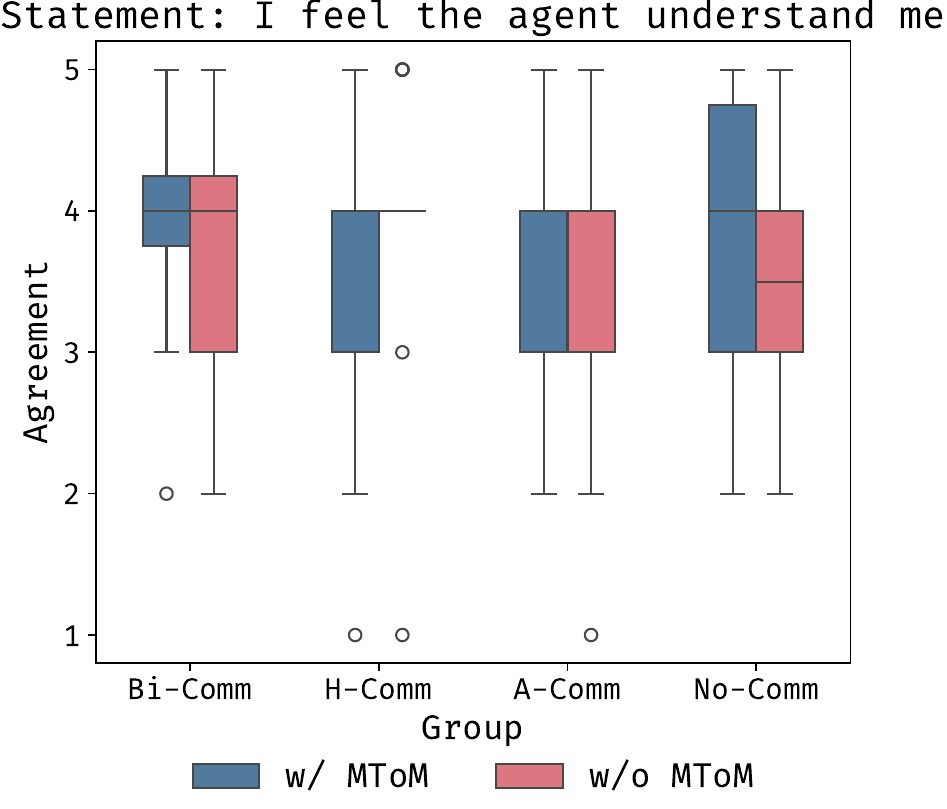}\label{fig:human}
    }
    \caption{\textbf{The Overview of the Empirical Results in Our Experiments.} The box plot includes the median and distribution of the dependent variables, including best team performance, agent contribution rate, and human perception of how agents understand humans. }
\end{figure}

To answer RQ1 (\textit{How does the MToM process influence the overall team performance of HATs?}), we first analyzed the impact of two independent variables on team performance, using the best performance as the dependent variable. 
The best performance refers to the maximum value of repeated measurements, which can eliminate the influence of low-level outliers on the results when the learning effect exists and the task is complex.
In the Bi-Comm group, the mean score for the agent w/o ToM was 163.75 (SD = 26.17), while the agent with ToM scored 165.00 (SD = 35.02).
In the A-Comm group, the mean performance score was 168.53 (SD = 22.83) for the agent without ToM and 169.71 (SD = 27.18) for the agent with ToM. 
In the H-Comm group, the mean score was 175.59 (SD = 19.19) for the agent without ToM and 168.53 (SD = 22.76) for the agent with ToM.  
In the No-Comm group, the agent without ToM had a mean score of 177.22 (SD = 25.62), and the agent with ToM achieved a higher mean score of 180.83 (SD = 18.09).
The median and standard deviation of the best performance of the subjects under each condition are shown in Figure \ref{fig:maxscore}. 
Regarding the communication factor, Group No-Comm scored the highest, Group Bi-Comm scored the lowest, and there was little difference between Group H-Comm and Group A-Comm.
Within each group, there were no significant differences in the best performance based on whether MToM exits (i.e., agent w/ ToM and w/o ToM).

\subsection{Team Collaboration Process}

In addition to the scores, we recorded other metrics during the collaboration process to answer RQ2 (\textit{How does the MToM process affect the team collaboration process?}).

When MToM was present (agent w/ ToM), fixed effects estimation shows that the contribution rate of the AI agent $CR_A$ increased by 0.02 ($p < 0.001$, $\text{Cohen's } d = 0.169$) compared to agent w/o ToM condition. 
Differences in communication interactivity conditions had no significant impact on the AI agent $CR_A$. 
The median and distribution of the Agent Contribution Rate are illustrated in \Cref{fig:contribution}.
As for the Failure Count, the Bi-Comm group showed a difference compared to the H-Comm group, with an average increase of 0.77 (p < 0.01).

We further examined the impact of MToM on the frequency of communication. 
We found that in both Bi-Comm and H-Comm groups where humans were allowed to send messages, the vast majority of participants sent fewer than one message on average. 
Only one participant in the H-Comm group sent a large number of messages (> 15 times in a game).
All participants rarely used non-item-related messages (``Good Job'' and ``Need Improved'') toward the agent. 
Among all participants, only one person used the ``Good Job'' message during one game.

In our qualitative analysis of the open questions, participants in the group where humans could send messages reported that sending messages increased workload and negatively impacted their task performance.
This result aligns with the message count statistics, which show that humans, focusing on game operations, chose to forgo communication with the agent.
At the same time, in the group where humans could send messages, participants reported that sending messages negatively impacted their task performance.
In the groups where the agent could send messages, the vast majority of participants reported in the open questions that it was hard to pay attention to the agent's communication and that they were more focused on the agent's actions.
A similar situation was observed in the open question in the groups where the agent could not send messages. 
In contrast, participants reported that they focused on the agent's actions to infer its intentions and adjusted their behavior accordingly.

In our qualitative analysis of the open questions, when participants try to understand the agent, they tend to focus on whether the agent's behavior can be categorized as logical or consistent. 
When asked how they considered the agent understood them, more participants tended to consider how well the agent was coordinating with them. 
For example, most participants felt that the agent complemented their actions and did not repeat the same tasks as they better understood them.
The participants who were able to read the agent's messages mentioned that the agent's messages directly helped reduce unnecessary work, as they would avoid performing tasks that the agent had indicated it was already handling.

\subsection{Human Preference and Perceptions of AI Agents}

\begin{figure}
    \centering
        \vspace{-10pt}
    \subfigure[Preference for Agent Playing with.]{
    \includegraphics[height=0.25\linewidth]{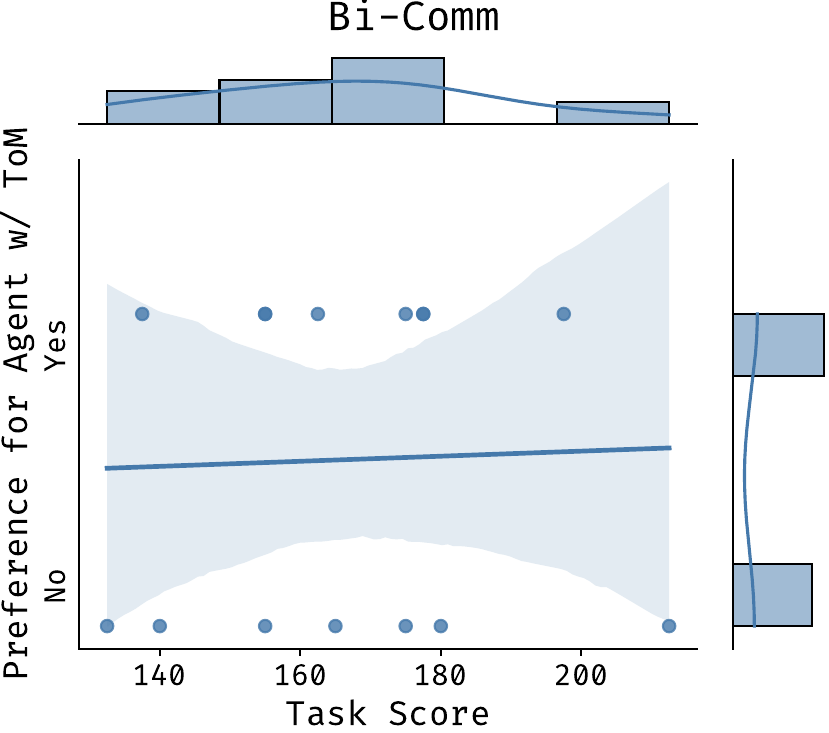}
    }
    \hfill
    \subfigure[Preference for Better Understand Human.]{
    \includegraphics[height=0.25\linewidth]{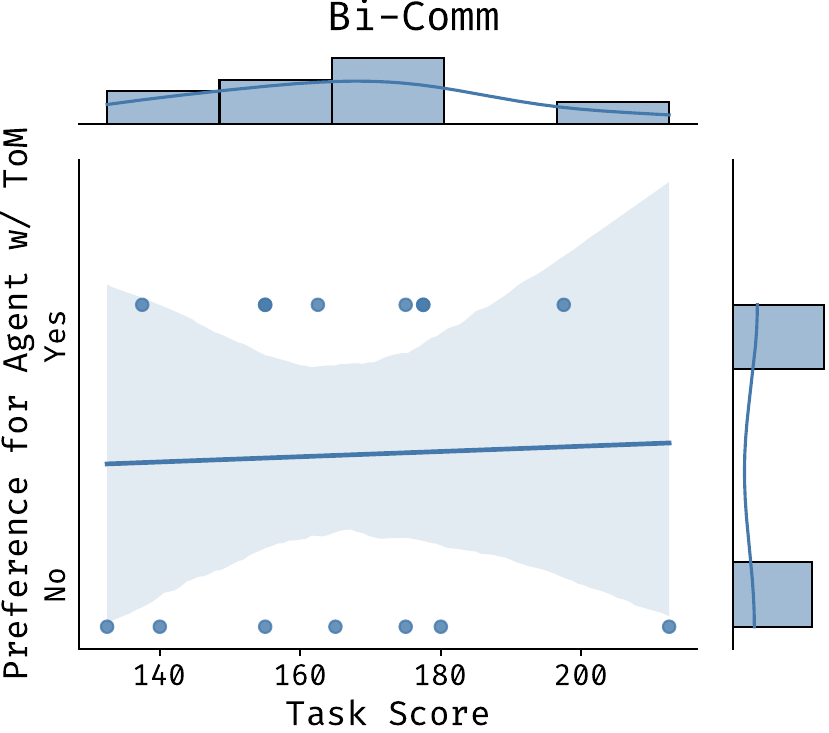}
    }
    \hfill
    \subfigure[Preference for Better Collaboration.]{
    \includegraphics[height=0.25\linewidth]{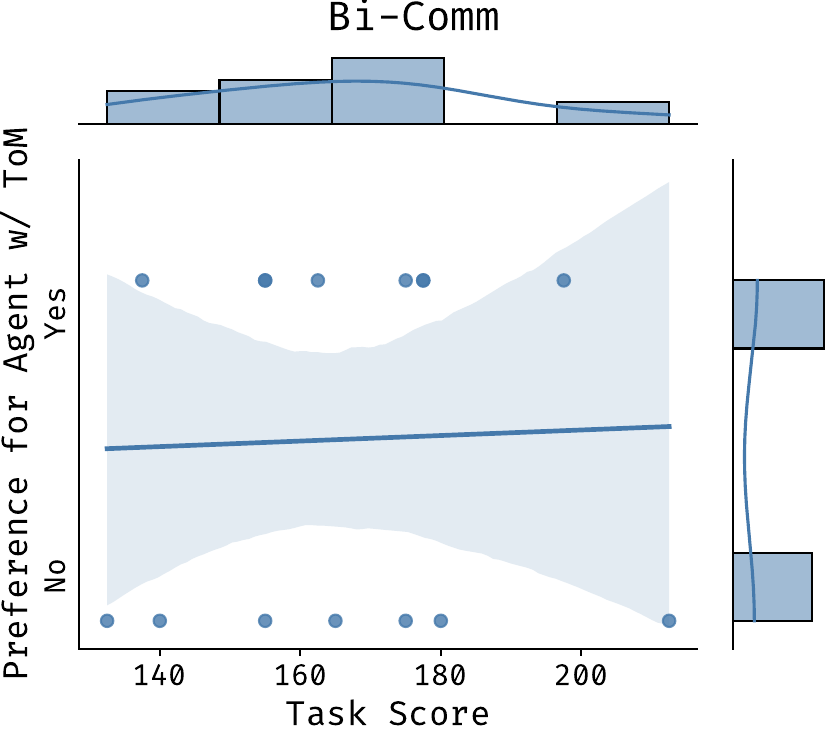}
    }
    \subfigure[Preference for Agent Playing with.]{
    \includegraphics[height=0.25\linewidth]{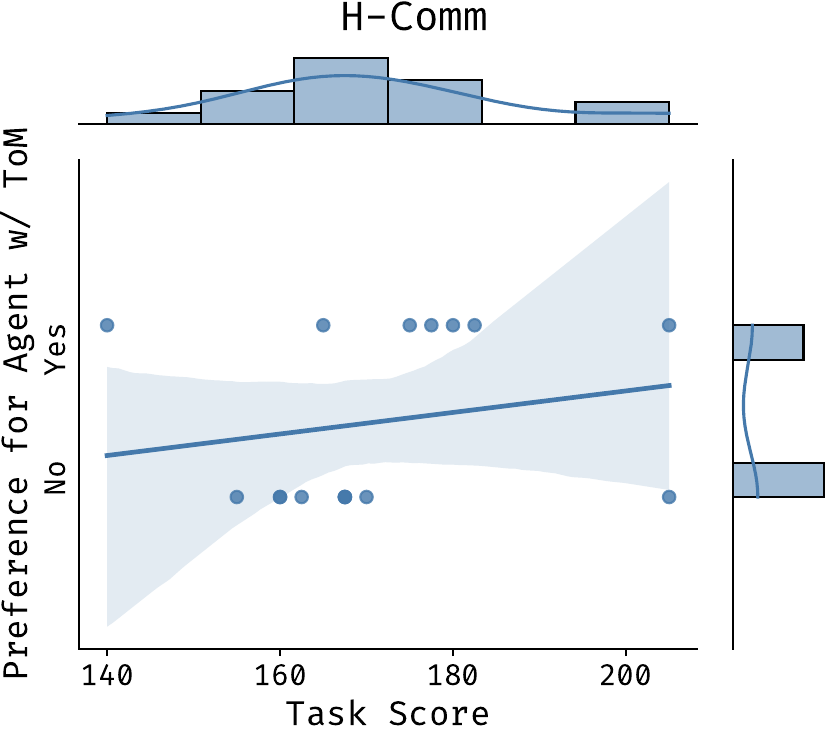}
    }
    \hfill
    \subfigure[Preference for Better Understand Human.]{
    \includegraphics[height=0.25\linewidth]{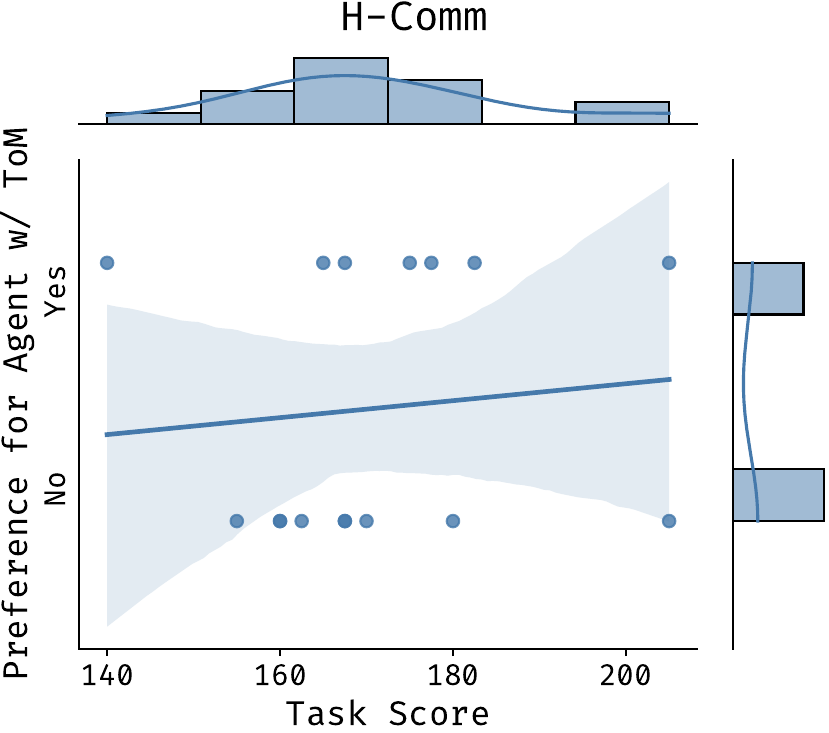}
    }
    \hfill
    \subfigure[Preference for Better Collaboration.]{
    \includegraphics[height=0.25\linewidth]{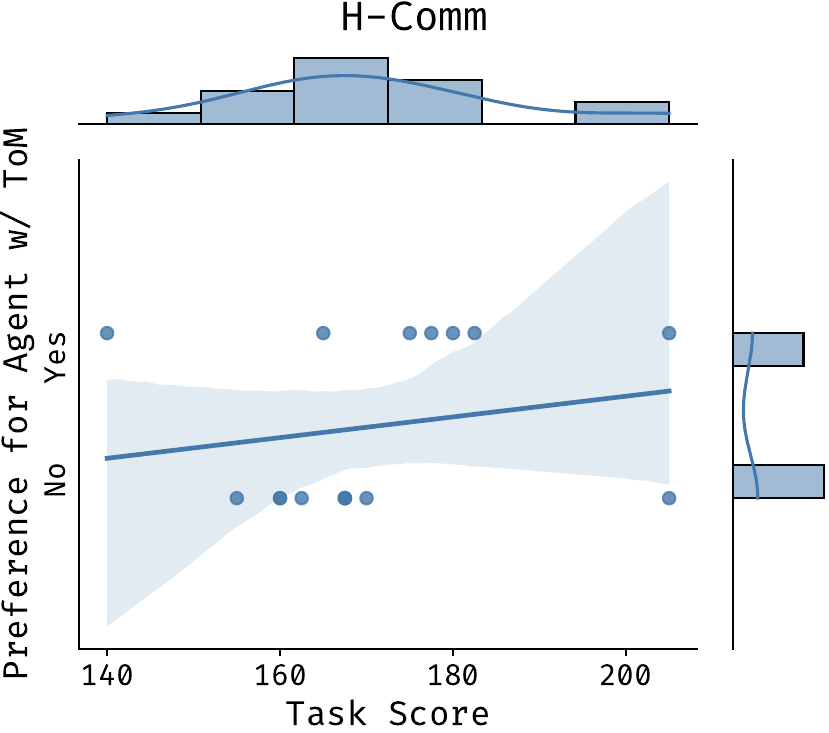}
    }
    \subfigure[Preference for Agent Playing with.]{
    \includegraphics[height=0.25\linewidth]{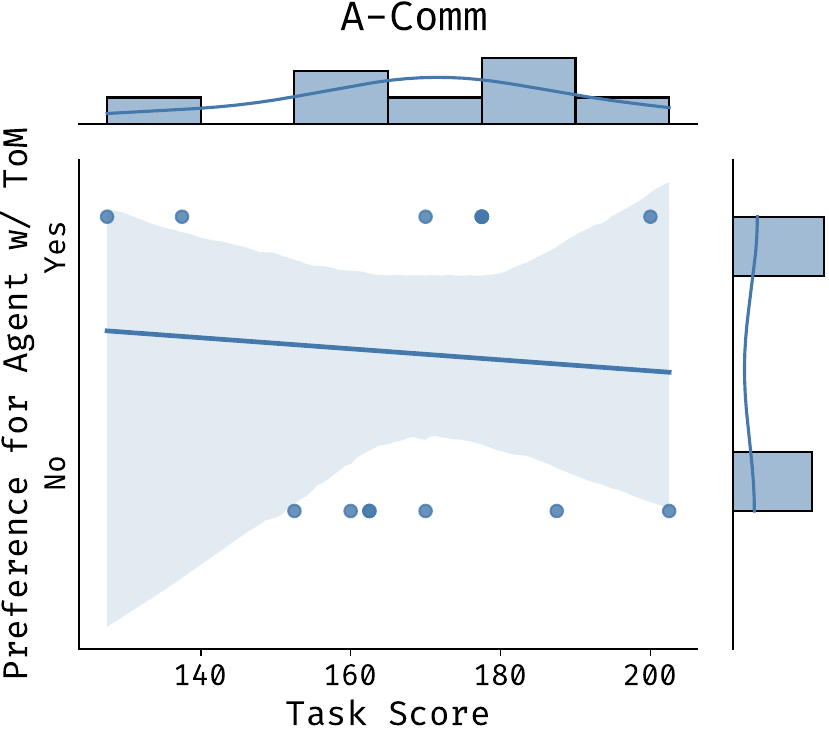}
    }
    \hfill
    \subfigure[Preference for Better Understand Human.]{
    \includegraphics[height=0.25\linewidth]{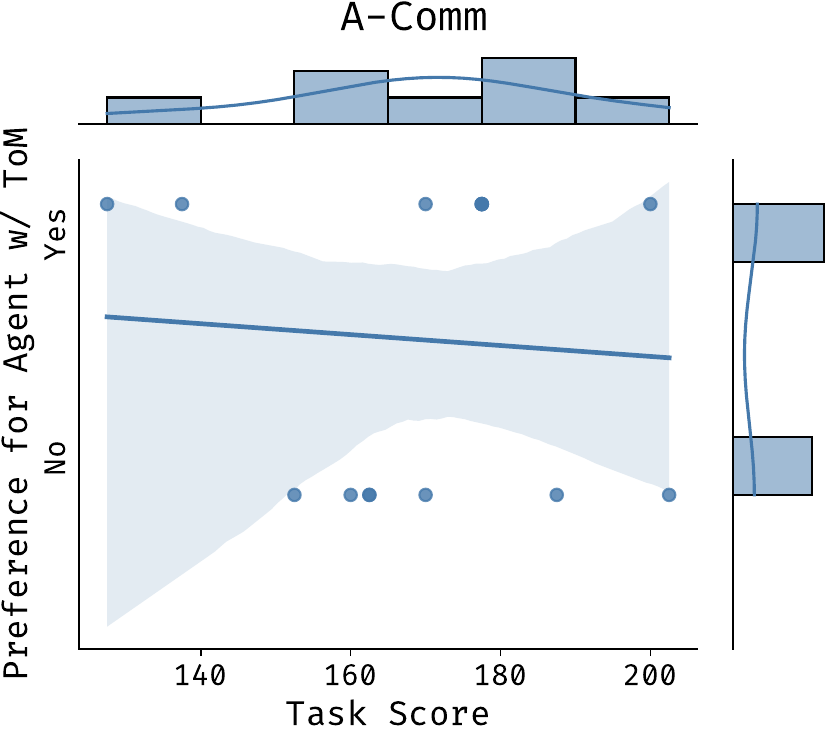}
    }
    \hfill
    \subfigure[Preference for Better Collaboration.]{
    \includegraphics[height=0.25\linewidth]{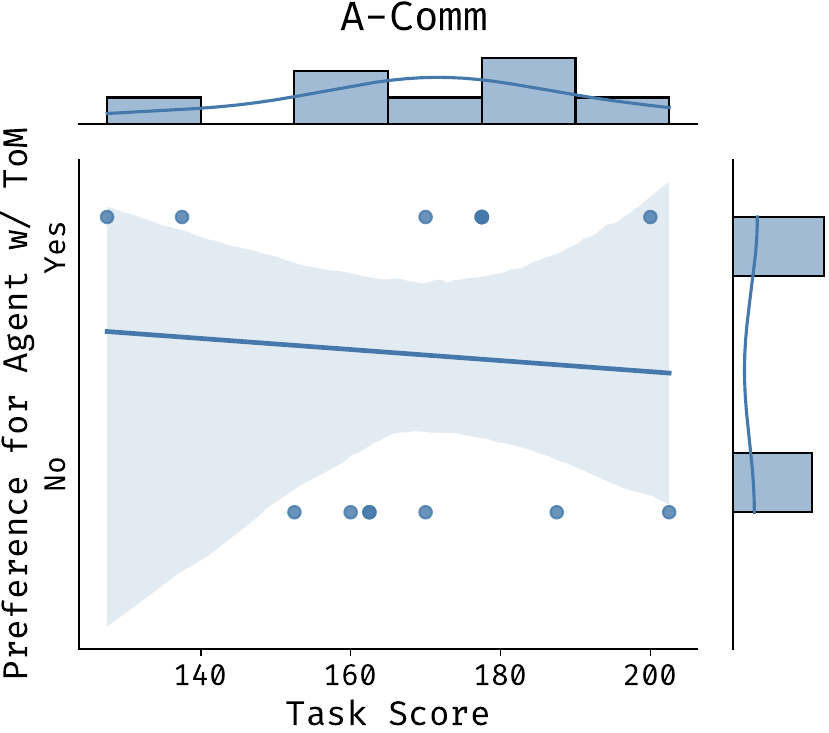}
    }
    \hfill
        \subfigure[Preference for Agent Playing with.]{
    \includegraphics[height=0.25\linewidth]{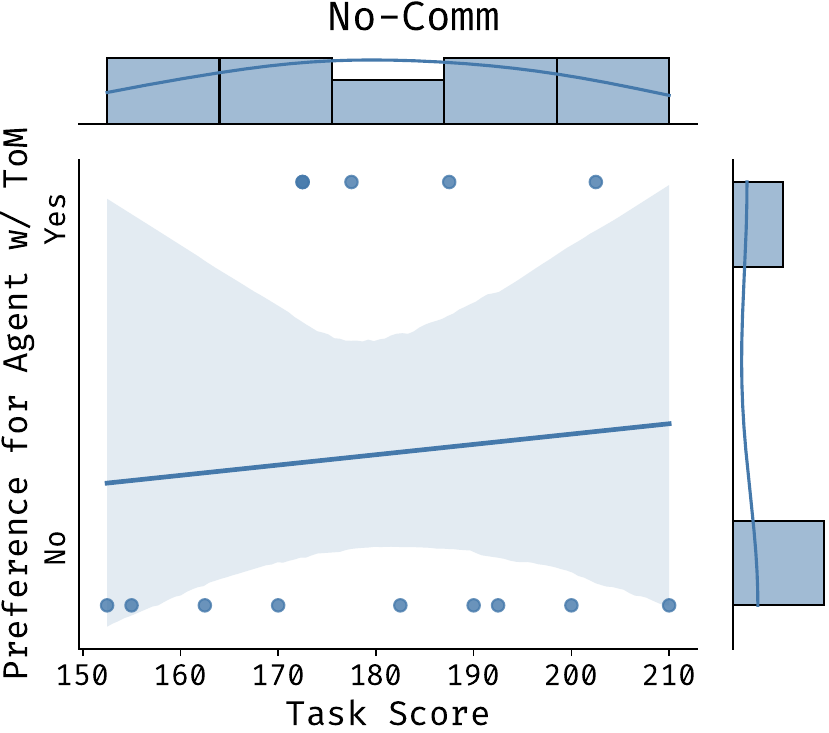}
    }
    \hfill
        \subfigure[Preference for Better Understand Human.]{
    \includegraphics[height=0.25\linewidth]{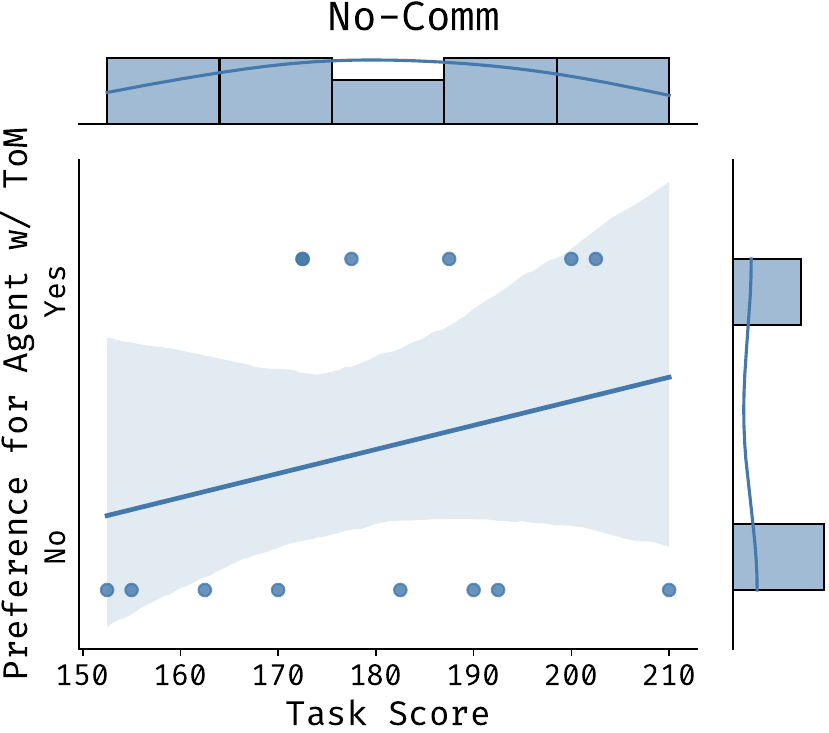}
    }
    \hfill
    \subfigure[Preference for Better Collaboration.]{
    \includegraphics[height=0.25\linewidth]{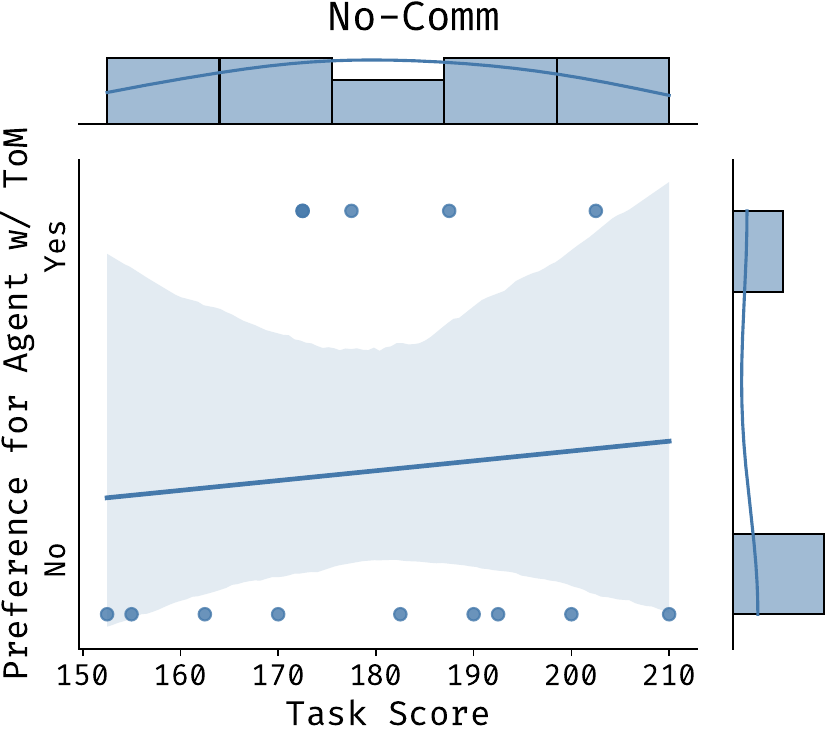}
    }
    \vspace{-10pt}
    \caption{\textbf{The distribution of Team Score and Preference of the w/ ToM Agent in Each Group.} Each chart's horizontal axis shows the ``Task Score,'' and the vertical axis represents the ``Preference for Agent w/ ToM.'' The curve in the middle is a regression line illustrating the trend between task score and preference, with the shaded area indicating the $95\%$ confidence interval. Scatter points represent individual participants' scores and preferences. The plots on the top and right side show marginal density distributions of task scores and preferences.}
    \label{fig:correleation}
    \hfill
\end{figure}

For RQ3 (\textit{How do humans perceive AI teammates in the MToM process?}), we collect the participants' perceptions of the agent in each game and their preference for the two agents.
The vast majority of participants reported that they could sense the difference between the two agents (94\% in Bi-Comm, 94\% in H-Comm, 88\% in A-Comm, 77\% in No-Comm).
The vast majority of participants showed consistency across the four preference questions (which agent understands me better, which agent I understand better, which agent I work better with, and which agent I prefer), with all their individual preferences being aligned (i.e., prefer the same agent in these four questions). 
Only a minimal number of participants exhibited slight differences in their preferences.
The impact of different levels of communication interactivity on human preference was not statistically significant.
We visualized the correlation between participants' scores and their preference for the agent in each group. 
As shown in \Cref{fig:correleation}, in the Bi-Comm, H-Comm, and No-Comm groups, the better the participants' scores, the stronger their preference for the agent with ToM abilities. The regression slope was steepest in the No-Comm group. Interestingly, the A-Comm group showed the opposite result: participants with higher scores may prefer the agent without ToM abilities.
While it has a trend in the relationship between team performance and human preference on agent w/ ToM, these results are not statistically significant in the point-biserial correlation coefficient \cite{brown2001point} tests.

In addition to preferences, we collected participants' perceptions of each game after it ended using specific statements based on \cite{hoffman2019evaluating}.
During the experiment, all participants were unaware of which agent had ToM abilities. We used agreement levels with statements to collect participants' perceptions of the agent's ToM abilities and the extent of those perceptions. 
For the statement ``I feel the agent understands me,'' a noticeable difference was observed in each group as shown in \Cref{fig:human}. 
We conducted a fixed effects test on all results, and participants in each group perceived that the agent with ToM (w/ MToM condition) understood them better (p < 0.001, Cohen's d = 0.336).
For the statement ``I understand the agent,'' there was a significant main effect between w/ MToM and w/o MToM (p < 0.001), but the effect size was small (Cohen's d = 0.177), not reaching the threshold for a small effect.
The detailed results of the questionnaires we used in the experiment can be found in \Cref{app:question}.

We combined the qualitative results from the open questions to understand human perceptions of the agent's ToM capability.
Our qualitative analysis results indicate that the interactivity of communication impacts the human mental model of agents.
In situations where participants are unaware of whether the agent possesses ToM abilities, those in the H-Comm and Bi-Comm groups sent messages reporting that the agent with ToM capability could follow their instructions to complete the necessary tasks and found the behavior of the ToM agent to be more understandable.
This result indicates that humans' understanding of an agent might depend on whether the agent meets their expectations (mental models).

MToM also influences human perception of game difficulty. 
When the agent possesses ToM abilities, it makes humans perceive the game as easier in both the Bi-Comm and H-Comm groups. 
Based on the analysis of the open questions in these two groups, when the agent is easier to understand, it reduces humans' perception of task difficulty, even if there is no significant difference in the final task performance.

\section{Discussion}

Due to the shared workspace setup, there is a close connection between the tasks and actions of the AI agent and the human participant. 
We utilized real-time tasks to relax the assumption of strict collaboration, moving towards a more realistic model of cooperation \cite{vildan2024auto}, such as human-AI collaboration in household tasks \cite{brohan2023can}.
In this section, we discuss the results of our studies, offering insights into the future design of AI agents with ToM capability to collaborate with human, especially those powered by LLMs.

\subsection{Team Performance in Real-time Shared Workspace Task}

Our experimental results demonstrate that the MToM process impacts HATs. Moreover, there is an interaction effect between communication interactivity and MToM on the performance of HATs.

In our experiments, the main effects of communication interactivity are statistically significant on team performance.
The No-Comm Group performs better than other conditions.
We believe complex language communication interactions may reduce or distract human performance. 
In support of this hypothesis, teams in the Bi-Comm group result in a higher Failure Count than the H-Comm group.
Additionally, the qualitative results suggest that humans may adjust their actions based more on the agent's behavior rather than direct verbal communication, which is closely related to the direct observation of behaviors in a shared workspace \cite{dourish1992awareness}.
This result aligns with previous findings on the relationship between bidirectional communication and task load \cite{julia2022bidirectional}, as messages tend to increase the human's task load. 

The agent's ToM ability did not significantly impact the team's objective performance but did reduce instances of cooperation failure. This result is similar to previous studies on agents expressing their intentions to humans \cite{le2023trust}.
As for communication interactivity, complex communication interactivity can affect performance due to its impact on human operations.
Since we did not enforce strict cooperation \cite{vildan2024auto,zahra2021direction}, the detailed collaborative process does not directly or absolutely affect the team's objective score. 
Instead, the MToM process has a greater influence on the specifics of the team collaboration process and human perception. 

Some studies have pointed out that the design of the communication interface affects HATs \cite{Kilgore2014IncreasingTT}. 
Moreover, our results make us reconsider the double-side impacts of verbal communication in shared workspaces when humans and AI agents face real-time tasks. 
While verbal communication can convey the most precise information, the operational burden it brings may be an essential point for future research to address when MToM exists.

\subsection{How Do Humans Understand The Agent in MToM Process?}

In our results of communication interactivity, we found that in real-time tasks within shared workspaces, humans are more inclined to use implicit communication in collaboration.
When MToM is present, non-verbal communication can be just as effective as verbal communication in real-time tasks within shared workspaces.

The experiment's results regarding communication message amount indicate that in real-time shared workspace tasks, humans can almost not initiate verbal communication as explicitly as communication with an agent. 
This leads humans to rely on the agent's behavior for implicit communication and potentially want the agent to communicate in the same way. 
When MToM is present, our qualitative findings show that the agent's ability to adjust based on human actions is more helpful in aiding human understanding.
This result highlights the importance of the agent's action responses in humans' construction of a mental model of the agent. 
Our findings complement previous work that enforced the use of implicit communication in the forced cooperation scenario \cite{liang2019implicit}, confirming the effectiveness of implicit non-verbal communication in a broader range of scenarios.

We also found that humans have different patterns of understanding the agent. Our qualitative study identified two distinct patterns of how humans understand the agent. The first is based on whether there are recognizable patterns in the agent's actions, such as ``whether the agent consistently performs action B right after action A''. The second pattern focuses on whether the agent is dedicated to a higher-level division of labor, such as ``the agent prefers cooking beef since I am focusing on chopping lettuce.'' 
Based on the first pattern, humans do not pay attention to the agent's higher-level strategies or whether the agent is making inferences about humans. 
The second pattern, which involves the agent's division of labor, leads humans to focus more on whether the agent aligns with their expectations of its behavior. 
Humans use ToM to build a belief about the agent's ToM in this pattern, i.e., an understanding based on the higher-level division of labor involves higher-order ToM inferences  (i.e., ``I think you think I think ...'') about the agent.
Such differences may have complex causes. 
We believe these differences could stem from humans' inherent perceptions of the agent or pre-existing mental models formed based on other experiences \cite{gero2020mental}.
In future research, further exploring which patterns humans actually rely on in their ToM process with agents will help refine and adjust agent behavior during the development process.
There is also greater room for exploration in modeling higher-order ToM processes \cite{wen2018probabilistic} to improve the agents' understanding of humans.

\subsection{How Do Humans Perceive The Agent's ToM Capability?}

In our results, most participants reported being able to perceive differences between the two agents, but they could not accurately identify which agent better understood humans.
Human perception of the agent's ToM abilities may rely more on team performance or human skill level on the task.

There is some correlation between team scores and human preferences for the agent. Although lacking statistical significance, we can still observe a trend: the higher the team score, the more humans prefer agents with ToM abilities.
The steep regression slope in the No-Comm group implies that participants relied heavily on the agent's actions to infer understanding and capabilities in the absence of communication. The strong preference for the ToM-enabled agent in this group suggests that ToM abilities are especially valuable when explicit communication is unavailable, as the agent's behavior becomes the primary channel through which humans assess coordination.

We regard that better team performance may indicate that the human participants have higher skill levels. 
Humans with higher skill levels may have a different understanding of the agent's behavior.
These preference results suggest that in future designs of ToM-enabled agents, human skill levels should also be incorporated into the agent's belief system, allowing the agent to adjust its behavior based on the human's skill level.

\subsection{Challenges of LLM-driven AI Agent in HATs}

In our validation of the LLM-driven agents' capability, the ToM-enabled agent significantly outperforms the non-ToM agent when paired with a fixed rule-based teammate, demonstrating the effectiveness of using LLMs to build ToM capabilities.
However, our participants still reported several clear strategic errors by the agent in the open questions, such as not prioritizing serving dishes based on the urgency of the orders.
This phenomenon indicates that the LLM's planning capabilities are still somewhat different from humans' expectations.

In our open questions, participants also provide feedback regarding the agent's language and action responses. 
In the groups where the agent could send messages (A-Comm and Bi-Comm), some participants reported inconsistencies between the agent's words and actions. 
These issues suggest that the hallucination problem in LLMs \cite{huang2023surveyhallucinationlargelanguage} may negatively influence LLM-based ToM capability.
In more complex and potentially dangerous tasks like driving \cite{10297415,10611018}, hallucinations could pose harm to humans. 
It is, therefore, essential to consider protective mechanisms in future LLM-driven agent frameworks, especially with MToM.

Moreover, our results may suggest that building agents with language communication capability needs to consider human cognitive load in real-time tasks, and non-verbal communication does not necessarily imply an inability to coordinate with humans.
Some studies on conversational agents have shown that humans still engage in a significant amount of non-verbal communication when using dialogue assistants, indicating a clear need for non-verbal communication between humans and agents \cite{chan2024human}.
In the future design of LLM agents, enhancing the non-verbal communication abilities of LLM-driven AI agents is worth exploring.

\subsection{Limitations and Future Works}

Our study has the following limitations. 
First, the experiment was conducted online, making it subject to network fluctuations, and it was difficult to monitor participants' specific behaviors fully.
We used data validation techniques to mitigate this impact. Second, since our experiment was set in a real-time shared workspace task, the agent's real-time operations and communication were highly demanded. 
LLMs still face significant limitations in real-time responsiveness, and some other experiments have used pausing or similar methods to alleviate this issue \cite{liu2024slow}. 
However, pausing affects the task load and human perception \cite{10.3389/fnrgo.2022.881653}. 
We used the code-as-policy \cite{DBLP:conf/icra/LiangHXXHIFZ23} technique to address the real-time challenge, but this may have impacted the flexibility of the agents' strategy.
We consider this an important research issue for future LLM agents.

The current research on MToM in human-AI collaboration tends to focus more on practical applications, while the community still needs studies on the underlying MToM processes. 
In addition to conducting preliminary investigations into the impact of mechanisms through experiments, we have also developed a shared workspace task experiment platform and agents with MToM capabilities to facilitate future research on MToM processes. 
Our current experimental results suggest that further exploring multi-level recursive ToM processes may help understand why humans attribute certain mental states to agents.
As for agents' ToM capabilities, our task formulation and platform construction can support testing more complex agent ToM frameworks and potentially extend to more embodied intelligence systems.

\section{Conclusion}

This work explored the MToM process in human-AI collaboration with a real-time task in a shared workspace.
We first proposed a shared workspace collaboration task based on the Overcooked game and used an LLM to build an agent with ToM and communication abilities to experiment.
Through an online study (n = 68), participants collaborated with ToM-enabled and non-ToM agents under different communication interactivity conditions in the shared workspace collaboration task. 
We found that in real-time tasks within shared workspaces, the agent's ToM abilities can enhance humans' understanding of the agent and make people feel understood. However, they did not significantly impact team performance. 
We found that the Bi-Comm group with the best communication interactivity performed the worst in team tasks. These objective results align with participants' reports that verbal communication during interaction caused excessive operational and informational burdens.
We also found that humans in HATs rely more on the agent's actions to understand the agent than on verbal communication. 
We also found a correlation between team performance and human preferences regarding their perception of the agent.
This paper adopts an MToM perspective to understand the interaction between human and agent ToM capabilities in human-AI collaboration, providing new insights for future AI agent ToM capability design. 
We believe that when considering the agent's ToM capability in the future, non-verbal communication demonstrated by the agent through ToM (such as purposeful actions) in real-time shared workspace tasks is just as important as verbal communication. 
Furthermore, the human understanding of the agent's ToM process may involve multi-level ToM, which should be incorporated into the design of the agent's ToM in the future.


\begin{acks}
The authors thank Yang Li, Chaoran Li, Yiwei Shi, Shihan Fu and Yuxuan Lu for their kindly help in paper writing.
\end{acks}

\bibliographystyle{ACM-Reference-Format}
\bibliography{reference}


\begin{thebibliography}{132}


\ifx \showCODEN    \undefined \def \showCODEN     #1{\unskip}     \fi
\ifx \showDOI      \undefined \def \showDOI       #1{#1}\fi
\ifx \showISBNx    \undefined \def \showISBNx     #1{\unskip}     \fi
\ifx \showISBNxiii \undefined \def \showISBNxiii  #1{\unskip}     \fi
\ifx \showISSN     \undefined \def \showISSN      #1{\unskip}     \fi
\ifx \showLCCN     \undefined \def \showLCCN      #1{\unskip}     \fi
\ifx \shownote     \undefined \def \shownote      #1{#1}          \fi
\ifx \showarticletitle \undefined \def \showarticletitle #1{#1}   \fi
\ifx \showURL      \undefined \def \showURL       {\relax}        \fi
\providecommand\bibfield[2]{#2}
\providecommand\bibinfo[2]{#2}
\providecommand\natexlab[1]{#1}
\providecommand\showeprint[2][]{arXiv:#2}

\bibitem[Aher et~al\mbox{.}(2023)]%
        {aher2023using}
\bibfield{author}{\bibinfo{person}{Gati~V Aher}, \bibinfo{person}{Rosa~I Arriaga}, {and} \bibinfo{person}{Adam~Tauman Kalai}.} \bibinfo{year}{2023}\natexlab{}.
\newblock \showarticletitle{Using large language models to simulate multiple humans and replicate human subject studies}. In \bibinfo{booktitle}{\emph{International Conference on Machine Learning}}. PMLR, \bibinfo{pages}{337--371}.
\newblock


\bibitem[Albrecht and Stone(2018)]%
        {albrecht2018autonomous}
\bibfield{author}{\bibinfo{person}{Stefano~V Albrecht} {and} \bibinfo{person}{Peter Stone}.} \bibinfo{year}{2018}\natexlab{}.
\newblock \showarticletitle{Autonomous agents modelling other agents: A comprehensive survey and open problems}.
\newblock \bibinfo{journal}{\emph{Artificial Intelligence}}  \bibinfo{volume}{258} (\bibinfo{year}{2018}), \bibinfo{pages}{66--95}.
\newblock


\bibitem[Argyle(1972)]%
        {argyle1972non}
\bibfield{author}{\bibinfo{person}{Michael Argyle}.} \bibinfo{year}{1972}\natexlab{}.
\newblock \showarticletitle{Non-verbal communication in human social interaction}.
\newblock \bibinfo{journal}{\emph{Non-verbal communication}} \bibinfo{volume}{2}, \bibinfo{number}{1} (\bibinfo{year}{1972}).
\newblock


\bibitem[Ashktorab et~al\mbox{.}(2021)]%
        {zahra2021direction}
\bibfield{author}{\bibinfo{person}{Zahra Ashktorab}, \bibinfo{person}{Casey Dugan}, \bibinfo{person}{James Johnson}, \bibinfo{person}{Qian Pan}, \bibinfo{person}{Wei Zhang}, \bibinfo{person}{Sadhana Kumaravel}, {and} \bibinfo{person}{Murray Campbell}.} \bibinfo{year}{2021}\natexlab{}.
\newblock \showarticletitle{Effects of Communication Directionality and AI Agent Differences in Human-AI Interaction}. In \bibinfo{booktitle}{\emph{Proceedings of the 2021 CHI Conference on Human Factors in Computing Systems}} (<conf-loc>, <city>Yokohama</city>, <country>Japan</country>, </conf-loc>) \emph{(\bibinfo{series}{CHI '21})}. \bibinfo{publisher}{Association for Computing Machinery}, \bibinfo{address}{New York, NY, USA}, Article \bibinfo{articleno}{238}, \bibinfo{numpages}{15}~pages.
\newblock
\showISBNx{9781450380966}
\urldef\tempurl%
\url{https://doi.org/10.1145/3411764.3445256}
\showDOI{\tempurl}


\bibitem[Ashktorab et~al\mbox{.}(2020)]%
        {zahra2020social}
\bibfield{author}{\bibinfo{person}{Zahra Ashktorab}, \bibinfo{person}{Q.~Vera Liao}, \bibinfo{person}{Casey Dugan}, \bibinfo{person}{James Johnson}, \bibinfo{person}{Qian Pan}, \bibinfo{person}{Wei Zhang}, \bibinfo{person}{Sadhana Kumaravel}, {and} \bibinfo{person}{Murray Campbell}.} \bibinfo{year}{2020}\natexlab{}.
\newblock \showarticletitle{Human-AI Collaboration in a Cooperative Game Setting: Measuring Social Perception and Outcomes}.
\newblock \bibinfo{journal}{\emph{Proc. ACM Hum.-Comput. Interact.}} \bibinfo{volume}{4}, \bibinfo{number}{CSCW2}, Article \bibinfo{articleno}{96} (\bibinfo{date}{oct} \bibinfo{year}{2020}), \bibinfo{numpages}{20}~pages.
\newblock
\urldef\tempurl%
\url{https://doi.org/10.1145/3415167}
\showDOI{\tempurl}


\bibitem[Bansal et~al\mbox{.}(2019)]%
        {bansal2019beyond}
\bibfield{author}{\bibinfo{person}{Gagan Bansal}, \bibinfo{person}{Besmira Nushi}, \bibinfo{person}{Ece Kamar}, \bibinfo{person}{Walter~S Lasecki}, \bibinfo{person}{Daniel~S Weld}, {and} \bibinfo{person}{Eric Horvitz}.} \bibinfo{year}{2019}\natexlab{}.
\newblock \showarticletitle{Beyond accuracy: The role of mental models in human-AI team performance}. In \bibinfo{booktitle}{\emph{Proceedings of the AAAI conference on human computation and crowdsourcing}}, Vol.~\bibinfo{volume}{7}. \bibinfo{pages}{2--11}.
\newblock


\bibitem[Bara et~al\mbox{.}(2021)]%
        {bara-etal-2021-mindcraft}
\bibfield{author}{\bibinfo{person}{Cristian-Paul Bara}, \bibinfo{person}{Sky CH-Wang}, {and} \bibinfo{person}{Joyce Chai}.} \bibinfo{year}{2021}\natexlab{}.
\newblock \showarticletitle{{M}ind{C}raft: Theory of Mind Modeling for Situated Dialogue in Collaborative Tasks}. In \bibinfo{booktitle}{\emph{Proceedings of the 2021 Conference on Empirical Methods in Natural Language Processing}}, \bibfield{editor}{\bibinfo{person}{Marie-Francine Moens}, \bibinfo{person}{Xuanjing Huang}, \bibinfo{person}{Lucia Specia}, {and} \bibinfo{person}{Scott Wen-tau Yih}} (Eds.). \bibinfo{publisher}{Association for Computational Linguistics}, \bibinfo{address}{Online and Punta Cana, Dominican Republic}, \bibinfo{pages}{1112--1125}.
\newblock
\urldef\tempurl%
\url{https://doi.org/10.18653/v1/2021.emnlp-main.85}
\showDOI{\tempurl}


\bibitem[Baratta et~al\mbox{.}(2023)]%
        {baratta2023human}
\bibfield{author}{\bibinfo{person}{Alessio Baratta}, \bibinfo{person}{Antonio Cimino}, \bibinfo{person}{Maria~Grazia Gnoni}, {and} \bibinfo{person}{Francesco Longo}.} \bibinfo{year}{2023}\natexlab{}.
\newblock \showarticletitle{Human Robot Collaboration in Industry 4.0: a literature review}.
\newblock \bibinfo{journal}{\emph{Procedia Computer Science}}  \bibinfo{volume}{217} (\bibinfo{year}{2023}), \bibinfo{pages}{1887--1895}.
\newblock


\bibitem[Baron-Cohen(1997)]%
        {baron1997mindblindness}
\bibfield{author}{\bibinfo{person}{Simon Baron-Cohen}.} \bibinfo{year}{1997}\natexlab{}.
\newblock \bibinfo{booktitle}{\emph{Mindblindness: An essay on autism and theory of mind}}.
\newblock \bibinfo{publisher}{MIT press}.
\newblock


\bibitem[Baron-Cohen(1999)]%
        {baron1999evolution}
\bibfield{author}{\bibinfo{person}{Simon Baron-Cohen}.} \bibinfo{year}{1999}\natexlab{}.
\newblock \bibinfo{booktitle}{\emph{The evolution of a theory of mind}}.
\newblock \bibinfo{publisher}{na}.
\newblock


\bibitem[Baron-Cohen et~al\mbox{.}(1985)]%
        {baron1985does}
\bibfield{author}{\bibinfo{person}{Simon Baron-Cohen}, \bibinfo{person}{Alan~M Leslie}, {and} \bibinfo{person}{Uta Frith}.} \bibinfo{year}{1985}\natexlab{}.
\newblock \showarticletitle{Does the autistic child have a “theory of mind”?}
\newblock \bibinfo{journal}{\emph{Cognition}} \bibinfo{volume}{21}, \bibinfo{number}{1} (\bibinfo{year}{1985}), \bibinfo{pages}{37--46}.
\newblock


\bibitem[Benninghoff et~al\mbox{.}(2013)]%
        {benninghoff2013theory}
\bibfield{author}{\bibinfo{person}{Brenda Benninghoff}, \bibinfo{person}{Philipp Kulms}, \bibinfo{person}{Laura Hoffmann}, {and} \bibinfo{person}{Nicole~C Kr{\"a}mer}.} \bibinfo{year}{2013}\natexlab{}.
\newblock \showarticletitle{Theory of mind in human-robot-communication: Appreciated or not?}
\newblock \bibinfo{journal}{\emph{Kognitive Systeme}} \bibinfo{volume}{2013}, \bibinfo{number}{1} (\bibinfo{year}{2013}).
\newblock


\bibitem[Bernstein et~al\mbox{.}(2002)]%
        {bernstein2002complexity}
\bibfield{author}{\bibinfo{person}{Daniel~S Bernstein}, \bibinfo{person}{Robert Givan}, \bibinfo{person}{Neil Immerman}, {and} \bibinfo{person}{Shlomo Zilberstein}.} \bibinfo{year}{2002}\natexlab{}.
\newblock \showarticletitle{The complexity of decentralized control of Markov decision processes}.
\newblock \bibinfo{journal}{\emph{Mathematics of operations research}} \bibinfo{volume}{27}, \bibinfo{number}{4} (\bibinfo{year}{2002}), \bibinfo{pages}{819--840}.
\newblock


\bibitem[Brohan et~al\mbox{.}(2023)]%
        {brohan2023can}
\bibfield{author}{\bibinfo{person}{Anthony Brohan}, \bibinfo{person}{Yevgen Chebotar}, \bibinfo{person}{Chelsea Finn}, \bibinfo{person}{Karol Hausman}, \bibinfo{person}{Alexander Herzog}, \bibinfo{person}{Daniel Ho}, \bibinfo{person}{Julian Ibarz}, \bibinfo{person}{Alex Irpan}, \bibinfo{person}{Eric Jang}, \bibinfo{person}{Ryan Julian}, {et~al\mbox{.}}} \bibinfo{year}{2023}\natexlab{}.
\newblock \showarticletitle{Do as i can, not as i say: Grounding language in robotic affordances}. In \bibinfo{booktitle}{\emph{Conference on robot learning}}. PMLR, \bibinfo{pages}{287--318}.
\newblock


\bibitem[Brown(2001)]%
        {brown2001point}
\bibfield{author}{\bibinfo{person}{James~Dean Brown}.} \bibinfo{year}{2001}\natexlab{}.
\newblock \showarticletitle{Point-biserial correlation coefficients}.
\newblock \bibinfo{journal}{\emph{Statistics}} \bibinfo{volume}{5}, \bibinfo{number}{3} (\bibinfo{year}{2001}), \bibinfo{pages}{12--6}.
\newblock


\bibitem[Carroll et~al\mbox{.}(2019)]%
        {carroll2019utility}
\bibfield{author}{\bibinfo{person}{Micah Carroll}, \bibinfo{person}{Rohin Shah}, \bibinfo{person}{Mark~K Ho}, \bibinfo{person}{Tom Griffiths}, \bibinfo{person}{Sanjit Seshia}, \bibinfo{person}{Pieter Abbeel}, {and} \bibinfo{person}{Anca Dragan}.} \bibinfo{year}{2019}\natexlab{}.
\newblock \showarticletitle{On the utility of learning about humans for human-ai coordination}.
\newblock \bibinfo{journal}{\emph{Advances in neural information processing systems}}  \bibinfo{volume}{32} (\bibinfo{year}{2019}).
\newblock


\bibitem[Carruthers and Smith(1996)]%
        {carruthers1996theories}
\bibfield{author}{\bibinfo{person}{Peter Carruthers} {and} \bibinfo{person}{Peter~K Smith}.} \bibinfo{year}{1996}\natexlab{}.
\newblock \bibinfo{booktitle}{\emph{Theories of theories of mind}}.
\newblock \bibinfo{publisher}{Cambridge university press}.
\newblock


\bibitem[Chan et~al\mbox{.}(2024)]%
        {chan2024human}
\bibfield{author}{\bibinfo{person}{Szeyi Chan}, \bibinfo{person}{Shihan Fu}, \bibinfo{person}{Jiachen Li}, \bibinfo{person}{Bingsheng Yao}, \bibinfo{person}{Smit Desai}, \bibinfo{person}{Mirjana Prpa}, {and} \bibinfo{person}{Dakuo Wang}.} \bibinfo{year}{2024}\natexlab{}.
\newblock \showarticletitle{Human and LLM-Based Voice Assistant Interaction: An Analytical Framework for User Verbal and Nonverbal Behaviors}.
\newblock \bibinfo{journal}{\emph{arXiv preprint arXiv:2408.16465}} (\bibinfo{year}{2024}).
\newblock


\bibitem[Chen et~al\mbox{.}(2024)]%
        {10611018}
\bibfield{author}{\bibinfo{person}{Long Chen}, \bibinfo{person}{Oleg Sinavski}, \bibinfo{person}{Jan Hünermann}, \bibinfo{person}{Alice Karnsund}, \bibinfo{person}{Andrew~James Willmott}, \bibinfo{person}{Danny Birch}, \bibinfo{person}{Daniel Maund}, {and} \bibinfo{person}{Jamie Shotton}.} \bibinfo{year}{2024}\natexlab{}.
\newblock \showarticletitle{Driving with LLMs: Fusing Object-Level Vector Modality for Explainable Autonomous Driving}. In \bibinfo{booktitle}{\emph{2024 IEEE International Conference on Robotics and Automation (ICRA)}}. \bibinfo{pages}{14093--14100}.
\newblock
\urldef\tempurl%
\url{https://doi.org/10.1109/ICRA57147.2024.10611018}
\showDOI{\tempurl}


\bibitem[Christoforou et~al\mbox{.}(2020)]%
        {christoforou2020overview}
\bibfield{author}{\bibinfo{person}{Eftychios~G Christoforou}, \bibinfo{person}{Andreas~S Panayides}, \bibinfo{person}{Sotiris Avgousti}, \bibinfo{person}{Panicos Masouras}, {and} \bibinfo{person}{Constantinos~S Pattichis}.} \bibinfo{year}{2020}\natexlab{}.
\newblock \showarticletitle{An overview of assistive robotics and technologies for elderly care}. In \bibinfo{booktitle}{\emph{XV Mediterranean Conference on Medical and Biological Engineering and Computing--MEDICON 2019: Proceedings of MEDICON 2019, September 26-28, 2019, Coimbra, Portugal}}. Springer, \bibinfo{pages}{971--976}.
\newblock


\bibitem[Cooke et~al\mbox{.}(2020)]%
        {cooke2020framework}
\bibfield{author}{\bibinfo{person}{Nancy Cooke}, \bibinfo{person}{Mustafa Demir}, {and} \bibinfo{person}{Lixiao Huang}.} \bibinfo{year}{2020}\natexlab{}.
\newblock \showarticletitle{A framework for human-autonomy team research}. In \bibinfo{booktitle}{\emph{Engineering Psychology and Cognitive Ergonomics. Cognition and Design: 17th International Conference, EPCE 2020, Held as Part of the 22nd HCI International Conference, HCII 2020, Copenhagen, Denmark, July 19--24, 2020, Proceedings, Part II 22}}. Springer, \bibinfo{pages}{134--146}.
\newblock


\bibitem[Corbin and Strauss(2014)]%
        {corbin2014basics}
\bibfield{author}{\bibinfo{person}{Juliet Corbin} {and} \bibinfo{person}{Anselm Strauss}.} \bibinfo{year}{2014}\natexlab{}.
\newblock \bibinfo{booktitle}{\emph{Basics of Qualitative Research: Techniques and Procedures for Developing Grounded Theory}}.
\newblock \bibinfo{publisher}{Sage publications}.
\newblock


\bibitem[Cui et~al\mbox{.}(2024)]%
        {10297415}
\bibfield{author}{\bibinfo{person}{Yaodong Cui}, \bibinfo{person}{Shucheng Huang}, \bibinfo{person}{Jiaming Zhong}, \bibinfo{person}{Zhenan Liu}, \bibinfo{person}{Yutong Wang}, \bibinfo{person}{Chen Sun}, \bibinfo{person}{Bai Li}, \bibinfo{person}{Xiao Wang}, {and} \bibinfo{person}{Amir Khajepour}.} \bibinfo{year}{2024}\natexlab{}.
\newblock \showarticletitle{DriveLLM: Charting the Path Toward Full Autonomous Driving With Large Language Models}.
\newblock \bibinfo{journal}{\emph{IEEE Transactions on Intelligent Vehicles}} \bibinfo{volume}{9}, \bibinfo{number}{1} (\bibinfo{year}{2024}), \bibinfo{pages}{1450--1464}.
\newblock
\urldef\tempurl%
\url{https://doi.org/10.1109/TIV.2023.3327715}
\showDOI{\tempurl}


\bibitem[Dehkordi et~al\mbox{.}(2021)]%
        {10.1016/j.procs.2021.09.122}
\bibfield{author}{\bibinfo{person}{Maryam~Banitalebi Dehkordi}, \bibinfo{person}{Reda Mansy}, \bibinfo{person}{Abolfazl Zaraki}, \bibinfo{person}{Arpit Singh}, {and} \bibinfo{person}{Rossitza Setchi}.} \bibinfo{year}{2021}\natexlab{}.
\newblock \showarticletitle{Explainability in Human-Robot Teaming}.
\newblock \bibinfo{journal}{\emph{Procedia Comput. Sci.}} \bibinfo{volume}{192}, \bibinfo{number}{C} (\bibinfo{date}{jan} \bibinfo{year}{2021}), \bibinfo{pages}{3487–3496}.
\newblock
\showISSN{1877-0509}
\urldef\tempurl%
\url{https://doi.org/10.1016/j.procs.2021.09.122}
\showDOI{\tempurl}


\bibitem[Demir et~al\mbox{.}(2016)]%
        {7497782}
\bibfield{author}{\bibinfo{person}{Mustafa Demir}, \bibinfo{person}{Nathan~J. McNeese}, {and} \bibinfo{person}{Nancy~J. Cooke}.} \bibinfo{year}{2016}\natexlab{}.
\newblock \showarticletitle{Team communication behaviors of the human-automation teaming}. In \bibinfo{booktitle}{\emph{2016 IEEE International Multi-Disciplinary Conference on Cognitive Methods in Situation Awareness and Decision Support (CogSIMA)}}. \bibinfo{pages}{28--34}.
\newblock
\urldef\tempurl%
\url{https://doi.org/10.1109/COGSIMA.2016.7497782}
\showDOI{\tempurl}


\bibitem[Demir et~al\mbox{.}(2017)]%
        {demir2017team}
\bibfield{author}{\bibinfo{person}{Mustafa Demir}, \bibinfo{person}{Nathan~J McNeese}, {and} \bibinfo{person}{Nancy~J Cooke}.} \bibinfo{year}{2017}\natexlab{}.
\newblock \showarticletitle{Team situation awareness within the context of human-autonomy teaming}.
\newblock \bibinfo{journal}{\emph{Cognitive Systems Research}}  \bibinfo{volume}{46} (\bibinfo{year}{2017}), \bibinfo{pages}{3--12}.
\newblock


\bibitem[Demir et~al\mbox{.}(2020)]%
        {DEMIR2020102436}
\bibfield{author}{\bibinfo{person}{Mustafa Demir}, \bibinfo{person}{Nathan~J. McNeese}, {and} \bibinfo{person}{Nancy~J. Cooke}.} \bibinfo{year}{2020}\natexlab{}.
\newblock \showarticletitle{Understanding human-robot teams in light of all-human teams: Aspects of team interaction and shared cognition}.
\newblock \bibinfo{journal}{\emph{International Journal of Human-Computer Studies}}  \bibinfo{volume}{140} (\bibinfo{year}{2020}), \bibinfo{pages}{102436}.
\newblock
\showISSN{1071-5819}
\urldef\tempurl%
\url{https://doi.org/10.1016/j.ijhcs.2020.102436}
\showDOI{\tempurl}


\bibitem[Devin and Alami(2016)]%
        {devin2016implemented}
\bibfield{author}{\bibinfo{person}{Sandra Devin} {and} \bibinfo{person}{Rachid Alami}.} \bibinfo{year}{2016}\natexlab{}.
\newblock \showarticletitle{An implemented theory of mind to improve human-robot shared plans execution}. In \bibinfo{booktitle}{\emph{2016 11th ACM/IEEE International Conference on Human-Robot Interaction (HRI)}}. IEEE, \bibinfo{pages}{319--326}.
\newblock


\bibitem[Dourish and Bellotti(1992)]%
        {dourish1992awareness}
\bibfield{author}{\bibinfo{person}{Paul Dourish} {and} \bibinfo{person}{Victoria Bellotti}.} \bibinfo{year}{1992}\natexlab{}.
\newblock \showarticletitle{Awareness and coordination in shared workspaces}. In \bibinfo{booktitle}{\emph{Proceedings of the 1992 ACM conference on Computer-supported cooperative work}}. \bibinfo{pages}{107--114}.
\newblock


\bibitem[DOWNES and McMILLAN(2000)]%
        {doi:10.1177/14614440022225751}
\bibfield{author}{\bibinfo{person}{EDWARD~J. DOWNES} {and} \bibinfo{person}{SALLY~J. McMILLAN}.} \bibinfo{year}{2000}\natexlab{}.
\newblock \showarticletitle{Defining Interactivity: A Qualitative Identification of Key Dimensions}.
\newblock \bibinfo{journal}{\emph{New Media \& Society}} \bibinfo{volume}{2}, \bibinfo{number}{2} (\bibinfo{year}{2000}), \bibinfo{pages}{157--179}.
\newblock
\urldef\tempurl%
\url{https://doi.org/10.1177/14614440022225751}
\showDOI{\tempurl}
\showeprint{https://doi.org/10.1177/14614440022225751}


\bibitem[Dragicevic(2015)]%
        {dragicevic2015hci}
\bibfield{author}{\bibinfo{person}{Pierre Dragicevic}.} \bibinfo{year}{2015}\natexlab{}.
\newblock \emph{\bibinfo{title}{HCI Statistics without p-values}}.
\newblock \bibinfo{thesistype}{Ph.\,D. Dissertation}. \bibinfo{school}{Inria}.
\newblock


\bibitem[Dragicevic(2016)]%
        {dragicevic2016fair}
\bibfield{author}{\bibinfo{person}{Pierre Dragicevic}.} \bibinfo{year}{2016}\natexlab{}.
\newblock \showarticletitle{Fair statistical communication in HCI}.
\newblock \bibinfo{journal}{\emph{Modern statistical methods for HCI}} (\bibinfo{year}{2016}), \bibinfo{pages}{291--330}.
\newblock


\bibitem[Faccio et~al\mbox{.}(2023)]%
        {faccio2023human}
\bibfield{author}{\bibinfo{person}{Maurizio Faccio}, \bibinfo{person}{Irene Granata}, \bibinfo{person}{Alberto Menini}, \bibinfo{person}{Mattia Milanese}, \bibinfo{person}{Chiara Rossato}, \bibinfo{person}{Matteo Bottin}, \bibinfo{person}{Riccardo Minto}, \bibinfo{person}{Patrik Pluchino}, \bibinfo{person}{Luciano Gamberini}, \bibinfo{person}{Giovanni Boschetti}, {et~al\mbox{.}}} \bibinfo{year}{2023}\natexlab{}.
\newblock \showarticletitle{Human factors in cobot era: a review of modern production systems features}.
\newblock \bibinfo{journal}{\emph{Journal of Intelligent Manufacturing}} \bibinfo{volume}{34}, \bibinfo{number}{1} (\bibinfo{year}{2023}), \bibinfo{pages}{85--106}.
\newblock


\bibitem[(FAIR)† et~al\mbox{.}(2022)]%
        {meta2022human}
\bibfield{author}{\bibinfo{person}{Meta Fundamental AI Research Diplomacy~Team (FAIR)†}, \bibinfo{person}{Anton Bakhtin}, \bibinfo{person}{Noam Brown}, \bibinfo{person}{Emily Dinan}, \bibinfo{person}{Gabriele Farina}, \bibinfo{person}{Colin Flaherty}, \bibinfo{person}{Daniel Fried}, \bibinfo{person}{Andrew Goff}, \bibinfo{person}{Jonathan Gray}, \bibinfo{person}{Hengyuan Hu}, {et~al\mbox{.}}} \bibinfo{year}{2022}\natexlab{}.
\newblock \showarticletitle{Human-level play in the game of Diplomacy by combining language models with strategic reasoning}.
\newblock \bibinfo{journal}{\emph{Science}} \bibinfo{volume}{378}, \bibinfo{number}{6624} (\bibinfo{year}{2022}), \bibinfo{pages}{1067--1074}.
\newblock


\bibitem[Feng and Feng(2024)]%
        {feng2024extremely}
\bibfield{author}{\bibinfo{person}{Shuang Feng} {and} \bibinfo{person}{Grace Feng}.} \bibinfo{year}{2024}\natexlab{}.
\newblock \showarticletitle{An Extremely Data-efficient and Generative LLM-based Reinforcement Learning Agent for Recommenders}.
\newblock \bibinfo{journal}{\emph{arXiv preprint arXiv:2408.16032}} (\bibinfo{year}{2024}).
\newblock


\bibitem[Fleming et~al\mbox{.}(1990)]%
        {fleming1990multiple}
\bibfield{author}{\bibinfo{person}{John~H Fleming}, \bibinfo{person}{John~M Darley}, \bibinfo{person}{James~L Hilton}, {and} \bibinfo{person}{Brian~A Kojetin}.} \bibinfo{year}{1990}\natexlab{}.
\newblock \showarticletitle{Multiple audience problem: A strategic communication perspective on social perception.}
\newblock \bibinfo{journal}{\emph{Journal of personality and social psychology}} \bibinfo{volume}{58}, \bibinfo{number}{4} (\bibinfo{year}{1990}), \bibinfo{pages}{593}.
\newblock


\bibitem[Fussell et~al\mbox{.}(1998)]%
        {fussell1998teamcomm}
\bibfield{author}{\bibinfo{person}{Susan~R. Fussell}, \bibinfo{person}{Robert~E. Kraut}, \bibinfo{person}{F.~Javier Lerch}, \bibinfo{person}{William~L. Scherlis}, \bibinfo{person}{Matthew~M. McNally}, {and} \bibinfo{person}{Jonathan~J. Cadiz}.} \bibinfo{year}{1998}\natexlab{}.
\newblock \showarticletitle{Coordination, overload and team performance: effects of team communication strategies}. In \bibinfo{booktitle}{\emph{Proceedings of the 1998 ACM Conference on Computer Supported Cooperative Work}} (Seattle, Washington, USA) \emph{(\bibinfo{series}{CSCW '98})}. \bibinfo{publisher}{Association for Computing Machinery}, \bibinfo{address}{New York, NY, USA}, \bibinfo{pages}{275–284}.
\newblock
\showISBNx{1581130090}
\urldef\tempurl%
\url{https://doi.org/10.1145/289444.289502}
\showDOI{\tempurl}


\bibitem[Gero et~al\mbox{.}(2020)]%
        {gero2020mental}
\bibfield{author}{\bibinfo{person}{Katy~Ilonka Gero}, \bibinfo{person}{Zahra Ashktorab}, \bibinfo{person}{Casey Dugan}, \bibinfo{person}{Qian Pan}, \bibinfo{person}{James Johnson}, \bibinfo{person}{Werner Geyer}, \bibinfo{person}{Maria Ruiz}, \bibinfo{person}{Sarah Miller}, \bibinfo{person}{David~R Millen}, \bibinfo{person}{Murray Campbell}, {et~al\mbox{.}}} \bibinfo{year}{2020}\natexlab{}.
\newblock \showarticletitle{Mental models of AI agents in a cooperative game setting}. In \bibinfo{booktitle}{\emph{Proceedings of the 2020 chi conference on human factors in computing systems}}. \bibinfo{pages}{1--12}.
\newblock


\bibitem[Gong et~al\mbox{.}(2024)]%
        {gong-etal-2024-mindagent}
\bibfield{author}{\bibinfo{person}{Ran Gong}, \bibinfo{person}{Qiuyuan Huang}, \bibinfo{person}{Xiaojian Ma}, \bibinfo{person}{Yusuke Noda}, \bibinfo{person}{Zane Durante}, \bibinfo{person}{Zilong Zheng}, \bibinfo{person}{Demetri Terzopoulos}, \bibinfo{person}{Li Fei-Fei}, \bibinfo{person}{Jianfeng Gao}, {and} \bibinfo{person}{Hoi Vo}.} \bibinfo{year}{2024}\natexlab{}.
\newblock \showarticletitle{{M}ind{A}gent: Emergent Gaming Interaction}. In \bibinfo{booktitle}{\emph{Findings of the Association for Computational Linguistics: NAACL 2024}}, \bibfield{editor}{\bibinfo{person}{Kevin Duh}, \bibinfo{person}{Helena Gomez}, {and} \bibinfo{person}{Steven Bethard}} (Eds.). \bibinfo{publisher}{Association for Computational Linguistics}, \bibinfo{address}{Mexico City, Mexico}, \bibinfo{pages}{3154--3183}.
\newblock
\urldef\tempurl%
\url{https://doi.org/10.18653/v1/2024.findings-naacl.200}
\showDOI{\tempurl}


\bibitem[Guan et~al\mbox{.}(2023)]%
        {guan2023efficient}
\bibfield{author}{\bibinfo{person}{Cong Guan}, \bibinfo{person}{Lichao Zhang}, \bibinfo{person}{Chunpeng Fan}, \bibinfo{person}{Yichen Li}, \bibinfo{person}{Feng Chen}, \bibinfo{person}{Lihe Li}, \bibinfo{person}{Yunjia Tian}, \bibinfo{person}{Lei Yuan}, {and} \bibinfo{person}{Yang Yu}.} \bibinfo{year}{2023}\natexlab{}.
\newblock \showarticletitle{Efficient Human-AI Coordination via Preparatory Language-based Convention}.
\newblock \bibinfo{journal}{\emph{arXiv preprint arXiv:2311.00416}} (\bibinfo{year}{2023}).
\newblock


\bibitem[Guo et~al\mbox{.}(2024)]%
        {guods}
\bibfield{author}{\bibinfo{person}{Siyuan Guo}, \bibinfo{person}{Cheng Deng}, \bibinfo{person}{Ying Wen}, \bibinfo{person}{Hechang Chen}, \bibinfo{person}{Yi Chang}, {and} \bibinfo{person}{Jun Wang}.} \bibinfo{year}{2024}\natexlab{}.
\newblock \showarticletitle{DS-Agent: Automated Data Science by Empowering Large Language Models with Case-Based Reasoning}. In \bibinfo{booktitle}{\emph{Forty-first International Conference on Machine Learning}}.
\newblock


\bibitem[Han et~al\mbox{.}(2024)]%
        {han2024llm}
\bibfield{author}{\bibinfo{person}{Dongge Han}, \bibinfo{person}{Trevor McInroe}, \bibinfo{person}{Adam Jelley}, \bibinfo{person}{Stefano~V Albrecht}, \bibinfo{person}{Peter Bell}, {and} \bibinfo{person}{Amos Storkey}.} \bibinfo{year}{2024}\natexlab{}.
\newblock \showarticletitle{LLM-Personalize: Aligning LLM Planners with Human Preferences via Reinforced Self-Training for Housekeeping Robots}.
\newblock \bibinfo{journal}{\emph{arXiv preprint arXiv:2404.14285}} (\bibinfo{year}{2024}).
\newblock


\bibitem[Harada et~al\mbox{.}(2023)]%
        {harada2023behavior}
\bibfield{author}{\bibinfo{person}{Tomohiro Harada}, \bibinfo{person}{Johei Matsuoka}, {and} \bibinfo{person}{Kiyohiko Hattori}.} \bibinfo{year}{2023}\natexlab{}.
\newblock \showarticletitle{Behavior analysis of emergent rule discovery for cooperative automated driving using deep reinforcement learning}.
\newblock \bibinfo{journal}{\emph{Artificial Life and Robotics}} \bibinfo{volume}{28}, \bibinfo{number}{1} (\bibinfo{year}{2023}), \bibinfo{pages}{31--42}.
\newblock


\bibitem[Hardy and Hinkin(2022)]%
        {10.3389/fnrgo.2022.881653}
\bibfield{author}{\bibinfo{person}{David~J. Hardy} {and} \bibinfo{person}{Charles~H. Hinkin}.} \bibinfo{year}{2022}\natexlab{}.
\newblock \showarticletitle{Mental Workload in Neuropsychology: An Example With the NASA-TLX in Adults With HIV}.
\newblock \bibinfo{journal}{\emph{Frontiers in Neuroergonomics}}  \bibinfo{volume}{3} (\bibinfo{year}{2022}).
\newblock
\showISSN{2673-6195}
\urldef\tempurl%
\url{https://doi.org/10.3389/fnrgo.2022.881653}
\showDOI{\tempurl}


\bibitem[Hart et~al\mbox{.}(1968)]%
        {hart1968formal}
\bibfield{author}{\bibinfo{person}{Peter~E Hart}, \bibinfo{person}{Nils~J Nilsson}, {and} \bibinfo{person}{Bertram Raphael}.} \bibinfo{year}{1968}\natexlab{}.
\newblock \showarticletitle{A formal basis for the heuristic determination of minimum cost paths}.
\newblock \bibinfo{journal}{\emph{IEEE transactions on Systems Science and Cybernetics}} \bibinfo{volume}{4}, \bibinfo{number}{2} (\bibinfo{year}{1968}), \bibinfo{pages}{100--107}.
\newblock


\bibitem[He et~al\mbox{.}(2023)]%
        {10.1145/3544548.3580983}
\bibfield{author}{\bibinfo{person}{Ziyao He}, \bibinfo{person}{Yunpeng Song}, \bibinfo{person}{Shurui Zhou}, {and} \bibinfo{person}{Zhongmin Cai}.} \bibinfo{year}{2023}\natexlab{}.
\newblock \showarticletitle{Interaction of Thoughts: Towards Mediating Task Assignment in Human-AI Cooperation with a Capability-Aware Shared Mental Model}. In \bibinfo{booktitle}{\emph{Proceedings of the 2023 CHI Conference on Human Factors in Computing Systems}} (Hamburg, Germany) \emph{(\bibinfo{series}{CHI '23})}. \bibinfo{publisher}{Association for Computing Machinery}, \bibinfo{address}{New York, NY, USA}, Article \bibinfo{articleno}{353}, \bibinfo{numpages}{18}~pages.
\newblock
\showISBNx{9781450394215}
\urldef\tempurl%
\url{https://doi.org/10.1145/3544548.3580983}
\showDOI{\tempurl}


\bibitem[Hiatt et~al\mbox{.}(2011)]%
        {hiatt2011accommodating}
\bibfield{author}{\bibinfo{person}{Laura~M Hiatt}, \bibinfo{person}{Anthony~M Harrison}, {and} \bibinfo{person}{J~Gregory Trafton}.} \bibinfo{year}{2011}\natexlab{}.
\newblock \showarticletitle{Accommodating human variability in human-robot teams through theory of mind}. In \bibinfo{booktitle}{\emph{Twenty-Second International Joint Conference on Artificial Intelligence}}.
\newblock


\bibitem[Hoffman(2019)]%
        {hoffman2019evaluating}
\bibfield{author}{\bibinfo{person}{Guy Hoffman}.} \bibinfo{year}{2019}\natexlab{}.
\newblock \showarticletitle{Evaluating fluency in human--robot collaboration}.
\newblock \bibinfo{journal}{\emph{IEEE Transactions on Human-Machine Systems}} \bibinfo{volume}{49}, \bibinfo{number}{3} (\bibinfo{year}{2019}), \bibinfo{pages}{209--218}.
\newblock


\bibitem[Huang et~al\mbox{.}(2023)]%
        {huang2023surveyhallucinationlargelanguage}
\bibfield{author}{\bibinfo{person}{Lei Huang}, \bibinfo{person}{Weijiang Yu}, \bibinfo{person}{Weitao Ma}, \bibinfo{person}{Weihong Zhong}, \bibinfo{person}{Zhangyin Feng}, \bibinfo{person}{Haotian Wang}, \bibinfo{person}{Qianglong Chen}, \bibinfo{person}{Weihua Peng}, \bibinfo{person}{Xiaocheng Feng}, \bibinfo{person}{Bing Qin}, {and} \bibinfo{person}{Ting Liu}.} \bibinfo{year}{2023}\natexlab{}.
\newblock \bibinfo{title}{A Survey on Hallucination in Large Language Models: Principles, Taxonomy, Challenges, and Open Questions}.
\newblock
\newblock
\showeprint[arxiv]{2311.05232}~[cs.CL]
\urldef\tempurl%
\url{https://arxiv.org/abs/2311.05232}
\showURL{%
\tempurl}


\bibitem[Iantovics(2008)]%
        {iantovics2008agent}
\bibfield{author}{\bibinfo{person}{Barna~L{\'a}szl{\'o} Iantovics}.} \bibinfo{year}{2008}\natexlab{}.
\newblock \showarticletitle{Agent-based medical diagnosis systems}.
\newblock \bibinfo{journal}{\emph{Computing and Informatics}} \bibinfo{volume}{27}, \bibinfo{number}{4} (\bibinfo{year}{2008}), \bibinfo{pages}{593--625}.
\newblock


\bibitem[Jakobson(1972)]%
        {jakobson1972verbal}
\bibfield{author}{\bibinfo{person}{Roman Jakobson}.} \bibinfo{year}{1972}\natexlab{}.
\newblock \showarticletitle{Verbal communication}.
\newblock \bibinfo{journal}{\emph{Scientific American}} \bibinfo{volume}{227}, \bibinfo{number}{3} (\bibinfo{year}{1972}), \bibinfo{pages}{72--81}.
\newblock


\bibitem[Jiang et~al\mbox{.}(2023)]%
        {jiang2023situation}
\bibfield{author}{\bibinfo{person}{Jinglu Jiang}, \bibinfo{person}{Alexander~J Karran}, \bibinfo{person}{Constantinos~K Coursaris}, \bibinfo{person}{Pierre-Majorique L{\'e}ger}, {and} \bibinfo{person}{Joerg Beringer}.} \bibinfo{year}{2023}\natexlab{}.
\newblock \showarticletitle{A situation awareness perspective on human-AI interaction: Tensions and opportunities}.
\newblock \bibinfo{journal}{\emph{International Journal of Human--Computer Interaction}} \bibinfo{volume}{39}, \bibinfo{number}{9} (\bibinfo{year}{2023}), \bibinfo{pages}{1789--1806}.
\newblock


\bibitem[Julia L.~Wright and Chen(2022)]%
        {julia2022bidirectional}
\bibfield{author}{\bibinfo{person}{Shan G.~Lakhmani Julia L.~Wright} {and} \bibinfo{person}{Jessie Y.~C. Chen}.} \bibinfo{year}{2022}\natexlab{}.
\newblock \showarticletitle{Bidirectional Communications in Human-Agent Teaming: The Effects of Communication Style and Feedback}.
\newblock \bibinfo{journal}{\emph{International Journal of Human–Computer Interaction}} \bibinfo{volume}{38}, \bibinfo{number}{18-20} (\bibinfo{year}{2022}), \bibinfo{pages}{1972--1985}.
\newblock
\urldef\tempurl%
\url{https://doi.org/10.1080/10447318.2022.2068744}
\showDOI{\tempurl}
\showeprint{https://doi.org/10.1080/10447318.2022.2068744}


\bibitem[Kaptein(2016)]%
        {kaptein2016using}
\bibfield{author}{\bibinfo{person}{Maurits Kaptein}.} \bibinfo{year}{2016}\natexlab{}.
\newblock \showarticletitle{Using generalized linear (mixed) models in HCI}.
\newblock \bibinfo{journal}{\emph{Modern Statistical Methods for HCI}} (\bibinfo{year}{2016}), \bibinfo{pages}{251--274}.
\newblock


\bibitem[Kenthapadi et~al\mbox{.}(2024)]%
        {kenthapadi2024grounding}
\bibfield{author}{\bibinfo{person}{Krishnaram Kenthapadi}, \bibinfo{person}{Mehrnoosh Sameki}, {and} \bibinfo{person}{Ankur Taly}.} \bibinfo{year}{2024}\natexlab{}.
\newblock \showarticletitle{Grounding and Evaluation for Large Language Models: Practical Challenges and Lessons Learned (Survey)}. In \bibinfo{booktitle}{\emph{Proceedings of the 30th ACM SIGKDD Conference on Knowledge Discovery and Data Mining}}. \bibinfo{pages}{6523--6533}.
\newblock


\bibitem[Key and Key(1980)]%
        {key1980relationship}
\bibfield{author}{\bibinfo{person}{Mary~Ritchie Key} {and} \bibinfo{person}{Mary~Ritchie Key}.} \bibinfo{year}{1980}\natexlab{}.
\newblock \bibinfo{booktitle}{\emph{The relationship of verbal and nonverbal communication}}.
\newblock \bibinfo{publisher}{Mouton The Hague}.
\newblock


\bibitem[Kilgore and Voshell(2014)]%
        {Kilgore2014IncreasingTT}
\bibfield{author}{\bibinfo{person}{Ryan~M. Kilgore} {and} \bibinfo{person}{Martin Voshell}.} \bibinfo{year}{2014}\natexlab{}.
\newblock \showarticletitle{Increasing the Transparency of Unmanned Systems: Applications of Ecological Interface Design}. In \bibinfo{booktitle}{\emph{Interacci{\'o}n}}.
\newblock
\urldef\tempurl%
\url{https://api.semanticscholar.org/CorpusID:31347319}
\showURL{%
\tempurl}


\bibitem[Kim et~al\mbox{.}(2023)]%
        {10.1145/3544548.3581340}
\bibfield{author}{\bibinfo{person}{Taenyun Kim}, \bibinfo{person}{Maria~D. Molina}, \bibinfo{person}{Minjin~(MJ) Rheu}, \bibinfo{person}{Emily~S. Zhan}, {and} \bibinfo{person}{Wei Peng}.} \bibinfo{year}{2023}\natexlab{}.
\newblock \showarticletitle{One AI Does Not Fit All: A Cluster Analysis of the Laypeople’s Perception of AI Roles}. In \bibinfo{booktitle}{\emph{Proceedings of the 2023 CHI Conference on Human Factors in Computing Systems}} (Hamburg, Germany) \emph{(\bibinfo{series}{CHI '23})}. \bibinfo{publisher}{Association for Computing Machinery}, \bibinfo{address}{New York, NY, USA}, Article \bibinfo{articleno}{29}, \bibinfo{numpages}{20}~pages.
\newblock
\showISBNx{9781450394215}
\urldef\tempurl%
\url{https://doi.org/10.1145/3544548.3581340}
\showDOI{\tempurl}


\bibitem[Krych-Appelbaum et~al\mbox{.}(2007)]%
        {krych2007think}
\bibfield{author}{\bibinfo{person}{Meredyth Krych-Appelbaum}, \bibinfo{person}{Julie~Banzon Law}, \bibinfo{person}{Dayna Jones}, \bibinfo{person}{Allyson Barnacz}, \bibinfo{person}{Amanda Johnson}, {and} \bibinfo{person}{Julian~Paul Keenan}.} \bibinfo{year}{2007}\natexlab{}.
\newblock \showarticletitle{“I think I know what you mean”: The role of theory of mind in collaborative communication}.
\newblock \bibinfo{journal}{\emph{Interaction Studies}} \bibinfo{volume}{8}, \bibinfo{number}{2} (\bibinfo{year}{2007}), \bibinfo{pages}{267--280}.
\newblock


\bibitem[Langley et~al\mbox{.}(2022)]%
        {langley2022theory}
\bibfield{author}{\bibinfo{person}{Christelle Langley}, \bibinfo{person}{Bogdan~Ionut Cirstea}, \bibinfo{person}{Fabio Cuzzolin}, {and} \bibinfo{person}{Barbara~J Sahakian}.} \bibinfo{year}{2022}\natexlab{}.
\newblock \showarticletitle{Theory of mind and preference learning at the interface of cognitive science, neuroscience, and AI: A review}.
\newblock \bibinfo{journal}{\emph{Frontiers in artificial intelligence}}  \bibinfo{volume}{5} (\bibinfo{year}{2022}), \bibinfo{pages}{778852}.
\newblock


\bibitem[Le~Guillou et~al\mbox{.}(2023)]%
        {le2023trust}
\bibfield{author}{\bibinfo{person}{Marin Le~Guillou}, \bibinfo{person}{Laurent Pr\'{e}vot}, {and} \bibinfo{person}{Bruno Berberian}.} \bibinfo{year}{2023}\natexlab{}.
\newblock \showarticletitle{Trusting Artificial Agents: Communication Trumps Performance}. In \bibinfo{booktitle}{\emph{Proceedings of the 2023 International Conference on Autonomous Agents and Multiagent Systems}} (London, United Kingdom) \emph{(\bibinfo{series}{AAMAS '23})}. \bibinfo{publisher}{International Foundation for Autonomous Agents and Multiagent Systems}, \bibinfo{address}{Richland, SC}, \bibinfo{pages}{299–306}.
\newblock
\showISBNx{9781450394321}


\bibitem[Lea and Spears(1992)]%
        {lea1992paralanguage}
\bibfield{author}{\bibinfo{person}{Martin Lea} {and} \bibinfo{person}{Russell Spears}.} \bibinfo{year}{1992}\natexlab{}.
\newblock \showarticletitle{Paralanguage and social perception in computer-mediated communication}.
\newblock \bibinfo{journal}{\emph{Journal of Organizational Computing and Electronic Commerce}} \bibinfo{volume}{2}, \bibinfo{number}{3-4} (\bibinfo{year}{1992}), \bibinfo{pages}{321--341}.
\newblock


\bibitem[Lee et~al\mbox{.}(2012)]%
        {lee2012loosely}
\bibfield{author}{\bibinfo{person}{Soyoung Lee}, \bibinfo{person}{Charlotte Tang}, \bibinfo{person}{Sun~Young Park}, {and} \bibinfo{person}{Yunan Chen}.} \bibinfo{year}{2012}\natexlab{}.
\newblock \showarticletitle{Loosely formed patient care teams: communication challenges and technology design}. In \bibinfo{booktitle}{\emph{Proceedings of the ACM 2012 conference on computer supported cooperative work}}. \bibinfo{pages}{867--876}.
\newblock


\bibitem[Li et~al\mbox{.}(2024a)]%
        {li2024llm}
\bibfield{author}{\bibinfo{person}{Wenhao Li}, \bibinfo{person}{Zhiyuan Yu}, \bibinfo{person}{Qijin She}, \bibinfo{person}{Zhinan Yu}, \bibinfo{person}{Yuqing Lan}, \bibinfo{person}{Chenyang Zhu}, \bibinfo{person}{Ruizhen Hu}, {and} \bibinfo{person}{Kai Xu}.} \bibinfo{year}{2024}\natexlab{a}.
\newblock \showarticletitle{LLM-enhanced Scene Graph Learning for Household Rearrangement}.
\newblock \bibinfo{journal}{\emph{arXiv preprint arXiv:2408.12093}} (\bibinfo{year}{2024}).
\newblock


\bibitem[Li et~al\mbox{.}(2023)]%
        {Yang23Cole}
\bibfield{author}{\bibinfo{person}{Yang Li}, \bibinfo{person}{Shao Zhang}, \bibinfo{person}{Jichen Sun}, \bibinfo{person}{Yali Du}, \bibinfo{person}{Ying Wen}, \bibinfo{person}{Xinbing Wang}, {and} \bibinfo{person}{Wei Pan}.} \bibinfo{year}{2023}\natexlab{}.
\newblock \showarticletitle{Cooperative Open-ended Learning Framework for Zero-Shot Coordination}. In \bibinfo{booktitle}{\emph{{ICML}}} \emph{(\bibinfo{series}{Proceedings of Machine Learning Research}, Vol.~\bibinfo{volume}{202})}. \bibinfo{publisher}{{PMLR}}, \bibinfo{pages}{20470--20484}.
\newblock


\bibitem[Li et~al\mbox{.}(2024b)]%
        {li2024tackling}
\bibfield{author}{\bibinfo{person}{Yang Li}, \bibinfo{person}{Shao Zhang}, \bibinfo{person}{Jichen Sun}, \bibinfo{person}{Wenhao Zhang}, \bibinfo{person}{Yali Du}, \bibinfo{person}{Ying Wen}, \bibinfo{person}{Xinbing Wang}, {and} \bibinfo{person}{Wei Pan}.} \bibinfo{year}{2024}\natexlab{b}.
\newblock \showarticletitle{Tackling cooperative incompatibility for zero-shot human-ai coordination}.
\newblock \bibinfo{journal}{\emph{Journal of Artificial Intelligence Research}}  \bibinfo{volume}{80} (\bibinfo{year}{2024}), \bibinfo{pages}{1139--1185}.
\newblock


\bibitem[Liang et~al\mbox{.}(2019)]%
        {liang2019implicit}
\bibfield{author}{\bibinfo{person}{Claire Liang}, \bibinfo{person}{Julia Proft}, \bibinfo{person}{Erik Andersen}, {and} \bibinfo{person}{Ross~A Knepper}.} \bibinfo{year}{2019}\natexlab{}.
\newblock \showarticletitle{Implicit communication of actionable information in human-ai teams}. In \bibinfo{booktitle}{\emph{Proceedings of the 2019 CHI conference on human factors in computing systems}}. \bibinfo{pages}{1--13}.
\newblock


\bibitem[Liang et~al\mbox{.}(2023)]%
        {DBLP:conf/icra/LiangHXXHIFZ23}
\bibfield{author}{\bibinfo{person}{Jacky Liang}, \bibinfo{person}{Wenlong Huang}, \bibinfo{person}{Fei Xia}, \bibinfo{person}{Peng Xu}, \bibinfo{person}{Karol Hausman}, \bibinfo{person}{Brian Ichter}, \bibinfo{person}{Pete Florence}, {and} \bibinfo{person}{Andy Zeng}.} \bibinfo{year}{2023}\natexlab{}.
\newblock \showarticletitle{Code as Policies: Language Model Programs for Embodied Control}. In \bibinfo{booktitle}{\emph{{IEEE} International Conference on Robotics and Automation, {ICRA} 2023, London, UK, May 29 - June 2, 2023}}. \bibinfo{publisher}{{IEEE}}, \bibinfo{pages}{9493--9500}.
\newblock
\urldef\tempurl%
\url{https://doi.org/10.1109/ICRA48891.2023.10160591}
\showDOI{\tempurl}


\bibitem[Lindner and El-Assady(2022)]%
        {lindner2022humans}
\bibfield{author}{\bibinfo{person}{David Lindner} {and} \bibinfo{person}{Mennatallah El-Assady}.} \bibinfo{year}{2022}\natexlab{}.
\newblock \showarticletitle{Humans are not boltzmann distributions: Challenges and opportunities for modelling human feedback and interaction in reinforcement learning}.
\newblock \bibinfo{journal}{\emph{arXiv preprint arXiv:2206.13316}} (\bibinfo{year}{2022}).
\newblock


\bibitem[Liu et~al\mbox{.}(2024)]%
        {liu2024slow}
\bibfield{author}{\bibinfo{person}{Jijia Liu}, \bibinfo{person}{Chao Yu}, \bibinfo{person}{Jiaxuan Gao}, \bibinfo{person}{Yuqing Xie}, \bibinfo{person}{Qingmin Liao}, \bibinfo{person}{Yi Wu}, {and} \bibinfo{person}{Yu Wang}.} \bibinfo{year}{2024}\natexlab{}.
\newblock \showarticletitle{LLM-Powered Hierarchical Language Agent for Real-time Human-AI Coordination}. In \bibinfo{booktitle}{\emph{Proceedings of the 23rd International Conference on Autonomous Agents and Multiagent Systems}} (Auckland, New Zealand) \emph{(\bibinfo{series}{AAMAS '24})}. \bibinfo{publisher}{International Foundation for Autonomous Agents and Multiagent Systems}, \bibinfo{address}{Richland, SC}, \bibinfo{pages}{1219–1228}.
\newblock
\showISBNx{9798400704864}


\bibitem[Lu et~al\mbox{.}(2021)]%
        {lu2021more}
\bibfield{author}{\bibinfo{person}{Zhicong Lu}, \bibinfo{person}{Chenxinran Shen}, \bibinfo{person}{Jiannan Li}, \bibinfo{person}{Hong Shen}, {and} \bibinfo{person}{Daniel Wigdor}.} \bibinfo{year}{2021}\natexlab{}.
\newblock \showarticletitle{More kawaii than a real-person live streamer: understanding how the otaku community engages with and perceives virtual YouTubers}. In \bibinfo{booktitle}{\emph{Proceedings of the 2021 CHI Conference on Human Factors in Computing Systems}}. \bibinfo{pages}{1--14}.
\newblock


\bibitem[M.~Bran et~al\mbox{.}(2024)]%
        {m2024augmenting}
\bibfield{author}{\bibinfo{person}{Andres M.~Bran}, \bibinfo{person}{Sam Cox}, \bibinfo{person}{Oliver Schilter}, \bibinfo{person}{Carlo Baldassari}, \bibinfo{person}{Andrew~D White}, {and} \bibinfo{person}{Philippe Schwaller}.} \bibinfo{year}{2024}\natexlab{}.
\newblock \showarticletitle{Augmenting large language models with chemistry tools}.
\newblock \bibinfo{journal}{\emph{Nature Machine Intelligence}} (\bibinfo{year}{2024}), \bibinfo{pages}{1--11}.
\newblock


\bibitem[Ma et~al\mbox{.}(2023)]%
        {ma2023laser}
\bibfield{author}{\bibinfo{person}{Kaixin Ma}, \bibinfo{person}{Hongming Zhang}, \bibinfo{person}{Hongwei Wang}, \bibinfo{person}{Xiaoman Pan}, {and} \bibinfo{person}{Dong Yu}.} \bibinfo{year}{2023}\natexlab{}.
\newblock \showarticletitle{Laser: Llm agent with state-space exploration for web navigation}.
\newblock \bibinfo{journal}{\emph{arXiv preprint arXiv:2309.08172}} (\bibinfo{year}{2023}).
\newblock


\bibitem[Ma et~al\mbox{.}(2022)]%
        {ma2022elign}
\bibfield{author}{\bibinfo{person}{Zixian Ma}, \bibinfo{person}{Rose Wang}, \bibinfo{person}{Fei-Fei Li}, \bibinfo{person}{Michael Bernstein}, {and} \bibinfo{person}{Ranjay Krishna}.} \bibinfo{year}{2022}\natexlab{}.
\newblock \showarticletitle{Elign: Expectation alignment as a multi-agent intrinsic reward}.
\newblock \bibinfo{journal}{\emph{Advances in Neural Information Processing Systems}}  \bibinfo{volume}{35} (\bibinfo{year}{2022}), \bibinfo{pages}{8304--8317}.
\newblock


\bibitem[McKenna et~al\mbox{.}(2023)]%
        {10.1145/3597512.3597514}
\bibfield{author}{\bibinfo{person}{Peter~E McKenna}, \bibinfo{person}{Marta Romeo}, \bibinfo{person}{Jhielson Pimentel}, \bibinfo{person}{Mohammed Diab}, \bibinfo{person}{Meriam Moujahid}, \bibinfo{person}{Helen Hastie}, {and} \bibinfo{person}{Yiannis Demiris}.} \bibinfo{year}{2023}\natexlab{}.
\newblock \showarticletitle{Theory of Mind and Trust in Human-Robot Navigation}. In \bibinfo{booktitle}{\emph{Proceedings of the First International Symposium on Trustworthy Autonomous Systems}} (Edinburgh, United Kingdom) \emph{(\bibinfo{series}{TAS '23})}. \bibinfo{publisher}{Association for Computing Machinery}, \bibinfo{address}{New York, NY, USA}, Article \bibinfo{articleno}{29}, \bibinfo{numpages}{5}~pages.
\newblock
\showISBNx{9798400707346}
\urldef\tempurl%
\url{https://doi.org/10.1145/3597512.3597514}
\showDOI{\tempurl}


\bibitem[McMillan and Hwang(2002)]%
        {doi:10.1080/00913367.2002.10673674}
\bibfield{author}{\bibinfo{person}{Sally~J. McMillan} {and} \bibinfo{person}{Jang-Sun Hwang}.} \bibinfo{year}{2002}\natexlab{}.
\newblock \showarticletitle{Measures of Perceived Interactivity: An Exploration of the Role of Direction of Communication, User Control, and Time in Shaping Perceptions of Interactivity}.
\newblock \bibinfo{journal}{\emph{Journal of Advertising}} \bibinfo{volume}{31}, \bibinfo{number}{3} (\bibinfo{year}{2002}), \bibinfo{pages}{29--42}.
\newblock
\urldef\tempurl%
\url{https://doi.org/10.1080/00913367.2002.10673674}
\showDOI{\tempurl}
\showeprint{https://doi.org/10.1080/00913367.2002.10673674}


\bibitem[McNeese et~al\mbox{.}(2021)]%
        {doi:10.1177/15553434211017354}
\bibfield{author}{\bibinfo{person}{Nathan~J. McNeese}, \bibinfo{person}{Mustafa Demir}, \bibinfo{person}{Nancy~J. Cooke}, {and} \bibinfo{person}{Manrong She}.} \bibinfo{year}{2021}\natexlab{}.
\newblock \showarticletitle{Team Situation Awareness and Conflict: A Study of Human–Machine Teaming}.
\newblock \bibinfo{journal}{\emph{Journal of Cognitive Engineering and Decision Making}} \bibinfo{volume}{15}, \bibinfo{number}{2-3} (\bibinfo{year}{2021}), \bibinfo{pages}{83--96}.
\newblock
\urldef\tempurl%
\url{https://doi.org/10.1177/15553434211017354}
\showDOI{\tempurl}
\showeprint{https://doi.org/10.1177/15553434211017354}


\bibitem[McNeese and Reddy(2015)]%
        {mcneese2015articulating}
\bibfield{author}{\bibinfo{person}{Nathan~J McNeese} {and} \bibinfo{person}{Madhu~C Reddy}.} \bibinfo{year}{2015}\natexlab{}.
\newblock \showarticletitle{Articulating and understanding the development of a team mental model in a distributed medium}. In \bibinfo{booktitle}{\emph{Proceedings of the Human Factors and Ergonomics Society Annual Meeting}}, Vol.~\bibinfo{volume}{59}. SAGE Publications Sage CA: Los Angeles, CA, \bibinfo{pages}{240--244}.
\newblock


\bibitem[Mehrabian(2017)]%
        {mehrabian2017nonverbal}
\bibfield{author}{\bibinfo{person}{Albert Mehrabian}.} \bibinfo{year}{2017}\natexlab{}.
\newblock \bibinfo{booktitle}{\emph{Nonverbal communication}}.
\newblock \bibinfo{publisher}{Routledge}.
\newblock


\bibitem[Meltzoff(1999)]%
        {meltzoff1999origins}
\bibfield{author}{\bibinfo{person}{Andrew~N Meltzoff}.} \bibinfo{year}{1999}\natexlab{}.
\newblock \showarticletitle{Origins of theory of mind, cognition and communication}.
\newblock \bibinfo{journal}{\emph{Journal of communication disorders}} \bibinfo{volume}{32}, \bibinfo{number}{4} (\bibinfo{year}{1999}), \bibinfo{pages}{251--269}.
\newblock


\bibitem[Mooney et~al\mbox{.}(1993)]%
        {mooney1993bootstrapping}
\bibfield{author}{\bibinfo{person}{Christopher~Z Mooney}, \bibinfo{person}{Robert~D Duval}, {and} \bibinfo{person}{Robert Duvall}.} \bibinfo{year}{1993}\natexlab{}.
\newblock \bibinfo{booktitle}{\emph{Bootstrapping: A nonparametric approach to statistical inference}}.
\newblock Number~95. \bibinfo{publisher}{sage}.
\newblock


\bibitem[Napierala(2012)]%
        {napierala2012bonferroni}
\bibfield{author}{\bibinfo{person}{Matthew~A Napierala}.} \bibinfo{year}{2012}\natexlab{}.
\newblock \showarticletitle{What is the Bonferroni correction?}
\newblock \bibinfo{journal}{\emph{Aaos Now}} (\bibinfo{year}{2012}), \bibinfo{pages}{40--41}.
\newblock


\bibitem[Oliver(2019)]%
        {oliver2019communication}
\bibfield{author}{\bibinfo{person}{Stephen Oliver}.} \bibinfo{year}{2019}\natexlab{}.
\newblock \showarticletitle{Communication and trust: rethinking the way construction industry professionals and software vendors utilise computer communication mediums}.
\newblock \bibinfo{journal}{\emph{Visualization in Engineering}} \bibinfo{volume}{7}, \bibinfo{number}{1} (\bibinfo{year}{2019}), \bibinfo{pages}{1}.
\newblock


\bibitem[OpenAI(2024)]%
        {openai2024gpt4omini}
\bibfield{author}{\bibinfo{person}{OpenAI}.} \bibinfo{year}{2024}\natexlab{}.
\newblock \bibinfo{title}{GPT-4o mini: Advancing Cost-Efficient Intelligence}.
\newblock \bibinfo{howpublished}{\url{https://openai.com/index/gpt-4o-mini-advancing-cost-efficient-intelligence/}}.
\newblock
\newblock
\shownote{Accessed: 2024-09-05}.


\bibitem[Oralbayeva et~al\mbox{.}(2022)]%
        {9889428}
\bibfield{author}{\bibinfo{person}{Nurziya Oralbayeva}, \bibinfo{person}{Aidar Shakerimov}, \bibinfo{person}{Shamil Sarmonov}, \bibinfo{person}{Kanagat Kantoreyeva}, \bibinfo{person}{Fatima Dadebayeva}, \bibinfo{person}{Nuray Serkali}, {and} \bibinfo{person}{Anara Sandygulova}.} \bibinfo{year}{2022}\natexlab{}.
\newblock \showarticletitle{K-Qbot: Language Learning Chatbot Based on Reinforcement Learning}. In \bibinfo{booktitle}{\emph{2022 17th ACM/IEEE International Conference on Human-Robot Interaction (HRI)}}. \bibinfo{pages}{963--967}.
\newblock
\urldef\tempurl%
\url{https://doi.org/10.1109/HRI53351.2022.9889428}
\showDOI{\tempurl}


\bibitem[O’Neill et~al\mbox{.}(2022)]%
        {o2022human}
\bibfield{author}{\bibinfo{person}{Thomas O’Neill}, \bibinfo{person}{Nathan McNeese}, \bibinfo{person}{Amy Barron}, {and} \bibinfo{person}{Beau Schelble}.} \bibinfo{year}{2022}\natexlab{}.
\newblock \showarticletitle{Human--autonomy teaming: A review and analysis of the empirical literature}.
\newblock \bibinfo{journal}{\emph{Human factors}} \bibinfo{volume}{64}, \bibinfo{number}{5} (\bibinfo{year}{2022}), \bibinfo{pages}{904--938}.
\newblock


\bibitem[Park et~al\mbox{.}(2023)]%
        {park2023generative}
\bibfield{author}{\bibinfo{person}{Joon~Sung Park}, \bibinfo{person}{Joseph O'Brien}, \bibinfo{person}{Carrie~Jun Cai}, \bibinfo{person}{Meredith~Ringel Morris}, \bibinfo{person}{Percy Liang}, {and} \bibinfo{person}{Michael~S Bernstein}.} \bibinfo{year}{2023}\natexlab{}.
\newblock \showarticletitle{Generative agents: Interactive simulacra of human behavior}. In \bibinfo{booktitle}{\emph{Proceedings of the 36th annual acm symposium on user interface software and technology}}. \bibinfo{pages}{1--22}.
\newblock


\bibitem[Peng et~al\mbox{.}(2023)]%
        {peng2023ecological}
\bibfield{author}{\bibinfo{person}{Jiankun Peng}, \bibinfo{person}{Weiqi Chen}, \bibinfo{person}{Yi Fan}, \bibinfo{person}{Hongwen He}, \bibinfo{person}{Zhongbao Wei}, {and} \bibinfo{person}{Chunye Ma}.} \bibinfo{year}{2023}\natexlab{}.
\newblock \showarticletitle{Ecological Driving Framework of Hybrid Electric Vehicle Based on Heterogeneous Multi-Agent Deep Reinforcement Learning}.
\newblock \bibinfo{journal}{\emph{IEEE Transactions on Transportation Electrification}} \bibinfo{volume}{10}, \bibinfo{number}{1} (\bibinfo{year}{2023}), \bibinfo{pages}{392--406}.
\newblock


\bibitem[Peters(2006)]%
        {peters2006designing}
\bibfield{author}{\bibinfo{person}{Christopher Peters}.} \bibinfo{year}{2006}\natexlab{}.
\newblock \showarticletitle{Designing synthetic memory systems for supporting autonomous embodied agent behaviour}. In \bibinfo{booktitle}{\emph{ROMAN 2006-The 15th IEEE International Symposium on Robot and Human Interactive Communication}}. IEEE, \bibinfo{pages}{14--19}.
\newblock


\bibitem[Pinheiro and Bates(2006)]%
        {pinheiro2006mixed}
\bibfield{author}{\bibinfo{person}{Jos{\'e} Pinheiro} {and} \bibinfo{person}{Douglas Bates}.} \bibinfo{year}{2006}\natexlab{}.
\newblock \bibinfo{booktitle}{\emph{Mixed-effects models in S and S-PLUS}}.
\newblock \bibinfo{publisher}{Springer science \& business media}.
\newblock


\bibitem[Pinski et~al\mbox{.}(2023)]%
        {10.1145/3544548.3580794}
\bibfield{author}{\bibinfo{person}{Marc Pinski}, \bibinfo{person}{Martin Adam}, {and} \bibinfo{person}{Alexander Benlian}.} \bibinfo{year}{2023}\natexlab{}.
\newblock \showarticletitle{AI Knowledge: Improving AI Delegation through Human Enablement}. In \bibinfo{booktitle}{\emph{Proceedings of the 2023 CHI Conference on Human Factors in Computing Systems}} (Hamburg, Germany) \emph{(\bibinfo{series}{CHI '23})}. \bibinfo{publisher}{Association for Computing Machinery}, \bibinfo{address}{New York, NY, USA}, Article \bibinfo{articleno}{25}, \bibinfo{numpages}{17}~pages.
\newblock
\showISBNx{9781450394215}
\urldef\tempurl%
\url{https://doi.org/10.1145/3544548.3580794}
\showDOI{\tempurl}


\bibitem[Premack and Woodruff(1978)]%
        {premack1978does}
\bibfield{author}{\bibinfo{person}{David Premack} {and} \bibinfo{person}{Guy Woodruff}.} \bibinfo{year}{1978}\natexlab{}.
\newblock \showarticletitle{Does the chimpanzee have a theory of mind?}
\newblock \bibinfo{journal}{\emph{Behavioral and brain sciences}} \bibinfo{volume}{1}, \bibinfo{number}{4} (\bibinfo{year}{1978}), \bibinfo{pages}{515--526}.
\newblock


\bibitem[Rabinowitz et~al\mbox{.}(2018)]%
        {neil2018tom}
\bibfield{author}{\bibinfo{person}{Neil Rabinowitz}, \bibinfo{person}{Frank Perbet}, \bibinfo{person}{Francis Song}, \bibinfo{person}{Chiyuan Zhang}, \bibinfo{person}{S.~M.~Ali Eslami}, {and} \bibinfo{person}{Matthew Botvinick}.} \bibinfo{year}{2018}\natexlab{}.
\newblock \showarticletitle{Machine Theory of Mind}. In \bibinfo{booktitle}{\emph{Proceedings of the 35th International Conference on Machine Learning}} \emph{(\bibinfo{series}{Proceedings of Machine Learning Research}, Vol.~\bibinfo{volume}{80})}, \bibfield{editor}{\bibinfo{person}{Jennifer Dy} {and} \bibinfo{person}{Andreas Krause}} (Eds.). \bibinfo{publisher}{PMLR}, \bibinfo{pages}{4218--4227}.
\newblock
\urldef\tempurl%
\url{https://proceedings.mlr.press/v80/rabinowitz18a.html}
\showURL{%
\tempurl}


\bibitem[Rato et~al\mbox{.}(2022)]%
        {10.1145/3527188.3561925}
\bibfield{author}{\bibinfo{person}{Diogo Rato}, \bibinfo{person}{Marta Couto}, {and} \bibinfo{person}{Rui Prada}.} \bibinfo{year}{2022}\natexlab{}.
\newblock \showarticletitle{Attributing Social Motivations to Changes in Agents’ Behavior and Appearance}. In \bibinfo{booktitle}{\emph{Proceedings of the 10th International Conference on Human-Agent Interaction}} (Christchurch, New Zealand) \emph{(\bibinfo{series}{HAI '22})}. \bibinfo{publisher}{Association for Computing Machinery}, \bibinfo{address}{New York, NY, USA}, \bibinfo{pages}{219–226}.
\newblock
\showISBNx{9781450393232}
\urldef\tempurl%
\url{https://doi.org/10.1145/3527188.3561925}
\showDOI{\tempurl}


\bibitem[Russell and Norvig(2016)]%
        {russell2016artificial}
\bibfield{author}{\bibinfo{person}{Stuart~J Russell} {and} \bibinfo{person}{Peter Norvig}.} \bibinfo{year}{2016}\natexlab{}.
\newblock \bibinfo{booktitle}{\emph{Artificial intelligence: a modern approach}}.
\newblock \bibinfo{publisher}{Pearson}.
\newblock


\bibitem[Salikutluk et~al\mbox{.}(2024)]%
        {vildan2024auto}
\bibfield{author}{\bibinfo{person}{Vildan Salikutluk}, \bibinfo{person}{Janik Sch\"{o}pper}, \bibinfo{person}{Franziska Herbert}, \bibinfo{person}{Katrin Scheuermann}, \bibinfo{person}{Eric Frodl}, \bibinfo{person}{Dirk Balfanz}, \bibinfo{person}{Frank J\"{a}kel}, {and} \bibinfo{person}{Dorothea Koert}.} \bibinfo{year}{2024}\natexlab{}.
\newblock \showarticletitle{An Evaluation of Situational Autonomy for Human-AI Collaboration in a Shared Workspace Setting}. In \bibinfo{booktitle}{\emph{Proceedings of the CHI Conference on Human Factors in Computing Systems}} (Honolulu, HI, USA) \emph{(\bibinfo{series}{CHI '24})}. \bibinfo{publisher}{Association for Computing Machinery}, \bibinfo{address}{New York, NY, USA}, Article \bibinfo{articleno}{300}, \bibinfo{numpages}{17}~pages.
\newblock
\showISBNx{9798400703300}
\urldef\tempurl%
\url{https://doi.org/10.1145/3613904.3642564}
\showDOI{\tempurl}


\bibitem[Schelble et~al\mbox{.}(2022)]%
        {10.1145/3492832}
\bibfield{author}{\bibinfo{person}{Beau~G. Schelble}, \bibinfo{person}{Christopher Flathmann}, \bibinfo{person}{Nathan~J. McNeese}, \bibinfo{person}{Guo Freeman}, {and} \bibinfo{person}{Rohit Mallick}.} \bibinfo{year}{2022}\natexlab{}.
\newblock \showarticletitle{Let's Think Together! Assessing Shared Mental Models, Performance, and Trust in Human-Agent Teams}.
\newblock \bibinfo{journal}{\emph{Proc. ACM Hum.-Comput. Interact.}} \bibinfo{volume}{6}, \bibinfo{number}{GROUP}, Article \bibinfo{articleno}{13} (\bibinfo{date}{jan} \bibinfo{year}{2022}), \bibinfo{numpages}{29}~pages.
\newblock
\urldef\tempurl%
\url{https://doi.org/10.1145/3492832}
\showDOI{\tempurl}


\bibitem[Shank et~al\mbox{.}(2019)]%
        {10.1016/j.chb.2019.04.001}
\bibfield{author}{\bibinfo{person}{Daniel~B. Shank}, \bibinfo{person}{Christopher Graves}, \bibinfo{person}{Alexander Gott}, \bibinfo{person}{Patrick Gamez}, {and} \bibinfo{person}{Sophia Rodriguez}.} \bibinfo{year}{2019}\natexlab{}.
\newblock \showarticletitle{Feeling our way to machine minds: People's emotions when perceiving mind in artificial intelligence}.
\newblock \bibinfo{journal}{\emph{Comput. Hum. Behav.}} \bibinfo{volume}{98}, \bibinfo{number}{C} (\bibinfo{date}{sep} \bibinfo{year}{2019}), \bibinfo{pages}{256–266}.
\newblock
\showISSN{0747-5632}
\urldef\tempurl%
\url{https://doi.org/10.1016/j.chb.2019.04.001}
\showDOI{\tempurl}


\bibitem[Sharma et~al\mbox{.}(2024)]%
        {sharma2024would}
\bibfield{author}{\bibinfo{person}{Manasi Sharma}, \bibinfo{person}{Ho~Chit Siu}, \bibinfo{person}{Rohan Paleja}, {and} \bibinfo{person}{Jaime~D Pe{\~n}a}.} \bibinfo{year}{2024}\natexlab{}.
\newblock \showarticletitle{Why Would You Suggest That? Human Trust in Language Model Responses}.
\newblock \bibinfo{journal}{\emph{arXiv preprint arXiv:2406.02018}} (\bibinfo{year}{2024}).
\newblock


\bibitem[Shen et~al\mbox{.}(2024)]%
        {10.1145/3613904.3642860}
\bibfield{author}{\bibinfo{person}{Chenxinran Shen}, \bibinfo{person}{Yan Xu}, \bibinfo{person}{Ray Lc}, {and} \bibinfo{person}{Zhicong Lu}.} \bibinfo{year}{2024}\natexlab{}.
\newblock \showarticletitle{Seeking Soulmate via Voice: Understanding Promises and Challenges of Online Synchronized Voice-Based Mobile Dating}. In \bibinfo{booktitle}{\emph{Proceedings of the CHI Conference on Human Factors in Computing Systems}} (Honolulu, HI, USA) \emph{(\bibinfo{series}{CHI '24})}. \bibinfo{publisher}{Association for Computing Machinery}, \bibinfo{address}{New York, NY, USA}, Article \bibinfo{articleno}{921}, \bibinfo{numpages}{14}~pages.
\newblock
\showISBNx{9798400703300}
\urldef\tempurl%
\url{https://doi.org/10.1145/3613904.3642860}
\showDOI{\tempurl}


\bibitem[Si et~al\mbox{.}(2010)]%
        {10.1007/s10458-009-9093-x}
\bibfield{author}{\bibinfo{person}{Mei Si}, \bibinfo{person}{Stacy~C. Marsella}, {and} \bibinfo{person}{David~V. Pynadath}.} \bibinfo{year}{2010}\natexlab{}.
\newblock \showarticletitle{Modeling appraisal in theory of mind reasoning}.
\newblock \bibinfo{journal}{\emph{Autonomous Agents and Multi-Agent Systems}} \bibinfo{volume}{20}, \bibinfo{number}{1} (\bibinfo{date}{jan} \bibinfo{year}{2010}), \bibinfo{pages}{14–31}.
\newblock
\showISSN{1387-2532}
\urldef\tempurl%
\url{https://doi.org/10.1007/s10458-009-9093-x}
\showDOI{\tempurl}


\bibitem[Sonlu et~al\mbox{.}(2024)]%
        {sonlu2024effects}
\bibfield{author}{\bibinfo{person}{Sinan Sonlu}, \bibinfo{person}{Bennie Bendiksen}, \bibinfo{person}{Funda Durupinar}, {and} \bibinfo{person}{U{\u{g}}ur G{\"u}d{\"u}kbay}.} \bibinfo{year}{2024}\natexlab{}.
\newblock \showarticletitle{The Effects of Embodiment and Personality Expression on Learning in LLM-based Educational Agents}.
\newblock \bibinfo{journal}{\emph{arXiv preprint arXiv:2407.10993}} (\bibinfo{year}{2024}).
\newblock


\bibitem[Steyvers et~al\mbox{.}(2022)]%
        {steyvers2022bayesian}
\bibfield{author}{\bibinfo{person}{Mark Steyvers}, \bibinfo{person}{Heliodoro Tejeda}, \bibinfo{person}{Gavin Kerrigan}, {and} \bibinfo{person}{Padhraic Smyth}.} \bibinfo{year}{2022}\natexlab{}.
\newblock \showarticletitle{Bayesian modeling of human--AI complementarity}.
\newblock \bibinfo{journal}{\emph{Proceedings of the National Academy of Sciences}} \bibinfo{volume}{119}, \bibinfo{number}{11} (\bibinfo{year}{2022}), \bibinfo{pages}{e2111547119}.
\newblock


\bibitem[Strachan et~al\mbox{.}(2024)]%
        {strachan2024testing}
\bibfield{author}{\bibinfo{person}{James~WA Strachan}, \bibinfo{person}{Dalila Albergo}, \bibinfo{person}{Giulia Borghini}, \bibinfo{person}{Oriana Pansardi}, \bibinfo{person}{Eugenio Scaliti}, \bibinfo{person}{Saurabh Gupta}, \bibinfo{person}{Krati Saxena}, \bibinfo{person}{Alessandro Rufo}, \bibinfo{person}{Stefano Panzeri}, \bibinfo{person}{Guido Manzi}, {et~al\mbox{.}}} \bibinfo{year}{2024}\natexlab{}.
\newblock \showarticletitle{Testing theory of mind in large language models and humans}.
\newblock \bibinfo{journal}{\emph{Nature Human Behaviour}} (\bibinfo{year}{2024}), \bibinfo{pages}{1--11}.
\newblock


\bibitem[Strouse et~al\mbox{.}(2021)]%
        {strouse2021fcp}
\bibfield{author}{\bibinfo{person}{DJ Strouse}, \bibinfo{person}{Kevin McKee}, \bibinfo{person}{Matt Botvinick}, \bibinfo{person}{Edward Hughes}, {and} \bibinfo{person}{Richard Everett}.} \bibinfo{year}{2021}\natexlab{}.
\newblock \showarticletitle{Collaborating with humans without human data}.
\newblock \bibinfo{journal}{\emph{Advances in Neural Information Processing Systems}}  \bibinfo{volume}{34} (\bibinfo{year}{2021}), \bibinfo{pages}{14502--14515}.
\newblock


\bibitem[Sumers et~al\mbox{.}(2023)]%
        {sumers2023cognitive}
\bibfield{author}{\bibinfo{person}{Theodore~R Sumers}, \bibinfo{person}{Shunyu Yao}, \bibinfo{person}{Karthik Narasimhan}, {and} \bibinfo{person}{Thomas~L Griffiths}.} \bibinfo{year}{2023}\natexlab{}.
\newblock \showarticletitle{Cognitive architectures for language agents}.
\newblock \bibinfo{journal}{\emph{arXiv preprint arXiv:2309.02427}} (\bibinfo{year}{2023}).
\newblock


\bibitem[Swan et~al\mbox{.}(2023)]%
        {swan2023math}
\bibfield{author}{\bibinfo{person}{Melanie Swan}, \bibinfo{person}{Takashi Kido}, \bibinfo{person}{Eric Roland}, {and} \bibinfo{person}{Renato P~dos Santos}.} \bibinfo{year}{2023}\natexlab{}.
\newblock \showarticletitle{Math agents: Computational infrastructure, mathematical embedding, and genomics}.
\newblock \bibinfo{journal}{\emph{arXiv preprint arXiv:2307.02502}} (\bibinfo{year}{2023}).
\newblock


\bibitem[Tudor~Car et~al\mbox{.}(2020)]%
        {tudor2020conversational}
\bibfield{author}{\bibinfo{person}{Lorainne Tudor~Car}, \bibinfo{person}{Dhakshenya~Ardhithy Dhinagaran}, \bibinfo{person}{Bhone~Myint Kyaw}, \bibinfo{person}{Tobias Kowatsch}, \bibinfo{person}{Shafiq Joty}, \bibinfo{person}{Yin-Leng Theng}, {and} \bibinfo{person}{Rifat Atun}.} \bibinfo{year}{2020}\natexlab{}.
\newblock \showarticletitle{Conversational agents in health care: scoping review and conceptual analysis}.
\newblock \bibinfo{journal}{\emph{Journal of medical Internet research}} \bibinfo{volume}{22}, \bibinfo{number}{8} (\bibinfo{year}{2020}), \bibinfo{pages}{e17158}.
\newblock


\bibitem[Vicari et~al\mbox{.}(2003)]%
        {vicari2003multi}
\bibfield{author}{\bibinfo{person}{Rosa~M Vicari}, \bibinfo{person}{Cecilia~D Flores}, \bibinfo{person}{Andre~M Silvestre}, \bibinfo{person}{Louise~J Seixas}, \bibinfo{person}{Marcelo Ladeira}, {and} \bibinfo{person}{Helder Coelho}.} \bibinfo{year}{2003}\natexlab{}.
\newblock \showarticletitle{A multi-agent intelligent environment for medical knowledge}.
\newblock \bibinfo{journal}{\emph{Artificial Intelligence in Medicine}} \bibinfo{volume}{27}, \bibinfo{number}{3} (\bibinfo{year}{2003}), \bibinfo{pages}{335--366}.
\newblock


\bibitem[Wang et~al\mbox{.}(2024b)]%
        {wang2024voyager}
\bibfield{author}{\bibinfo{person}{Guanzhi Wang}, \bibinfo{person}{Yuqi Xie}, \bibinfo{person}{Yunfan Jiang}, \bibinfo{person}{Ajay Mandlekar}, \bibinfo{person}{Chaowei Xiao}, \bibinfo{person}{Yuke Zhu}, \bibinfo{person}{Linxi Fan}, {and} \bibinfo{person}{Anima Anandkumar}.} \bibinfo{year}{2024}\natexlab{b}.
\newblock \showarticletitle{Voyager: An Open-Ended Embodied Agent with Large Language Models}.
\newblock \bibinfo{journal}{\emph{Transactions on Machine Learning Research}} (\bibinfo{year}{2024}).
\newblock
\showISSN{2835-8856}
\urldef\tempurl%
\url{https://openreview.net/forum?id=ehfRiF0R3a}
\showURL{%
\tempurl}


\bibitem[Wang and Goel(2022)]%
        {wang2022mutual}
\bibfield{author}{\bibinfo{person}{Qiaosi Wang} {and} \bibinfo{person}{Ashok~K Goel}.} \bibinfo{year}{2022}\natexlab{}.
\newblock \showarticletitle{Mutual theory of mind for human-ai communication}.
\newblock \bibinfo{journal}{\emph{arXiv preprint arXiv:2210.03842}} (\bibinfo{year}{2022}).
\newblock


\bibitem[Wang et~al\mbox{.}(2021)]%
        {wang2021mtom}
\bibfield{author}{\bibinfo{person}{Qiaosi Wang}, \bibinfo{person}{Koustuv Saha}, \bibinfo{person}{Eric Gregori}, \bibinfo{person}{David Joyner}, {and} \bibinfo{person}{Ashok Goel}.} \bibinfo{year}{2021}\natexlab{}.
\newblock \showarticletitle{Towards Mutual Theory of Mind in Human-AI Interaction: How Language Reflects What Students Perceive About a Virtual Teaching Assistant}. In \bibinfo{booktitle}{\emph{Proceedings of the 2021 CHI Conference on Human Factors in Computing Systems}} (Yokohama, Japan) \emph{(\bibinfo{series}{CHI '21})}. \bibinfo{publisher}{Association for Computing Machinery}, \bibinfo{address}{New York, NY, USA}, Article \bibinfo{articleno}{384}, \bibinfo{numpages}{14}~pages.
\newblock
\showISBNx{9781450380966}
\urldef\tempurl%
\url{https://doi.org/10.1145/3411764.3445645}
\showDOI{\tempurl}


\bibitem[Wang et~al\mbox{.}(2024a)]%
        {wang2024tom}
\bibfield{author}{\bibinfo{person}{Qiaosi Wang}, \bibinfo{person}{Sarah Walsh}, \bibinfo{person}{Mei Si}, \bibinfo{person}{Jeffrey Kephart}, \bibinfo{person}{Justin~D. Weisz}, {and} \bibinfo{person}{Ashok~K. Goel}.} \bibinfo{year}{2024}\natexlab{a}.
\newblock \showarticletitle{Theory of Mind in Human-AI Interaction}. In \bibinfo{booktitle}{\emph{Extended Abstracts of the 2024 CHI Conference on Human Factors in Computing Systems}} \emph{(\bibinfo{series}{CHI EA '24})}. \bibinfo{publisher}{Association for Computing Machinery}, \bibinfo{address}{New York, NY, USA}, Article \bibinfo{articleno}{493}, \bibinfo{numpages}{6}~pages.
\newblock
\showISBNx{9798400703317}
\urldef\tempurl%
\url{https://doi.org/10.1145/3613905.3636308}
\showDOI{\tempurl}


\bibitem[Wang et~al\mbox{.}(2023)]%
        {wang2023order}
\bibfield{author}{\bibinfo{person}{Xihuai Wang}, \bibinfo{person}{Zheng Tian}, \bibinfo{person}{Ziyu Wan}, \bibinfo{person}{Ying Wen}, \bibinfo{person}{Jun Wang}, {and} \bibinfo{person}{Weinan Zhang}.} \bibinfo{year}{2023}\natexlab{}.
\newblock \showarticletitle{Order Matters: Agent-by-agent Policy Optimization}. In \bibinfo{booktitle}{\emph{The Eleventh International Conference on Learning Representations}}.
\newblock
\urldef\tempurl%
\url{https://openreview.net/forum?id=Q-neeWNVv1}
\showURL{%
\tempurl}


\bibitem[Wang et~al\mbox{.}(2024c)]%
        {wang2024zscevalevaluationtoolkitbenchmark}
\bibfield{author}{\bibinfo{person}{Xihuai Wang}, \bibinfo{person}{Shao Zhang}, \bibinfo{person}{Wenhao Zhang}, \bibinfo{person}{Wentao Dong}, \bibinfo{person}{Jingxiao Chen}, \bibinfo{person}{Ying Wen}, {and} \bibinfo{person}{Weinan Zhang}.} \bibinfo{year}{2024}\natexlab{c}.
\newblock \bibinfo{title}{ZSC-Eval: An Evaluation Toolkit and Benchmark for Multi-agent Zero-shot Coordination}.
\newblock
\newblock
\showeprint[arxiv]{2310.05208}~[cs.AI]
\urldef\tempurl%
\url{https://arxiv.org/abs/2310.05208}
\showURL{%
\tempurl}


\bibitem[Wei et~al\mbox{.}(2022)]%
        {wei2022chain}
\bibfield{author}{\bibinfo{person}{Jason Wei}, \bibinfo{person}{Xuezhi Wang}, \bibinfo{person}{Dale Schuurmans}, \bibinfo{person}{Maarten Bosma}, \bibinfo{person}{Fei Xia}, \bibinfo{person}{Ed Chi}, \bibinfo{person}{Quoc~V Le}, \bibinfo{person}{Denny Zhou}, {et~al\mbox{.}}} \bibinfo{year}{2022}\natexlab{}.
\newblock \showarticletitle{Chain-of-thought prompting elicits reasoning in large language models}.
\newblock \bibinfo{journal}{\emph{Advances in neural information processing systems}}  \bibinfo{volume}{35} (\bibinfo{year}{2022}), \bibinfo{pages}{24824--24837}.
\newblock


\bibitem[Weisz et~al\mbox{.}(2024)]%
        {weisz2024expedient}
\bibfield{author}{\bibinfo{person}{Justin~D Weisz}, \bibinfo{person}{Michael Muller}, \bibinfo{person}{Arielle Goldberg}, {and} \bibinfo{person}{Dario Andres~Silva Moran}.} \bibinfo{year}{2024}\natexlab{}.
\newblock \showarticletitle{Expedient Assistance and Consequential Misunderstanding: Envisioning an Operationalized Mutual Theory of Mind}.
\newblock \bibinfo{journal}{\emph{arXiv preprint arXiv:2406.11946}} (\bibinfo{year}{2024}).
\newblock


\bibitem[Wen et~al\mbox{.}(2019)]%
        {wen2018probabilistic}
\bibfield{author}{\bibinfo{person}{Ying Wen}, \bibinfo{person}{Yaodong Yang}, \bibinfo{person}{Rui Luo}, \bibinfo{person}{Jun Wang}, {and} \bibinfo{person}{Wei Pan}.} \bibinfo{year}{2019}\natexlab{}.
\newblock \showarticletitle{Probabilistic Recursive Reasoning for Multi-Agent Reinforcement Learning}. In \bibinfo{booktitle}{\emph{International Conference on Learning Representations}}.
\newblock
\urldef\tempurl%
\url{https://openreview.net/forum?id=rkl6As0cF7}
\showURL{%
\tempurl}


\bibitem[Wester et~al\mbox{.}(2024)]%
        {wester2024theory}
\bibfield{author}{\bibinfo{person}{Joel Wester}, \bibinfo{person}{Rune~M{\o}berg Jacobsen}, \bibinfo{person}{Sander de Jong}, \bibinfo{person}{Naja Kathrine~Kollerup Als}, \bibinfo{person}{Helena~B{\o}jer Djern{\ae}s}, {and} \bibinfo{person}{Niels van Berkel}.} \bibinfo{year}{2024}\natexlab{}.
\newblock \showarticletitle{Theory of Mind and Self-Presentation in Human-LLM Interactions}. In \bibinfo{booktitle}{\emph{Adjunct Proceedings of the ACM SIGCHI Conference on Human Factors in Computing Systems}}.
\newblock


\bibitem[Wu et~al\mbox{.}(2023)]%
        {wu2023tidybot}
\bibfield{author}{\bibinfo{person}{Jimmy Wu}, \bibinfo{person}{Rika Antonova}, \bibinfo{person}{Adam Kan}, \bibinfo{person}{Marion Lepert}, \bibinfo{person}{Andy Zeng}, \bibinfo{person}{Shuran Song}, \bibinfo{person}{Jeannette Bohg}, \bibinfo{person}{Szymon Rusinkiewicz}, {and} \bibinfo{person}{Thomas Funkhouser}.} \bibinfo{year}{2023}\natexlab{}.
\newblock \showarticletitle{Tidybot: Personalized robot assistance with large language models}.
\newblock \bibinfo{journal}{\emph{Autonomous Robots}} \bibinfo{volume}{47}, \bibinfo{number}{8} (\bibinfo{year}{2023}), \bibinfo{pages}{1087--1102}.
\newblock


\bibitem[Wu et~al\mbox{.}(2021)]%
        {gymcooking}
\bibfield{author}{\bibinfo{person}{Sarah~A. Wu}, \bibinfo{person}{Rose~E. Wang}, \bibinfo{person}{James~A. Evans}, \bibinfo{person}{Joshua~B. Tenenbaum}, \bibinfo{person}{David~C. Parkes}, {and} \bibinfo{person}{Max Kleiman-Weiner}.} \bibinfo{year}{2021}\natexlab{}.
\newblock \showarticletitle{Too Many Cooks: Bayesian Inference for Coordinating Multi-Agent Collaboration}.
\newblock \bibinfo{journal}{\emph{Topics in Cognitive Science}} \bibinfo{volume}{13}, \bibinfo{number}{2} (\bibinfo{year}{2021}), \bibinfo{pages}{414--432}.
\newblock
\urldef\tempurl%
\url{https://doi.org/10.1111/tops.12525}
\showDOI{\tempurl}
\showeprint{https://onlinelibrary.wiley.com/doi/pdf/10.1111/tops.12525}


\bibitem[Yao et~al\mbox{.}(2022a)]%
        {yao2022webshop}
\bibfield{author}{\bibinfo{person}{Shunyu Yao}, \bibinfo{person}{Howard Chen}, \bibinfo{person}{John Yang}, {and} \bibinfo{person}{Karthik Narasimhan}.} \bibinfo{year}{2022}\natexlab{a}.
\newblock \showarticletitle{Webshop: Towards scalable real-world web interaction with grounded language agents}.
\newblock \bibinfo{journal}{\emph{Advances in Neural Information Processing Systems}}  \bibinfo{volume}{35} (\bibinfo{year}{2022}), \bibinfo{pages}{20744--20757}.
\newblock


\bibitem[Yao et~al\mbox{.}(2022b)]%
        {yao2022react}
\bibfield{author}{\bibinfo{person}{Shunyu Yao}, \bibinfo{person}{Jeffrey Zhao}, \bibinfo{person}{Dian Yu}, \bibinfo{person}{Nan Du}, \bibinfo{person}{Izhak Shafran}, \bibinfo{person}{Karthik Narasimhan}, {and} \bibinfo{person}{Yuan Cao}.} \bibinfo{year}{2022}\natexlab{b}.
\newblock \showarticletitle{React: Synergizing reasoning and acting in language models}.
\newblock \bibinfo{journal}{\emph{arXiv preprint arXiv:2210.03629}} (\bibinfo{year}{2022}).
\newblock


\bibitem[Yu et~al\mbox{.}(2023)]%
        {yu23hsp}
\bibfield{author}{\bibinfo{person}{Chao Yu}, \bibinfo{person}{Jiaxuan Gao}, \bibinfo{person}{Weilin Liu}, \bibinfo{person}{Botian Xu}, \bibinfo{person}{Hao Tang}, \bibinfo{person}{Jiaqi Yang}, \bibinfo{person}{Yu Wang}, {and} \bibinfo{person}{Yi Wu}.} \bibinfo{year}{2023}\natexlab{}.
\newblock \showarticletitle{Learning Zero-Shot Cooperation with Humans, Assuming Humans Are Biased}. In \bibinfo{booktitle}{\emph{{ICLR}}}. \bibinfo{publisher}{OpenReview.net}.
\newblock


\bibitem[Zhang et~al\mbox{.}(2023a)]%
        {zhang2023comm}
\bibfield{author}{\bibinfo{person}{Rui Zhang}, \bibinfo{person}{Wen Duan}, \bibinfo{person}{Christopher Flathmann}, \bibinfo{person}{Nathan McNeese}, \bibinfo{person}{Guo Freeman}, {and} \bibinfo{person}{Alyssa Williams}.} \bibinfo{year}{2023}\natexlab{a}.
\newblock \showarticletitle{Investigating AI Teammate Communication Strategies and Their Impact in Human-AI Teams for Effective Teamwork}.
\newblock \bibinfo{journal}{\emph{Proc. ACM Hum.-Comput. Interact.}} \bibinfo{volume}{7}, \bibinfo{number}{CSCW2}, Article \bibinfo{articleno}{281} (\bibinfo{date}{oct} \bibinfo{year}{2023}), \bibinfo{numpages}{31}~pages.
\newblock
\urldef\tempurl%
\url{https://doi.org/10.1145/3610072}
\showDOI{\tempurl}


\bibitem[Zhang et~al\mbox{.}(2021a)]%
        {zhang2020ideal}
\bibfield{author}{\bibinfo{person}{Rui Zhang}, \bibinfo{person}{Nathan~J. McNeese}, \bibinfo{person}{Guo Freeman}, {and} \bibinfo{person}{Geoff Musick}.} \bibinfo{year}{2021}\natexlab{a}.
\newblock \showarticletitle{"An Ideal Human": Expectations of AI Teammates in Human-AI Teaming}.
\newblock \bibinfo{journal}{\emph{Proc. ACM Hum.-Comput. Interact.}} \bibinfo{volume}{4}, \bibinfo{number}{CSCW3}, Article \bibinfo{articleno}{246} (\bibinfo{date}{jan} \bibinfo{year}{2021}), \bibinfo{numpages}{25}~pages.
\newblock
\urldef\tempurl%
\url{https://doi.org/10.1145/3432945}
\showDOI{\tempurl}


\bibitem[Zhang et~al\mbox{.}(2023b)]%
        {zhang2023geodeepshovel}
\bibfield{author}{\bibinfo{person}{Shao Zhang}, \bibinfo{person}{Hui Xu}, \bibinfo{person}{Yuting Jia}, \bibinfo{person}{Ying Wen}, \bibinfo{person}{Dakuo Wang}, \bibinfo{person}{Luoyi Fu}, \bibinfo{person}{Xinbing Wang}, {and} \bibinfo{person}{Chenghu Zhou}.} \bibinfo{year}{2023}\natexlab{b}.
\newblock \showarticletitle{GeoDeepShovel: A platform for building scientific database from geoscience literature with AI assistance}.
\newblock \bibinfo{journal}{\emph{Geoscience Data Journal}} \bibinfo{volume}{10}, \bibinfo{number}{4} (\bibinfo{year}{2023}), \bibinfo{pages}{519--537}.
\newblock


\bibitem[Zhang et~al\mbox{.}(2024b)]%
        {10.1145/3613904.3642343}
\bibfield{author}{\bibinfo{person}{Shao Zhang}, \bibinfo{person}{Jianing Yu}, \bibinfo{person}{Xuhai Xu}, \bibinfo{person}{Changchang Yin}, \bibinfo{person}{Yuxuan Lu}, \bibinfo{person}{Bingsheng Yao}, \bibinfo{person}{Melanie Tory}, \bibinfo{person}{Lace~M. Padilla}, \bibinfo{person}{Jeffrey Caterino}, \bibinfo{person}{Ping Zhang}, {and} \bibinfo{person}{Dakuo Wang}.} \bibinfo{year}{2024}\natexlab{b}.
\newblock \showarticletitle{Rethinking Human-AI Collaboration in Complex Medical Decision Making: A Case Study in Sepsis Diagnosis}. In \bibinfo{booktitle}{\emph{Proceedings of the CHI Conference on Human Factors in Computing Systems}} (Honolulu, HI, USA) \emph{(\bibinfo{series}{CHI '24})}. \bibinfo{publisher}{Association for Computing Machinery}, \bibinfo{address}{New York, NY, USA}, Article \bibinfo{articleno}{445}, \bibinfo{numpages}{18}~pages.
\newblock
\showISBNx{9798400703300}
\urldef\tempurl%
\url{https://doi.org/10.1145/3613904.3642343}
\showDOI{\tempurl}


\bibitem[Zhang et~al\mbox{.}(2024a)]%
        {DBLP:conf/acl/ZhangTWW0HTLZ024}
\bibfield{author}{\bibinfo{person}{Wenqi Zhang}, \bibinfo{person}{Ke Tang}, \bibinfo{person}{Hai Wu}, \bibinfo{person}{Mengna Wang}, \bibinfo{person}{Yongliang Shen}, \bibinfo{person}{Guiyang Hou}, \bibinfo{person}{Zeqi Tan}, \bibinfo{person}{Peng Li}, \bibinfo{person}{Yueting Zhuang}, {and} \bibinfo{person}{Weiming Lu}.} \bibinfo{year}{2024}\natexlab{a}.
\newblock \showarticletitle{Agent-Pro: Learning to Evolve via Policy-Level Reflection and Optimization}. In \bibinfo{booktitle}{\emph{Proceedings of the 62nd Annual Meeting of the Association for Computational Linguistics (Volume 1: Long Papers), {ACL} 2024, Bangkok, Thailand, August 11-16, 2024}}, \bibfield{editor}{\bibinfo{person}{Lun{-}Wei Ku}, \bibinfo{person}{Andre Martins}, {and} \bibinfo{person}{Vivek Srikumar}} (Eds.). \bibinfo{publisher}{Association for Computational Linguistics}, \bibinfo{pages}{5348--5375}.
\newblock
\urldef\tempurl%
\url{https://aclanthology.org/2024.acl-long.292}
\showURL{%
\tempurl}


\bibitem[Zhang et~al\mbox{.}(2021b)]%
        {ijcai2021p0466}
\bibfield{author}{\bibinfo{person}{Weinan Zhang}, \bibinfo{person}{Xihuai Wang}, \bibinfo{person}{Jian Shen}, {and} \bibinfo{person}{Ming Zhou}.} \bibinfo{year}{2021}\natexlab{b}.
\newblock \showarticletitle{Model-based Multi-agent Policy Optimization with Adaptive Opponent-wise Rollouts}. In \bibinfo{booktitle}{\emph{Proceedings of the Thirtieth International Joint Conference on Artificial Intelligence, {IJCAI-21}}}, \bibfield{editor}{\bibinfo{person}{Zhi-Hua Zhou}} (Ed.). \bibinfo{publisher}{International Joint Conferences on Artificial Intelligence Organization}, \bibinfo{pages}{3384--3391}.
\newblock
\urldef\tempurl%
\url{https://doi.org/10.24963/ijcai.2021/466}
\showDOI{\tempurl}
\newblock
\shownote{Main Track}.


\bibitem[Zhang et~al\mbox{.}(2023c)]%
        {zhang2023fast}
\bibfield{author}{\bibinfo{person}{Ziqian Zhang}, \bibinfo{person}{Lei Yuan}, \bibinfo{person}{Lihe Li}, \bibinfo{person}{Ke Xue}, \bibinfo{person}{Chengxing Jia}, \bibinfo{person}{Cong Guan}, \bibinfo{person}{Chao Qian}, {and} \bibinfo{person}{Yang Yu}.} \bibinfo{year}{2023}\natexlab{c}.
\newblock \showarticletitle{Fast teammate adaptation in the presence of sudden policy change}. In \bibinfo{booktitle}{\emph{Uncertainty in Artificial Intelligence}}. PMLR, \bibinfo{pages}{2465--2476}.
\newblock


\bibitem[Zhang et~al\mbox{.}(2024c)]%
        {zhang2024simulating}
\bibfield{author}{\bibinfo{person}{Zheyuan Zhang}, \bibinfo{person}{Daniel Zhang-Li}, \bibinfo{person}{Jifan Yu}, \bibinfo{person}{Linlu Gong}, \bibinfo{person}{Jinchang Zhou}, \bibinfo{person}{Zhiyuan Liu}, \bibinfo{person}{Lei Hou}, {and} \bibinfo{person}{Juanzi Li}.} \bibinfo{year}{2024}\natexlab{c}.
\newblock \showarticletitle{Simulating Classroom Education with LLM-Empowered Agents}.
\newblock \bibinfo{journal}{\emph{arXiv preprint arXiv:2406.19226}} (\bibinfo{year}{2024}).
\newblock


\end{thebibliography}

\appendix

\newpage

\section{The Detailed Results of Questionnaires in Each Condition}\label{app:question}

In \Cref{tab:ingame}, we provide the items of the questionnaire used after each game, along with the corresponding mean human agreement levels in each group and condition, with standard deviations shown in parentheses.
\begin{table}[H]
\caption{\textbf{The Results of the Questionnaire after Each Game.} We report the mean values of the agreement the participants indicated and the standard deviation in brackets.}\label{tab:ingame}
\resizebox{\linewidth}{!}{
\begin{tabular}{l|llllllll}
\toprule
\multicolumn{1}{c}{\multirow{2}{*}{\diagbox[width=8.5cm]{Statement}{Condition}}} &
  \multicolumn{2}{c}{Bi-Comm} &
  \multicolumn{2}{c}{H-Comm} &
  \multicolumn{2}{c}{A-Comm} &
  \multicolumn{2}{c}{No-Comm} \\
\multicolumn{1}{c}{}                                & w/ MToM    & w/o MToM   & w/ MToM    & w/o MToM   & w/ MToM    & w/o MToM   & w/ MToM    & w/o MToM   \\
\midrule
The agent is good at playing this game.             & 4.25 (1.13) & 4.38 (0.65) & 4.18 (0.53) & 4.18 (0.53) & 4.06 (0.56) & 4.06 (0.56) & 4.22 (0.42) & 4.11 (0.58) \\
The agent and I are working fluently together       & 3.75  (1.27) & 3.81  (0.70) & 3.94 (0.43) & 4.24 (0.57) & 3.65 (0.87) & 3.59 (1.63) & 3.83 (0.62) & 3.78 (0.77) \\
The agent and I contributed equally to the success. & 3.75  (1.13) & 3.81  (1.23) & 4.12 (0.49) & 3.94 (0.56) & 3.94 (1.06) & 4.00 (1.13) & 4.00 (0.47) & 3.72 (0.68) \\
I was the most important team member on the team.   & 3.81  (1.23) & 3.88  (1.18) & 3.70 (0.60) & 3.70 (0.85) & 4.06 (0.93) & 3.94 (1.18) & 3.72 (0.45) & 3.94 (0.53) \\
The agent was the most important team member on the team &
  3.75 (0.87) &
  4.00 (0.80) &
  3.88 (0.61) &
  3.65 (0.74) &
  3.88 (0.99) &
  3.88 (1.23) &
  3.83 (0.50) &
  3.67 (0.82) \\
I can understand agent                              & 3.81  (1.36) & 3.81  (0.43) & 4.29 (0.47) & 4.24 (0.44) & 3.82 (1.03) & 3.65 (1.37) & 4.00 (0.47) & 3.83 (0.85) \\
I feel agent understand me                          & 3.94  (0.73) & 3.75  (1.00) & 3.59 (1.38) & 4.00 (0.88) & 3.59 (1.01) & 3.47 (1.39) & 3.83 (0.97) & 3.44 (0.85) \\
The agent and I are working on the same goal        & 4.18  (0.96) & 4.44  (0.53) & 4.24 (0.44) & 4.58 (0.38) & 4.00 (1.00) & 4.12 (1.36) & 4.22 (0.54) & 4.00 (0.59) \\
I feel that it is easy to score high in this game.  & 3.56  (1.60) & 3.81  (0.96) & 3.94 (0.43) & 4.18 (0.53) & 3.47 (1.64) & 3.47 (1.26) & 3.94 (0.76) & 3.94 (0.76)\\
\bottomrule
\end{tabular}
}
\end{table}

\section{The Open Questions in Each Condition }\label{app:open}

To gather participants' perspectives on the agents' capabilities and their communication process during the experiments, we design open questions tailored to each group’s different levels of communication interactivity.

For Bi-Comm Group:
\begin{itemize}
    \item Have you sent messages to the agents? How do you think that sending messages to the agents has affected your own actions?
    \item Do you spend time paying attention to the messages sent by the agents? How do you think reading messages to the agents has affected your own actions?
    \item How do you feel about the messages from the agents? Have you responded to the agents’ messages with your actions? If so, please provide examples.
    \item Which agent (please use hat color for reference) do you believe is able to understand and respond to your messages? Please provide examples.
        \item How do you feel that your AI teammate understands you? Please explain which AI agent and why.
    \item How do you feel that you understand your AI teammate? Please explain which AI agent and why.
\end{itemize}

For H-Comm Group:
\begin{itemize}
    \item Have you sent messages to the agents? How do you think that sending messages to the agents has affected your own actions?
    \item Which agent (please use hat color for reference) do you believe is able to understand and respond to your messages? Please provide examples.
        \item How do you feel that your AI teammate understands you? Please explain which AI agent and why.
    \item How do you feel that you understand your AI teammate? Please explain which AI agent and why.
\end{itemize}

For A-Comm Group:
\begin{itemize}
    \item Do you spend time paying attention to the messages sent by the agents? How do you think reading messages to the agents has affected your own actions?
    \item How do you feel about the messages from the agents? Have you responded to the agents’ messages with your actions? If so, please provide examples.
    \item How do you feel that your AI teammate understands you? Please explain which AI agent and why.
    \item How do you feel that you understand your AI teammate? Please explain which AI agent and why.
\end{itemize}

For No-Comm Group:
\begin{itemize}
    \item How do you feel that your AI teammate understands you? Please explain which AI agent and why.
    \item How do you feel that you understand your AI teammate? Please explain which AI agent and why.

\end{itemize}

\end{document}